\documentclass[aps,prd,reprint,onecolumn,superscriptaddress,nobibnotes,nofootinbib,floatfix,preprintnumbers,showpacs]{revtex4-1}
\usepackage{stmaryrd}
\usepackage{amssymb}
\usepackage{amsmath}

\usepackage{graphicx}
\usepackage{wrapfig}
\usepackage{psfrag}
\usepackage{dsfont}

\newcommand{\bc}{\begin{center}}
\newcommand{\ec}{\end{center}}
\newcommand{\be}{\begin{equation}}
\newcommand{\ee}{\end{equation}}
\newcommand{\bea}{\vspace{-0mm}\begin{eqnarray}}
\newcommand{\eea}{\end{eqnarray}}

\def\t0{t\text{{\ttfamily =}}0}
\def\Kappa{\text{K}}
\def\ix0{\xi\text{{\ttfamily =}}0}

\def\eps{\epsilon}

\def\lat{\text{lat}}
\def\phys{\text{phys}}
\def\GeV{\text{ GeV}}
\def\MeV{\text{ MeV}}
\def\fm{\text{ fm}}

\def\snk{\text{snk}}

\def\deltam{\delta m}

\def\t0{t\text{$=$}0}
\def\ix0{\xi\text{$=$}0}

\begin{document}

\preprint{DESY 11-102}
\preprint{Edinburgh 2011/18}
\preprint{MKPH-T-11-13}

\title{Dirac and Pauli form factors from lattice QCD}

\author{S.~Collins}  
\affiliation{Institut f\"ur Theoretische Physik, Universit\"at Regensburg,
              93040 Regensburg, Germany}
\author{M.~G\"{o}ckeler}        
\affiliation{Institut f\"ur Theoretische Physik, Universit\"at Regensburg,
              93040 Regensburg, Germany}
\author{Ph.~H\"{a}gler}
\email[]{haegler@kph.uni-mainz.de} 
\altaffiliation{Current address: Deutsches Elektronen-Synchrotron DESY, 22603 Hamburg, Germany}    
\affiliation{Institut f\"ur Kernphysik,
Johannes Gutenberg-Universit\"at Mainz,
%Johann-Joachim-Becher-Weg 45,
55128 Mainz, Germany}
\affiliation{Institut f\"ur Theoretische Physik, Universit\"at Regensburg,
              93040 Regensburg, Germany} 
\author{R.~Horsley}  
\affiliation{School of Physics, University of Edinburgh, Edinburgh EH9 3JZ,
              UK}                                       
\author{Y.~Nakamura}  
\affiliation{RIKEN Advanced Institute for Computational Science, Kobe, Hyogo 650-0047, Japan}
\author{A.~Nobile}  
\affiliation{Institut f\"ur Theoretische Physik, Universit\"at Regensburg,
              93040 Regensburg, Germany}
\author{D.~Pleiter}  
\email[]{d.pleiter@fz-juelich.de}
\affiliation{JSC, Research Center J\"{u}lich, 52425 J\"{u}lich, Germany}
\affiliation{Institut f\"ur Theoretische Physik, Universit\"at Regensburg,
              93040 Regensburg, Germany} 
\author{P.E.L.~Rakow}  
\affiliation{Theoretical Physics Division, Department of Mathematical Sciences,
              University of Liverpool, Liverpool L69 3BX, UK}
\author{A.~Sch\"{a}fer}  
\affiliation{Institut f\"ur Theoretische Physik, Universit\"at Regensburg,
              93040 Regensburg, Germany}  
\author{G.~Schierholz}  
\affiliation{Institut f\"ur Theoretische Physik, Universit\"at Regensburg,
              93040 Regensburg, Germany} 
\affiliation{Deutsches Elektronen-Synchrotron DESY, 22603 Hamburg, Germany}     
\author{W.~Schroers}  
\affiliation{Numerik \& Analyse Schroers, Stubenrauchstr.~3, 12357 Berlin, Germany}           
\author{H.~St\"{u}ben}  
\affiliation{Konrad-Zuse-Zentrum f\"ur Informationstechnik Berlin,
              14195 Berlin, Germany} 
\author{F.~Winter} 
\affiliation{School of Physics, University of Edinburgh, Edinburgh EH9 3JZ,
              UK} 
\author{J.M.~Zanotti} 
\email[]{jzanotti@ph.ed.ac.uk} 
\affiliation{School of Physics, University of Edinburgh, Edinburgh EH9 3JZ,
              UK}    

\collaboration{QCDSF/UKQCD Collaboration}

%\date{\today}

\begin{abstract}
We present a comprehensive analysis of the electromagnetic form
factors of the nucleon from a lattice simulation with two flavors of
dynamical ${\cal O}(a)$-improved Wilson fermions.
A key feature of our calculation is that we make use of an extensive
ensemble of lattice gauge field configurations with four different
lattice spacings, multiple volumes, and pion masses down to
$m_\pi\sim 180$~MeV.
We find that by employing Kelly-inspired parametrizations for the
$Q^2$-dependence of the form factors, we are able to obtain stable
fits over our complete ensemble.
Dirac and Pauli radii and the anomalous magnetic moments of the nucleon
are extracted and results at light quark masses provide evidence
for chiral non-analytic behavior in these fundamental observables.
\end{abstract}


\maketitle

\section{Introduction}
Nucleon electromagnetic form factors are fundamental quantities and
reveal important information on the spatial distribution of charge and
magnetization within a nucleon
\cite{Gao:2003ag,HydeWright:2004gh,Arrington:2006zm,Perdrisat:2006hj,Arrington:2011kb}.
Understanding the nucleon electromagnetic structure in terms of the
underlying quark and gluon degrees of freedom of quantum
chromodynamics is a challenging task which has attracted the attention
of both theory and experiment for many years.

For a long time, the overall trend of the experimental results for
small and moderate values of the momentum transfer $Q^2=-q^2$ could be
described reasonably well by phenomenological (dipole) fits
\begin{eqnarray}
  G_E^p(Q^2) &\sim& \frac{G_M^p(Q^2)}{\mu^p} \sim
  \frac{G_M^n(Q^2)}{\mu^n} \nonumber\\
  &\sim& (1 + Q^2/m^2_D)^{-2}\ ,\nonumber\\
  G_E^n(Q^2) &\sim& 0\ ,
  \label{eq:3}
\end{eqnarray}
with $m_D\sim 0.84$~GeV and the magnetic moments
\begin{equation}
  \mu^p\sim 2.79\ ,\ \mu^n\sim -1.91\ ,
  \label{eq:1}  
\end{equation}
in units of nuclear magnetons.
More recently, the improved accuracy of the experimental data allows
us to see clear deviations from this dipole behavior in the region of
low and intermediate $Q^2$.
This has led to a significant amount of theoretical work aimed at
describing these form factors, such as dispersion theory analysis
\cite{Hohler:1976ax,Mergell:1995bf,Belushkin:2006qa}, vector meson
exchange/dominance \cite{Bijker:2004yu,Crawford:2010gv} and polynomial
fits based on the Kelly parametrization
\cite{Kelly:2004hm}.
We will apply a simplified variant of the latter to lattice results in this
paper.
Interest in these form factors has been revived over the last 10 years
by experiments at Jefferson Laboratory which found an unexpected
dependence of the nucleon's electric and magnetic form factors on the
momentum transferred to the target nucleon 
\cite{Jones:1999rz,Zhan:2011ji,Ron:2011rd}.
More recently, a measurement of the Lamb shift in muonic hydrogen \cite{Pohl:2010zz} has
produced a result for the electric radius of the proton that is several sigma below
the PDG (CODATA) value \cite{Nakamura:2010zzi,Mohr:2008fa}.
At the same time, a new high-precision determination of the proton form factors 
from $ep$-scattering experiments at MAMI has been reported \cite{Bernauer:2010wm}, which confirms
the traditional results for the electric (and magnetic) mean square radius.

From a lattice perspective, it is common to evaluate fundamental
observables like the charge radii and anomalous magnetic moments in
order to make a comparison with experimental and theoretical results.
A feature of any lattice simulation is that the quark mass is an input
parameter, hence it is possible to map out the form factors not only
as a function of $Q^2$, but also $m_\pi^2$.
Baryon charge radii and magnetic moments are of particular interest in
this case as predictions from chiral perturbation theory (ChPT)
indicate that these quantities should provide an excellent opportunity
to observe the chiral non-analytic behavior of QCD
\cite{Bernard:1992qa,Bernard:1995dp,Bernard:1998gv,Hemmert:2002uh,Gockeler:2003ay,Young:2004tb}.
An additional advantage of a lattice simulation of nucleon
electromagnetic form factors is that since they are performed at the
quark level, it is possible to determine the individual up and
down quark contributions, providing valuable insights into the
distribution of charge and magnetization within a nucleon.

These issues are now beginning to be addressed in modern lattice simulations 
\cite{Gockeler:2003ay,Boinepalli:2006xd,Syritsyn:2009mx,Yamazaki:2009zq,Bratt:2010jn,Alexandrou:2011db}
(see also \cite{Hagler:2009ni} for a review).
A common feature of present lattice simulations with unphysical quark masses is that they 
tend to underestimate the experimental and phenomenological results for the
radii and magnetic moments of the nucleon.
As mentioned above, predictions from ChPT indicate that these
observables should exhibit a dramatic non-analytic dependence on the
quark mass close to the chiral limit, however such features have yet
to be seen clearly in a lattice simulation with dynamical quarks.

In this paper, we will confront these issues through simulations with
pion masses as low as $m_\pi\sim 180$~MeV.
In our analysis, we place a strong emphasis on addressing the
systematic errors present in a lattice simulation, such as finite
volume and lattice spacing effects.
We also consider the effects of finite momentum resolution in lattice
determinations of form factors through the use of several
parametrizations of the momentum dependence.

\section{Lattice setup and methods}

Below we briefly describe our lattice setup and the methods that we have used to compute the nucleon form factors.

\subsection{Simulation parameters}

We perform our simulations with two flavors of non-perturbatively
${\mathcal{O}(a)}$-improved Wilson (Clover) fermions and Wilson glue.
Using these actions, we have generated gauge field configurations with
the parameters given in Table~\ref{tab:params}, where we have used the
Sommer parameter with $r_0=0.5$~fm to set the physical scale
\cite{Najjar:2011}.
Summarizing these parameters, we see that our four values of
$\beta=5.20,\,5.25\,,5.29\,,5.40$, correspond to lattice spacings in
the range $0.06<a<0.1$~fm, allowing for the approach to the continuum
limit to be assessed, while a range of lattice volumes
($0.9<L<3.0$~fm) enable us to study finite size effects in our
simulations.
Finally, our pion masses now reach well into the chiral regime, down
to $m_\pi\sim 180$~MeV, allowing us to investigate the
applicability of different ChPT approaches around and above the
physical pion mass, and to search for chiral non-analytic behavior in
our results.
\begin{table}
\begin{tabular}{c|c|c|c|c|c|c|c}
\hline
$\beta$ & \# & $\kappa$ & $N^3\times T$ & $m_\pi$ [GeV] & $a$ [fm] &
$L$ [fm] & $N_{\mathrm{traj}}$ \\
\hline
 5.20 &  1 & 0.13420 & $16^3\times 32$ & 1.40 & 0.083 & 1.3 & ${\mathcal{O}(5000)}$ \\
 5.20 &  2 & 0.13500 & $16^3\times 32$ & 0.99 &       & 1.3 & ${\mathcal{O}(8000)}$ \\
 5.20 &  3 & 0.13550 & $16^3\times 32$ & 0.69 &       & 1.3 & ${\mathcal{O}(8000)}$ \\
\hline
 5.25 &  4 & 0.13460 & $16^3\times 32$ & 1.29 & 0.076 & 1.2 & ${\mathcal{O}(6000)}$ \\
 5.25 &  5 & 0.13520 & $16^3\times 32$ & 1.00 &       & 1.2 & ${\mathcal{O}(8000)}$ \\
 5.25 &  6 & 0.13575 & $24^3\times 48$ & 0.67 &       & 1.8 & ${\mathcal{O}(6000)}$ \\
 5.25 &  7 & 0.13600 & $24^3\times 48$ & 0.48 &       & 1.8 & ${\mathcal{O}(5000)}$ \\
\hline
 5.29 &  8 & 0.13400 & $16^3\times 32$ & 1.59 & 0.072 & 1.1 & ${\mathcal{O}(4000)}$ \\
 5.29 &  9 & 0.13500 & $16^3\times 32$ & 1.16 &       & 1.1 & ${\mathcal{O}(5500)}$ \\
 5.29 & 10 & 0.13550 & $12^3\times 32$ & 0.99 &       & 0.9 & ${\mathcal{O}(4500)}$ \\
 5.29 & 11 & 0.13550 & $16^3\times 32$ & 0.92 &       & 1.1 & ${\mathcal{O}(5000)}$ \\
 5.29 & 12 & 0.13550 & $24^3\times 48$ & 0.90 &       & 1.7 & ${\mathcal{O}(2000)}$ \\
 5.29 & 13 & 0.13590 & $12^3\times 32$ & 0.93 &       & 0.9 & ${\mathcal{O}(5500)}$ \\
 5.29 & 14 & 0.13590 & $16^3\times 32$ & 0.69 &       & 1.1 & ${\mathcal{O}(7000)}$ \\
 5.29 & 15 & 0.13590 & $24^3\times 48$ & 0.66 &       & 1.7 & ${\mathcal{O}(6000)}$ \\
 5.29 & 15 & 0.13620 & $24^3\times 48$ & 0.43 &       & 1.7 & ${\mathcal{O}(5500)}$ \\
 5.29 & 17 & 0.13632 & $24^3\times 48$ & 0.31 &       & 1.7 & ${\mathcal{O}(7000)}$ \\
 5.29 & 18 & 0.13632 & $32^3\times 64$ & 0.30 &       & 2.3 & ${\mathcal{O}(2700)}$ \\
 5.29 & 19 & 0.13632 & $40^3\times 64$ & 0.29 &       & 2.9 & ${\mathcal{O}(2000)}$ \\
 5.29 & 20 & 0.13640 & $40^3\times 64$ & 0.18 &       & 2.9 & ${\mathcal{O}(1000)}$ \\
\hline
 5.40 & 21 & 0.13500 & $24^3\times 48$ & 1.32 & 0.060 & 1.4 & ${\mathcal{O}(3500)}$ \\
 5.40 & 22 & 0.13560 & $24^3\times 48$ & 1.02 &       & 1.4 & ${\mathcal{O}(3500)}$ \\
 5.40 & 23 & 0.13610 & $24^3\times 48$ & 0.72 &       & 1.4 & ${\mathcal{O}(4000)}$ \\
 5.40 & 24 & 0.13625 & $24^3\times 48$ & 0.62 &       & 1.4 & ${\mathcal{O}(6000)}$ \\
 5.40 & 25 & 0.13640 & $24^3\times 48$ & 0.50 &       & 1.4 & ${\mathcal{O}(2500)}$ \\
 5.40 & 26 & 0.13640 & $32^3\times 64$ & 0.49 &       & 1.9 & ${\mathcal{O}(2500)}$ \\
 5.40 & 27 & 0.13660 & $32^3\times 64$ & 0.28 &       & 1.9 & ${\mathcal{O}(2800)}$ \\
 5.40 & 28 & 0.13660 & $48^3\times 64$ & 0.26 &       & 2.9 & ${\mathcal{O}(2200)}$ \\
\hline
\end{tabular}
\caption{Overview of our simulation parameters where we have used the
  Sommer parameter with $r_0=0.5\fm$ to set the physical scale. 
  %The error on $m_\pi$ is purely statistical.
  }
\label{tab:params}
\end{table}
When computing correlation functions on these configurations, we
generally over-sample using up to 4 different locations of the
fermion source on a single configuration.
We then use binning to obtain an effective distance of 20
trajectories.
We find that beyond this, the size of the bins has little effect on
the error, which indicates residual auto-correlations are small.

\subsection{Extraction of form factors}

On the lattice, we determine the form factors $F_1(Q^2)$ and
$F_2(Q^2)$ by calculating the following matrix element of the
electromagnetic current
\be
\langle p',\,s'| j^{\mu}(0)|p,\,s\rangle
\, = 
 \bar{U}(p',\,s')
 \left[ \gamma^\mu F_1(q^2) +
       i\sigma^{\mu\nu}\frac{q_\nu}{2m_N}F_2(q^2) \right] 
 U(p,\,s) \, ,
\label{eq:em-me}
\ee
where $U(p,\,s)$ is a Dirac spinor with momentum $p$ and spin
polarization $s$, $q = p' - p$ is the momentum transfer, $m_N$ is
the nucleon mass and $j_\mu$ is the electromagnetic current.
The Dirac, $F_1$, and Pauli, $F_2$, form factors of the proton are
obtained by using 
\begin{equation}
  \label{eq:pcur}
j_\mu^{(p)} = \frac{2}{3}\bar{u}\gamma_\mu u -
\frac{1}{3}\bar{d}\gamma_\mu d,  
\end{equation}
between proton states.
The isovector form factors are also obtained from proton states, but
with the current
\begin{equation}
  \label{eq:isovcur}
j_\mu^v = \bar{u}\gamma_\mu u - \bar{d}\gamma_\mu d\ .  
\end{equation}
Similarly, we used the isoscalar current $\bar{u}\gamma_\mu u +
\bar{d}\gamma_\mu d$ for the computation of isoscalar form factors.

In electron scattering, it is common to rewrite the form factors $F_1$
and $F_2$ in terms of the electric and magnetic Sachs form factors,
\begin{eqnarray}
G_E(Q^2)&=& F_1(Q^2) - \frac{Q^2}{(2m_N)^2}\, F_2(Q^2)\,, \nonumber\\ 
G_M(Q^2)&=& F_1(Q^2) + F_2(Q^2) \,,
  \label{eq:sachs}
\end{eqnarray}
as then the (unpolarized) cross section becomes a linear combination
of squares of the form factors.

For, e.g., the proton
$F_1^{(p)}(0) = G_E^{(p)}(0) =1$ gives the electric charge,
while $G_M^{(p)}(0) = \mu^{(p)} = 1 + \kappa^{(p)}$
gives the magnetic moment, where $F_2^{(p)}(0) = \kappa^{(p)}$ is the
anomalous magnetic moment.
For a classical point particle, both form factors are independent of
$Q^2$, so deviations from this behavior tell us something about the
extended nature of the nucleon.

In our lattice study, we use the standard proton interpolating field for a
proton with momentum $\vec{p}$
\begin{eqnarray}
B_\alpha(t,\vec{p}) &=& \sum_{\vec x,x_4=t}
 e^{-\mathrm{i}\vec{p}\cdot\vec{x}} \epsilon_{ijk} 
 u_\alpha^i(x)u_\beta^j(x) (C\gamma_5)_{\beta\gamma} d_\gamma^k(x)\,,
 \nonumber\\
\bar{B}_\alpha(t,\vec{p}) &=& \sum_{\vec x,x_4=t}
 e^{\mathrm{i}\vec{p}\cdot\vec{x}} \epsilon_{ijk} 
 \bar{d}_\beta^i(x) (C\gamma_5)_{\beta\gamma} \bar{u}_\gamma^j(x)
 \bar{u}_\alpha^k(x)\,,
  \label{eq:interpolators}  
\end{eqnarray}
where $C$ is the charge conjugation matrix, $i\,,j\,,k$ are color
indices and $\alpha\,,\beta\,,\gamma$ are Dirac indices.

In order to improve the overlap of these interpolating fields with the
ground state proton, we employ two improvements: Jacobi smearing and
non-relativistic projection.
The latter of these has the additional advantage that we only need to perform
$2 \times 3$ inversions rather than the usual $4 \times 3$, since we only 
consider the first two Dirac components.

The matrix elements in Eq.~(\ref{eq:em-me}) are obtained from ratios
of three-point to two-point functions,
\begin{equation}
R(t,\tau;\vec{p}\,',\vec{p};{\cal O})\, =
 \frac{C_{3pt}^\Gamma (t,\tau;\vec{p}\,', \vec{p},{\cal O})} 
        {C_{2pt}(t,\vec{p}\,')}
 \left[
  \frac{C_{2pt}(\tau,\vec{p}\,') C_{2pt}(t,\vec{p}\,') C_{2pt}(t-\tau,\vec{p})}
  {C_{2pt}(\tau,\vec{p}) C_{2pt}(t,\vec{p}) C_{2pt}(t-\tau,\vec{p}\,')}
\right]^{\frac{1}{2} } \, ,
\label{eq:ratio}
\end{equation}
for large time separations, $0 \ll \tau \ll t \lesssim
\frac{1}{2} L_T $, where $L_T$ is the temporal extent of our lattice.
The nucleon two- and three-point functions are given,
respectively, by
\begin{eqnarray}
C_{2pt}(\tau,\vec{p}) &=& {\rm Tr}\left[\Gamma_{\rm unpol} 
\langle B(\tau,\vec{p}) \overline{B}(0,\vec{p})\rangle\right] \,,
\nonumber \\
C_{3pt}^\Gamma (t,\tau;\vec{p}\,', \vec{p},{\cal O}) &=& {\rm
  Tr}\left[\Gamma
\langle B(t,\vec{p}\,') {\cal O}(\vec{q},\tau)
\overline{B}(0,\vec{p})\rangle\right] \,. 
\label{eq:2-3ptfns}
\end{eqnarray}
Here $t$ and $\tau$ are the Euclidean times of the nucleon sink and
operator insertion, respectively, $\vec{p}\,'\ (\vec{p})$ is the nucleon
momentum at the sink (source), and ${\cal O}$ is the local vector current
\begin{equation}
{\cal O}_\mu(\vec{q},\tau) = \sum_{\vec{x}} e^{i\vec{q}\cdot\vec{x}}\bar q(\vec{x},\tau)\gamma_\mu q(\vec{x},\tau)\ .
\end{equation}
The trace in Eq.~(\ref{eq:2-3ptfns}) is over spinor indices and the
$\Gamma$ matrix in the three-point function is one
of
\begin{eqnarray}
\Gamma_{\rm unpol} &=& \frac{1}{2}(1+\gamma_4)\ , \\
\Gamma_1 &=& \frac{1}{2}(1+\gamma_4){\rm i}\gamma_5\gamma_1\ , \\
\Gamma_2 &=& \frac{1}{2}(1+\gamma_4){\rm i}\gamma_5\gamma_2\ .
\end{eqnarray}
We simulate with three different sink momenta $\vec{p}\,'$
\begin{equation}
\frac{L}{2\pi}\vec{p}\,' = (0,\,0,\,0),\ (1,\,0,\,0),\
 (0,\,1,\,0)\ .
\end{equation}
Finally, we use 17 different momentum transfers
$\vec{q}=\vec{p}\,'-\vec{p}$.
Equations with identical values of virtual momentum transfer $q^2$ are
combined to return the optimal statistics available at each working
point.
This procedure is outlined in more detail in
Ref.~\cite{Gockeler:2003ay}.

Note that quark line disconnected contributions to the three-point function in Eq.~(\ref{eq:2-3ptfns}),
which are relevant for the flavor singlet observables but cancel out in the isovector case,
have not been included in our study.

\section{Numerical results}
In the following, we present and discuss in some detail our numerical results for the Dirac and Pauli nucleon form factors.

\begin{figure}[t]
    \begin{minipage}{0.48\textwidth}
        \centering
          \includegraphics[angle=0,width=0.9\textwidth,clip=true,angle=0]{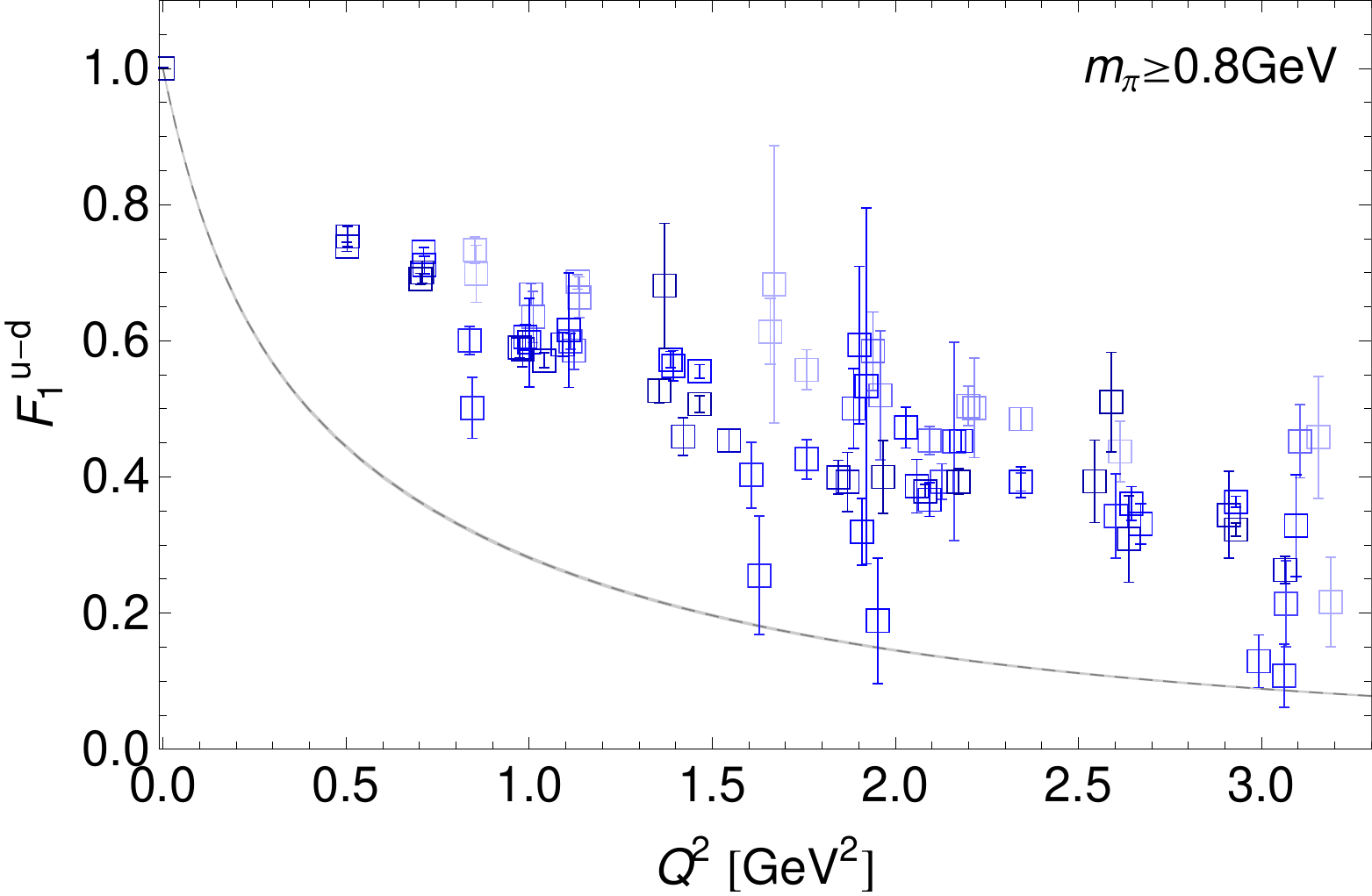}
          \includegraphics[angle=0,width=0.9\textwidth,clip=true,angle=0]{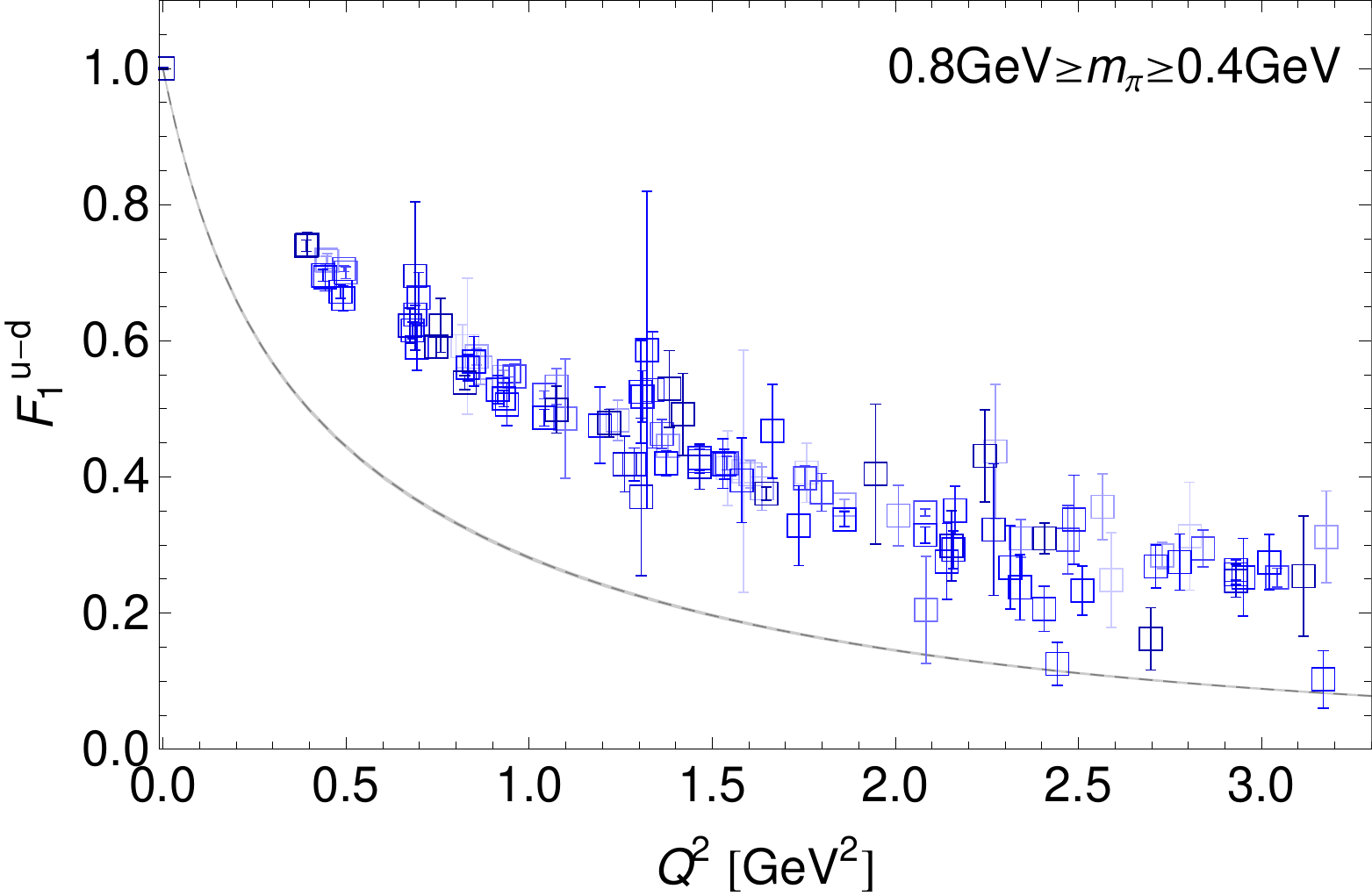}
          \includegraphics[angle=0,width=0.9\textwidth,clip=true,angle=0]{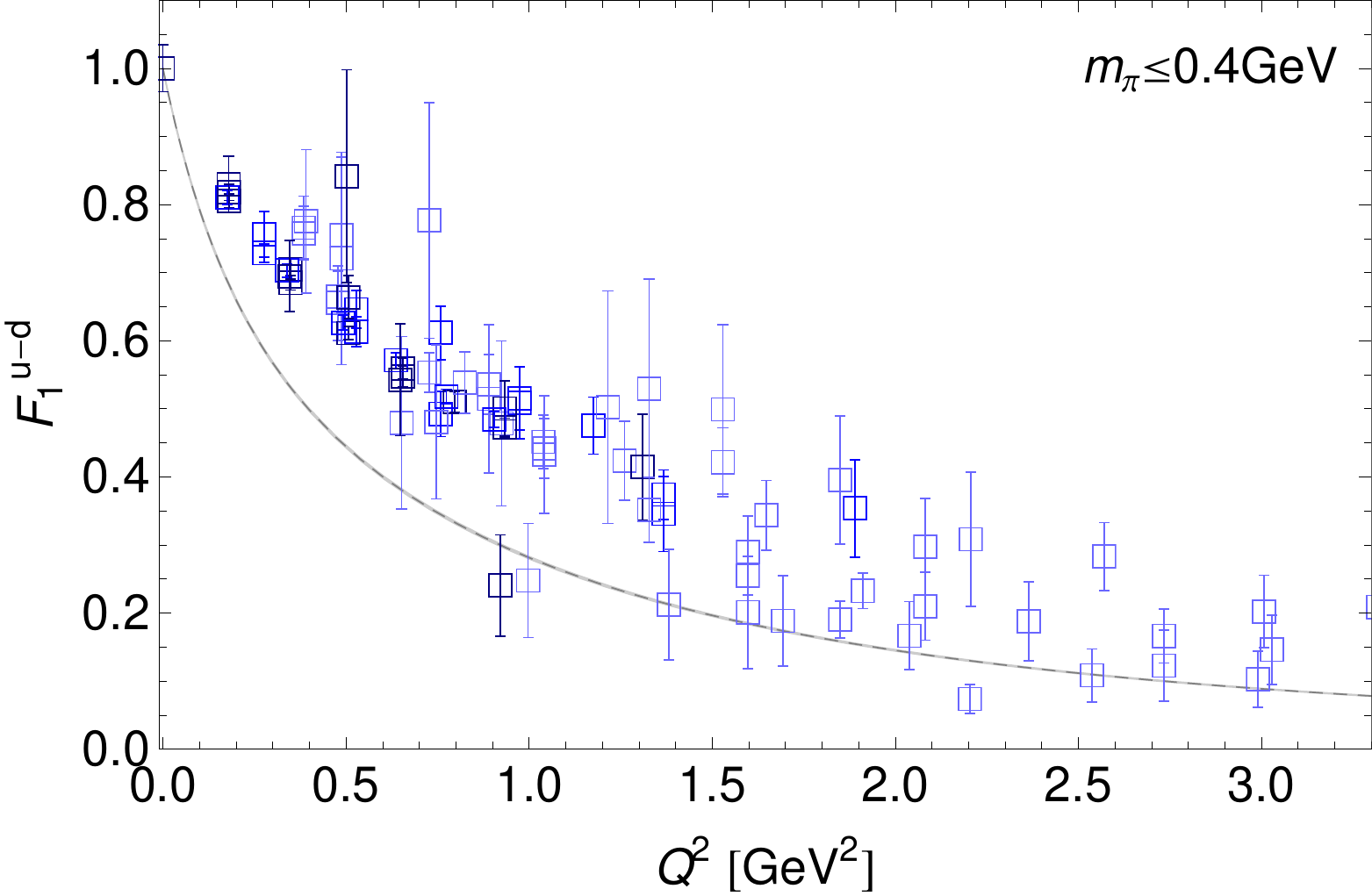}
  \caption{Dirac form factor $F_1(Q^2)$ in the isovector channel. 
  All ensembles are included, and darker colors correspond to lighter pion masses.
  The gray shaded band represents the parametrization by Alberico et al. \cite{Alberico:2008sz} of the experimental data.
  %A legend is provided in Fig.~\ref{F2v}.
  }
  \label{F1v}
     \end{minipage} 
         \hspace{0.2cm}
    \begin{minipage}{0.48\textwidth}
      \centering
          \includegraphics[angle=0,width=0.89\textwidth,clip=true,angle=0]{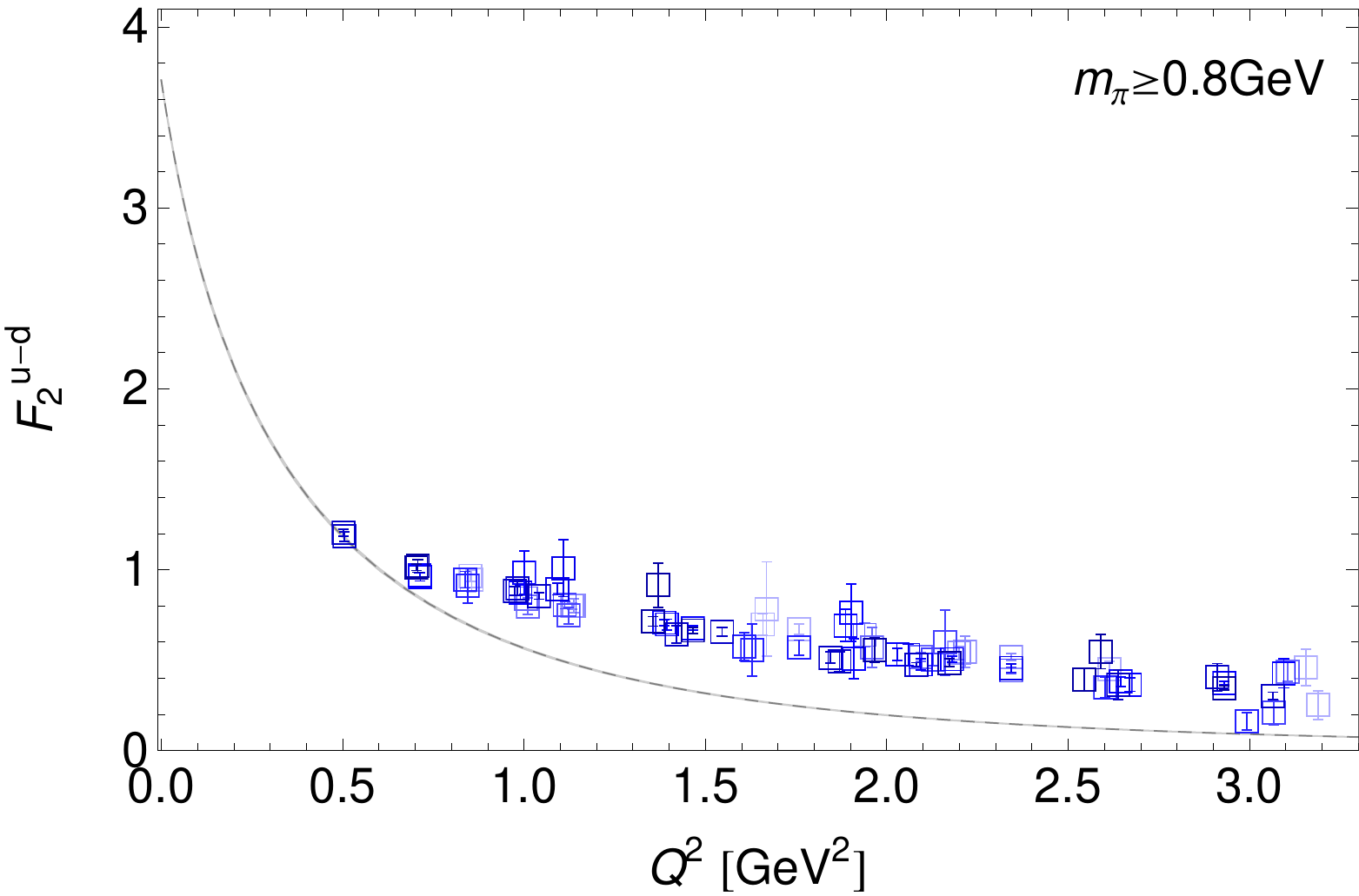}
          \includegraphics[angle=0,width=0.89\textwidth,clip=true,angle=0]{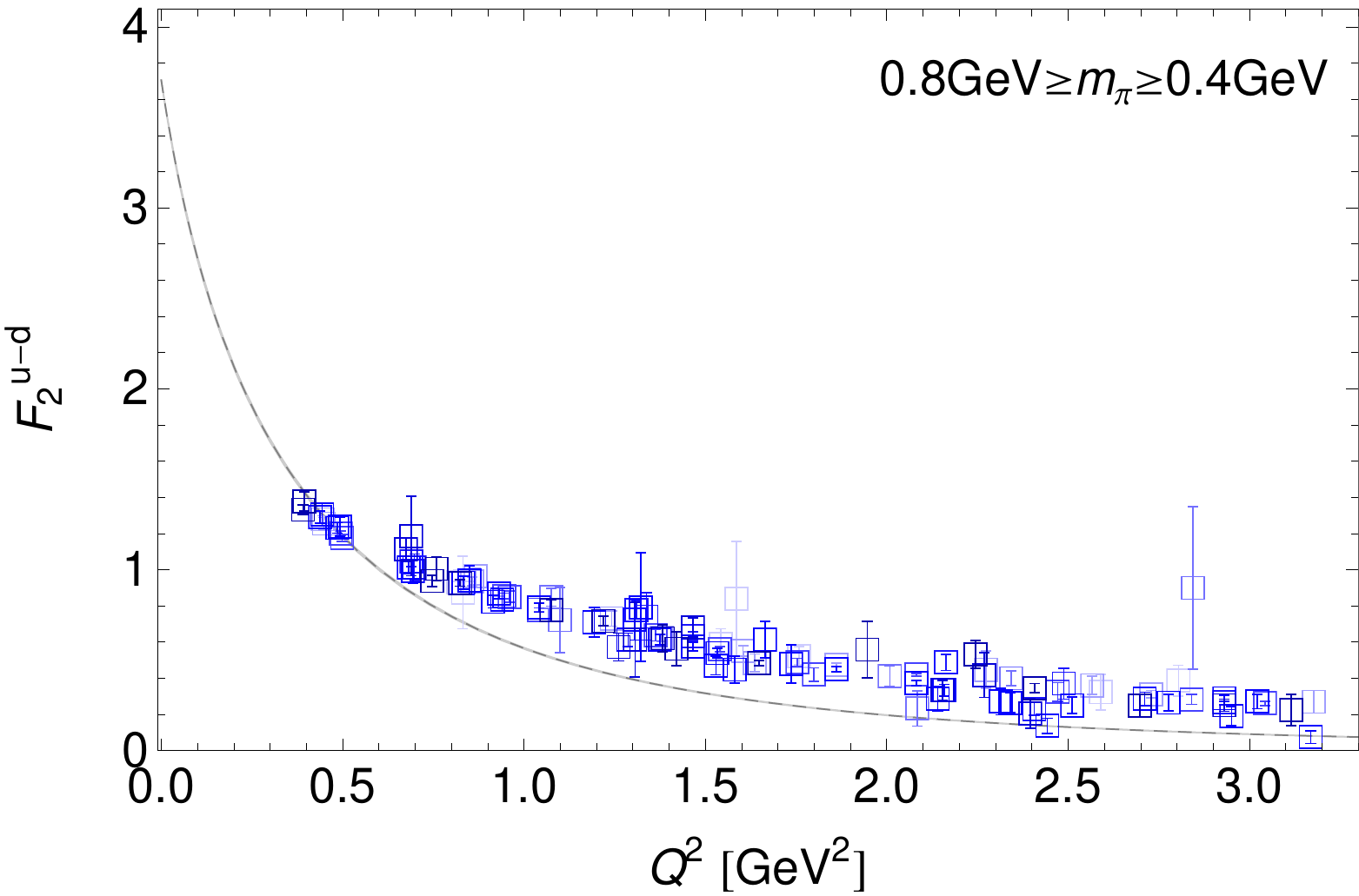}
          \includegraphics[angle=0,width=0.89\textwidth,clip=true,angle=0]{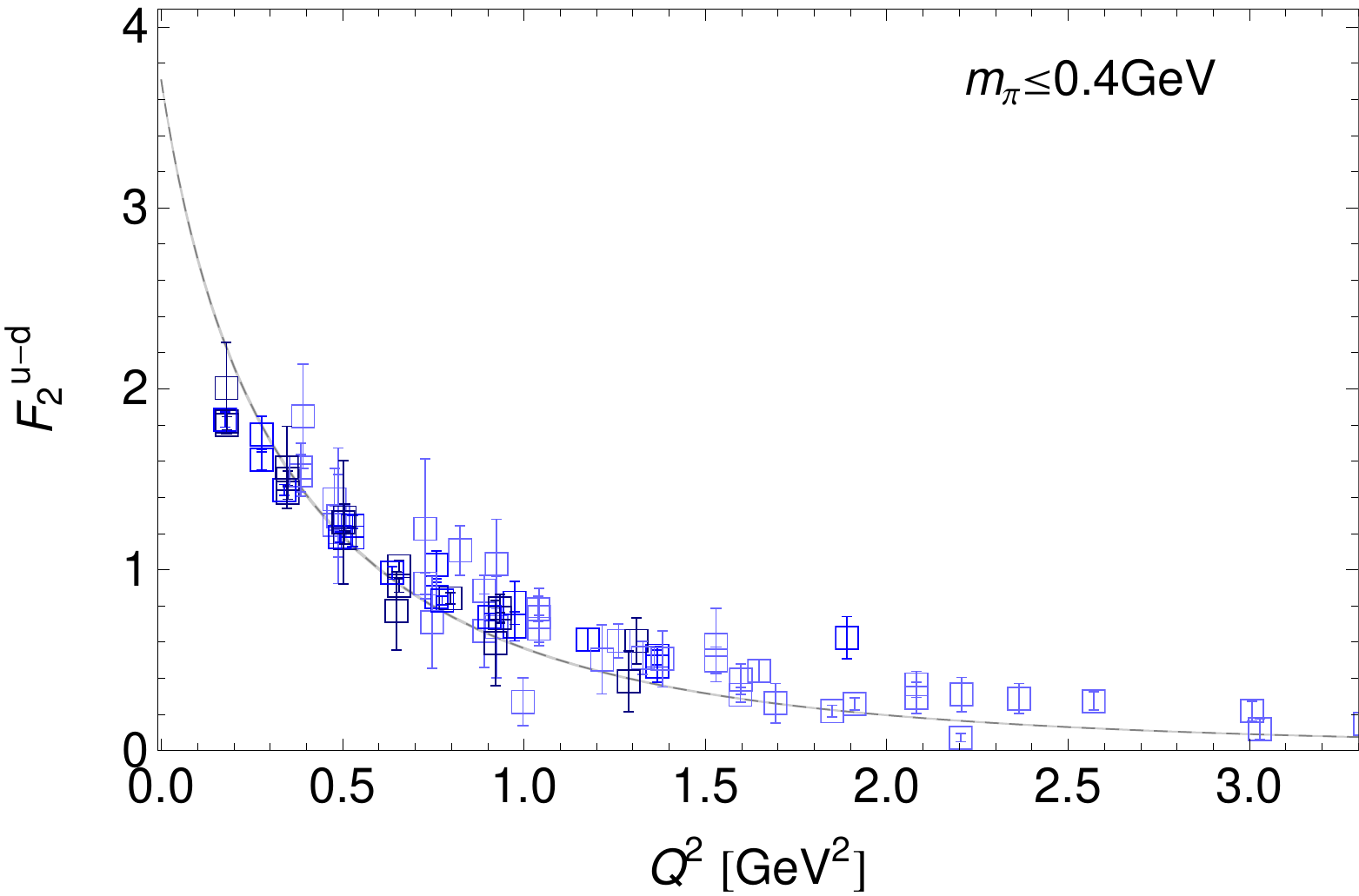}
\caption{Pauli form factor $F_2(Q^2)$ in the isovector channel. 
  All ensembles are included, and darker colors correspond to lighter pion masses.
  The gray shaded band represents the parametrization of Ref.~\cite{Alberico:2008sz} of the experimental data.\newline}
  \label{F2v}
     \end{minipage}
 \end{figure}

\subsection{$Q^2$-dependence of $F_1$ and $F_2$}

\begin{figure}[t]
    \begin{minipage}{0.48\textwidth}
        \centering
          \includegraphics[angle=0,width=0.9\textwidth,clip=true,angle=0]{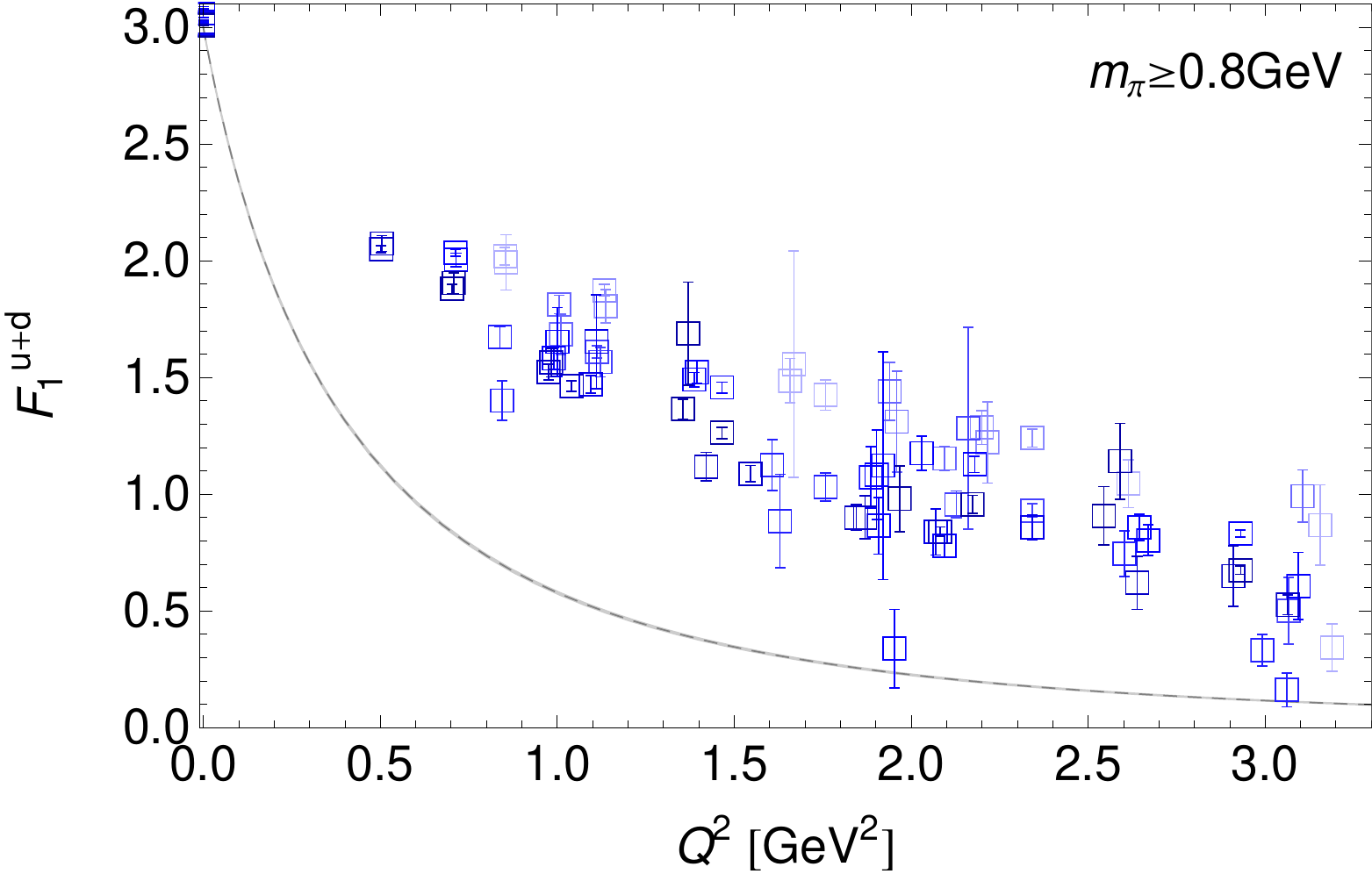}
          \includegraphics[angle=0,width=0.9\textwidth,clip=true,angle=0]{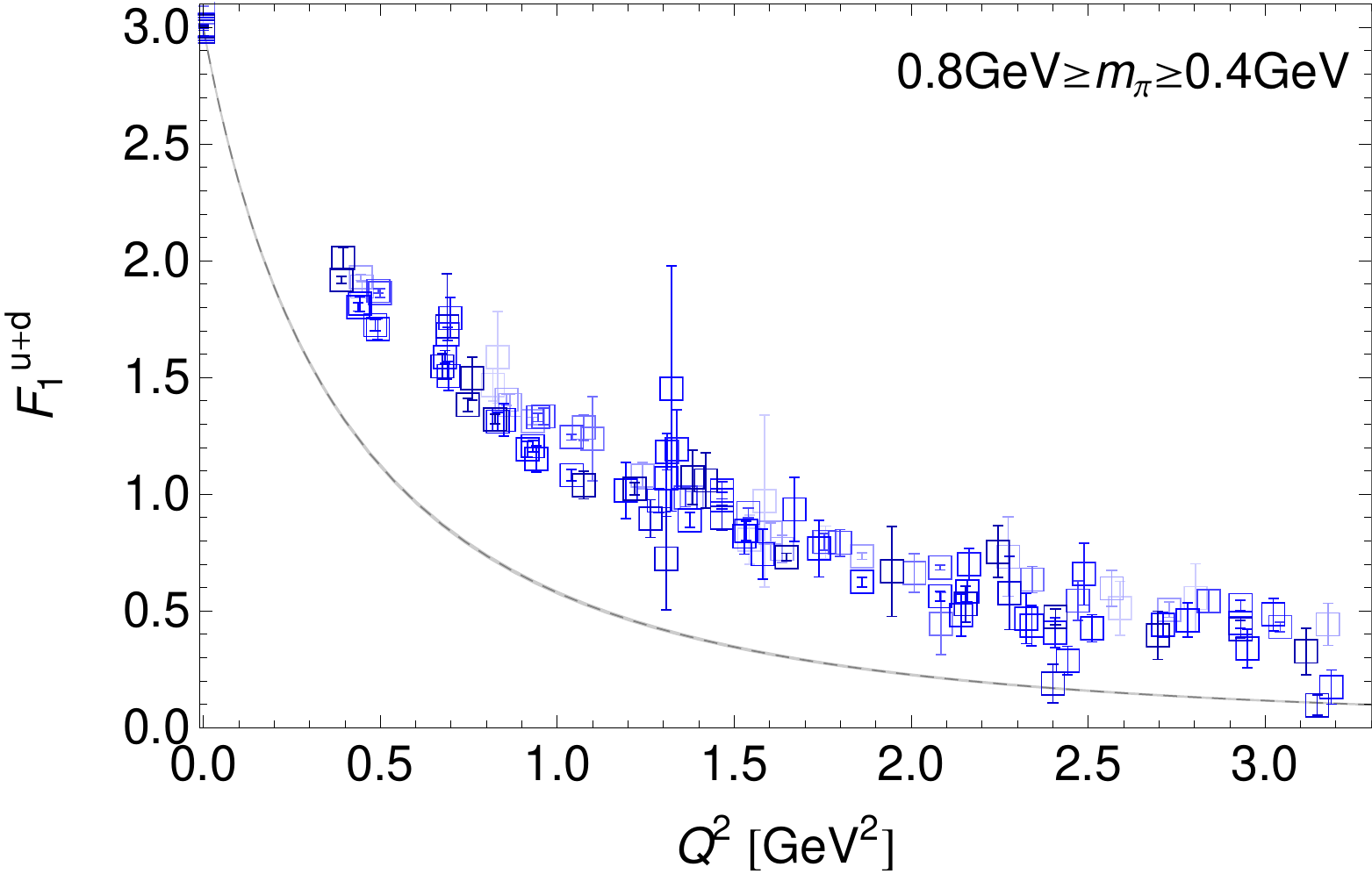}
          \includegraphics[angle=0,width=0.9\textwidth,clip=true,angle=0]{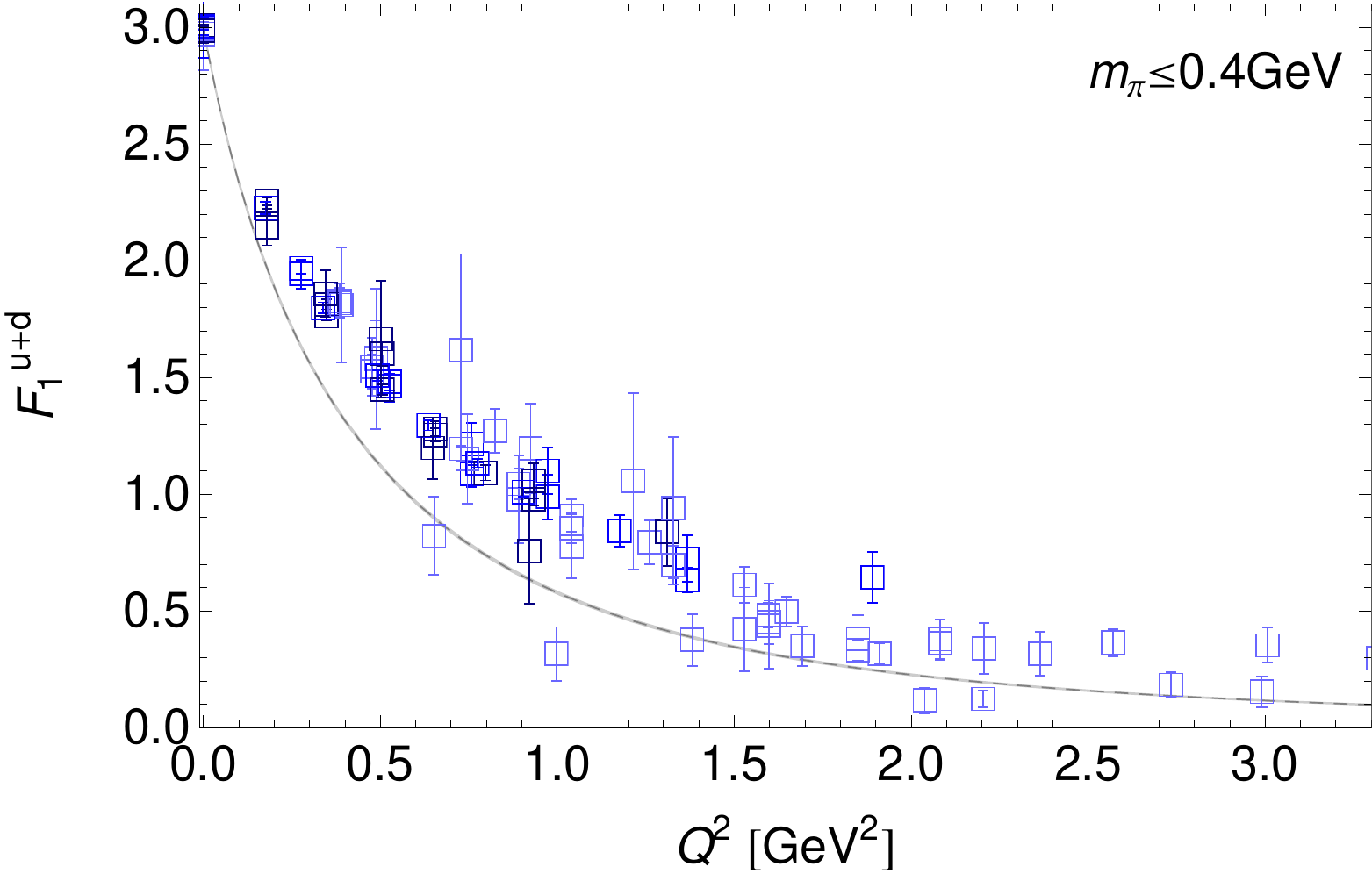}
  \caption{Dirac form factor $F_1(Q^2)$ in the isosinglet ($u+d$) channel. 
  All ensembles are included, and darker colors correspond to lighter pion masses.
  The gray shaded band represents the parametrization by Alberico et al. \cite{Alberico:2008sz} of the experimental data.
  %A legend is provided in Fig.~\ref{F2v}.
  }
  \label{F1s}
     \end{minipage} 
         \hspace{0.2cm}
    \begin{minipage}{0.48\textwidth}
      \centering
            \vspace*{0.2cm}
          \includegraphics[angle=0,width=0.89\textwidth,clip=true,angle=0]{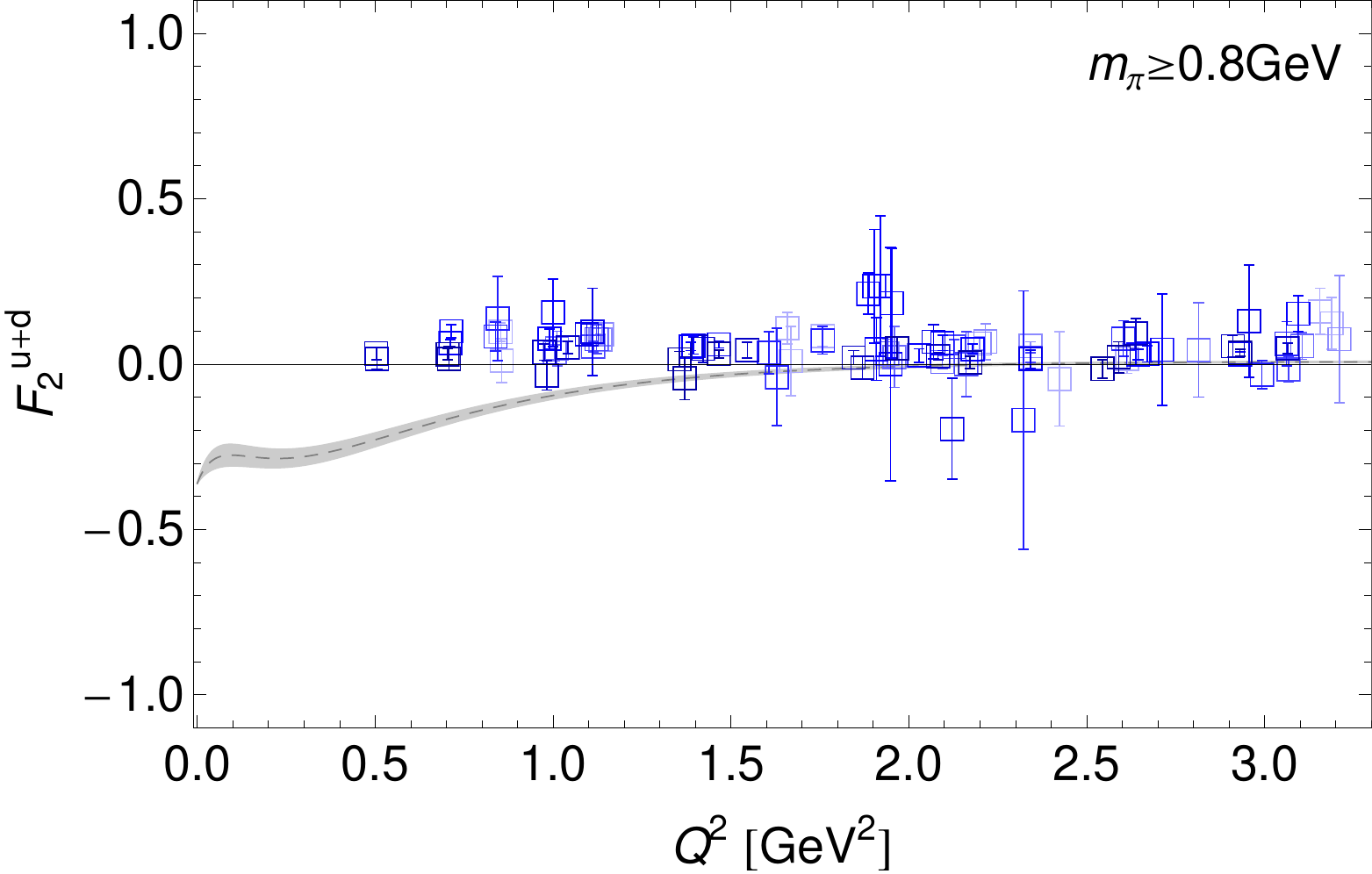}
          \includegraphics[angle=0,width=0.89\textwidth,clip=true,angle=0]{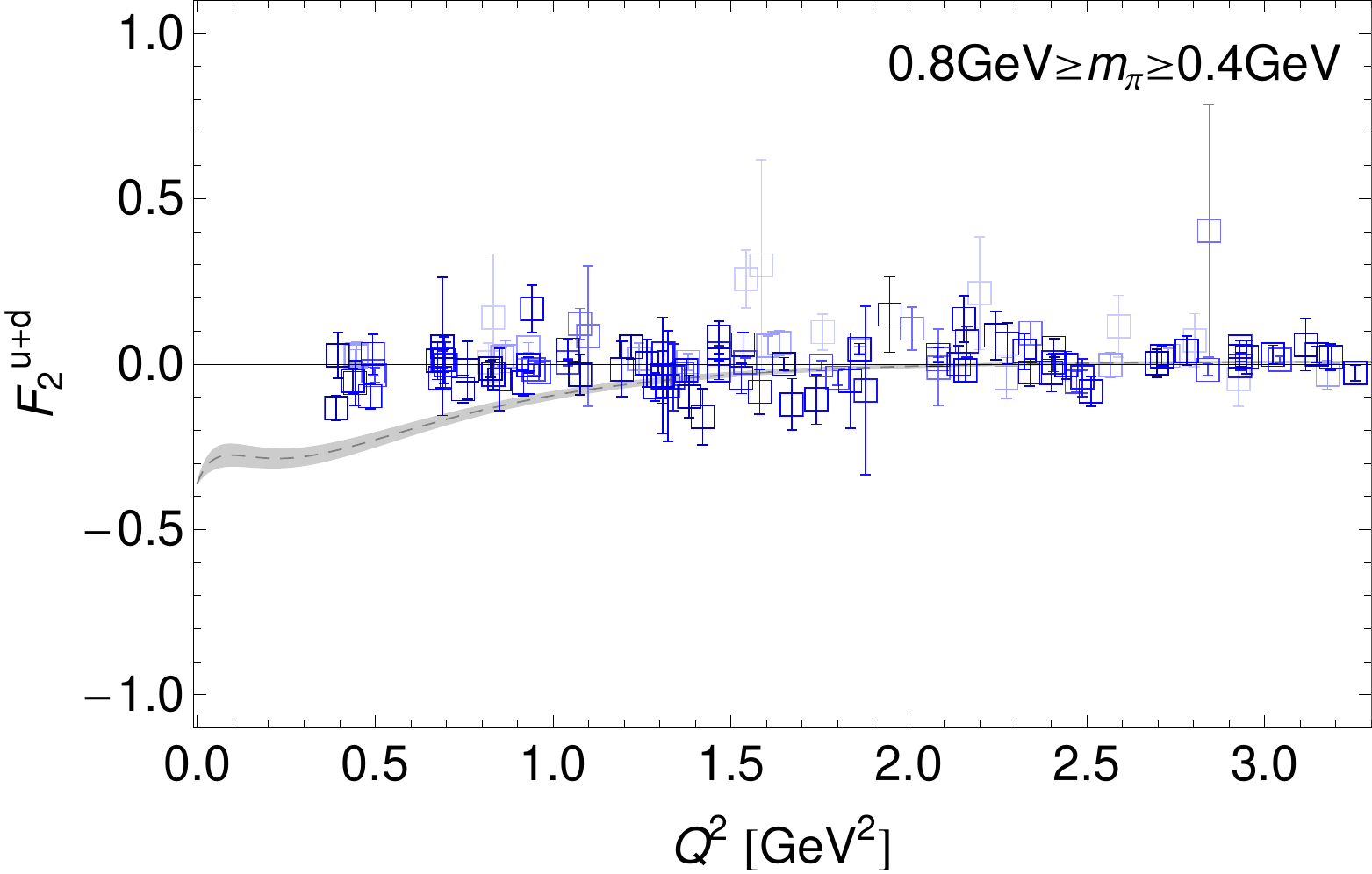}
          \includegraphics[angle=0,width=0.89\textwidth,clip=true,angle=0]{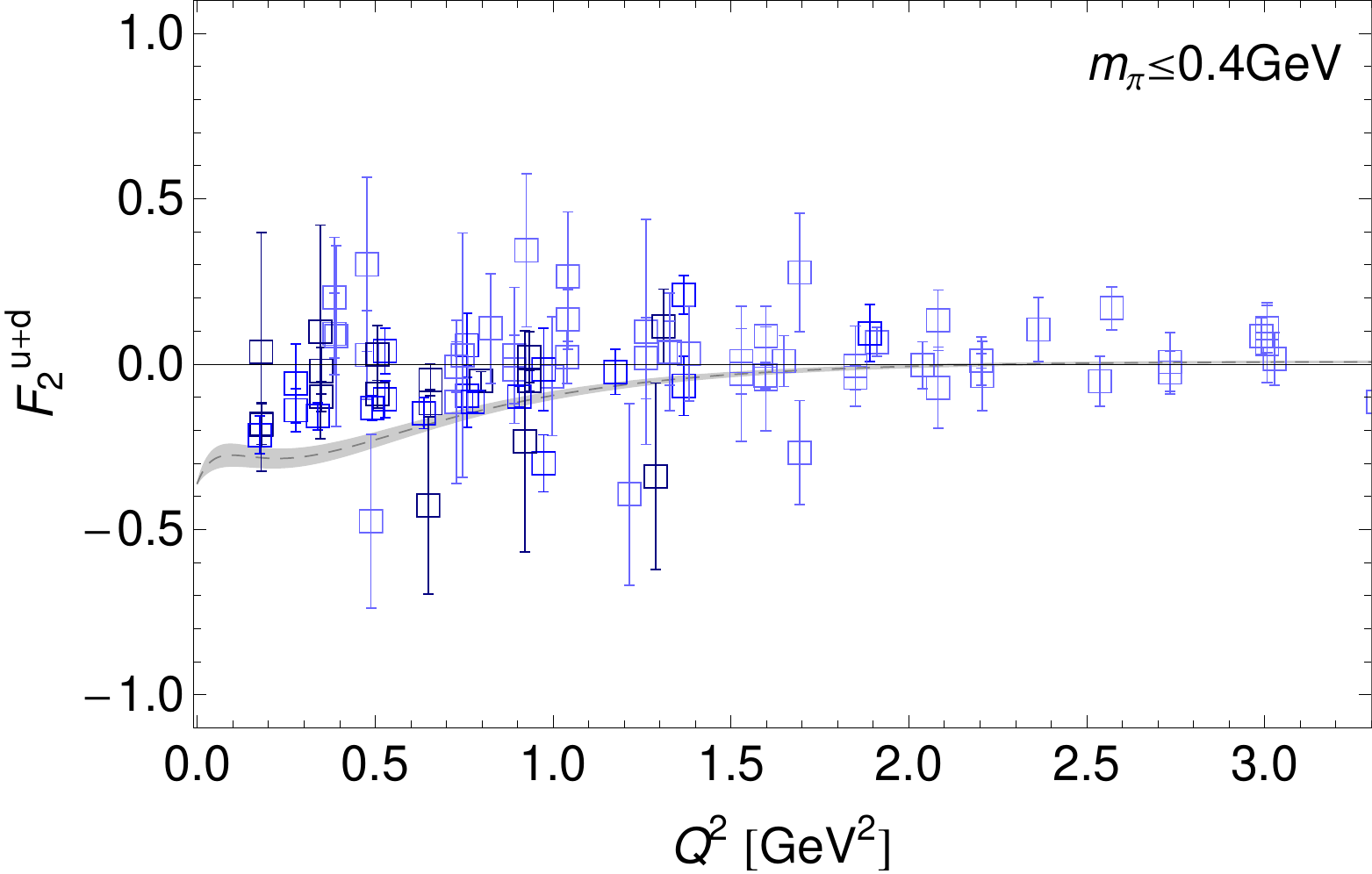}
\caption{Pauli form factor $F_2(Q^2)$ in the isosinglet ($u+d$) channel. 
  All ensembles are included, and darker colors correspond to lighter pion masses.
  The gray shaded band represents the parametrization of Ref.~\cite{Alberico:2008sz} of the experimental data.\newline}
  \label{F2s}
     \end{minipage}
 \end{figure}

In Figs.~\ref{F1v} to \ref{F2s}, we provide an overview of our results for the Dirac and 
Pauli form factors\footnote{As in our earlier study in \cite{Gockeler:2003ay}, we have normalized the results for $F_2$ 
such that the anomalous magnetic moment is given in units of the physical nuclear magneton, 
$e/(2m_N^{\phys})$.} in the isovector ($u-d$) and
isosinglet ($u+d$) channels, for three different ranges of $m_\pi$, including all ensembles specified in Table \ref{tab:params}.
Lighter pion masses are indicated by darker colored points.

For comparison, we also show in each case the parametrization of the $Q^2$-dependence of the experimental data obtained by 
Alberico et al. \cite{Alberico:2008sz} as gray error bands. 
This parametrization has been originally performed for Sachs electric and magnetic form factors
for the proton and the neutron, $G_{E,M}^{p,n}$, using Kelly's parametrization ansatz \cite{Kelly:2004hm} for $G_{E,M}^{p}$ and $G_{M}^{n}$,
and a Galster parametrization for $G_{E}^{n}$, with in total 14 parameters. 
Since the parameters are strongly correlated,
we have employed the full error correlation matrices provided at \cite{Alberico:2008szWEB} 
for the error propagation.
For our purposes, we consider the resulting parametrization of the $Q^2$-dependence of $F_{1,2}^{u\pm d}$ as a 
reasonably faithful representation of the experimental data, at least for not too large $Q^2$, and will use it as such throughout this work.
We note that since the quality and availability of experimental results at larger values of the momentum transfer above $\sim 1\GeV^2$
decreases, in particular for $G_{E}^{n}$, the shown error bands might significantly underestimate the actual uncertainties in this region.
%In the following, we will concentrate on the proton form factors and omit the superscript $p$ for notational simplicity.
If not stated otherwise we will consider the proton form factors in the following and omit the superscript $p$ for notational simplicity.
\begin{figure}[t]
    \begin{minipage}{0.48\textwidth}
        \centering
          \includegraphics[angle=0,width=0.9\textwidth,clip=true,angle=0]{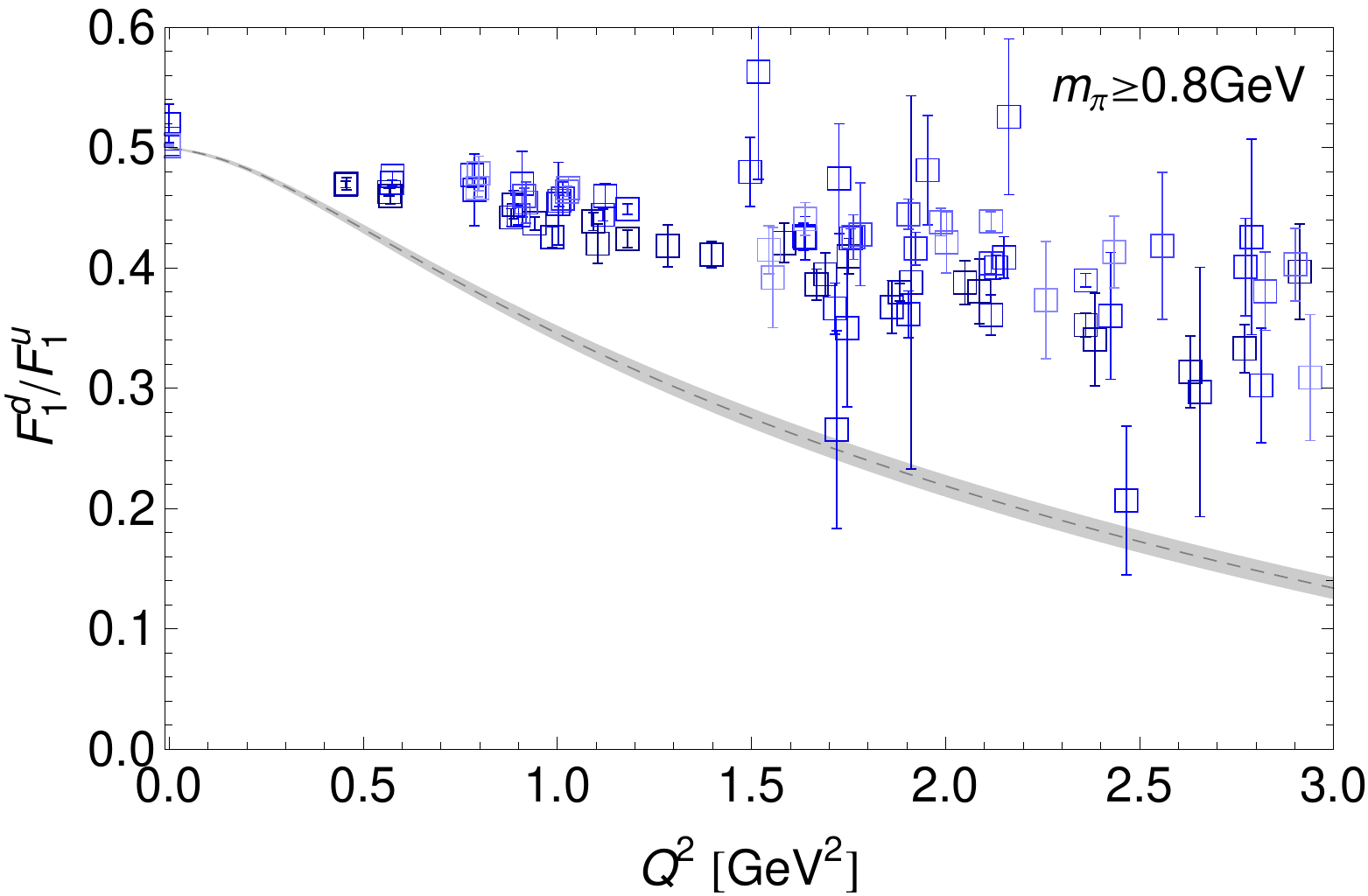}
     \end{minipage}
         \begin{minipage}{0.48\textwidth}
        \centering
          \includegraphics[angle=0,width=0.9\textwidth,clip=true,angle=0]{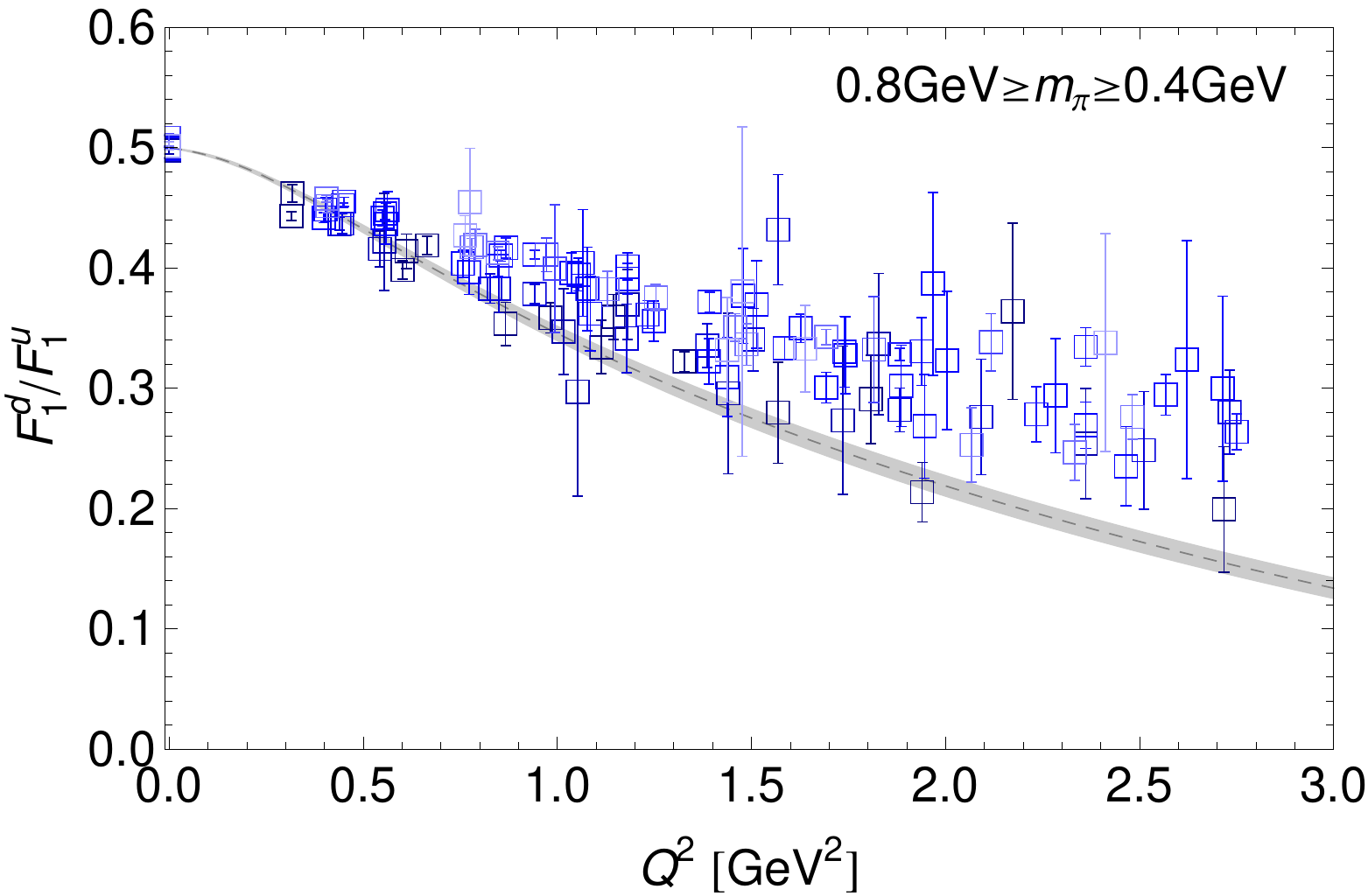}
     \end{minipage} 
         \hspace{0.2cm}
    \begin{minipage}{0.48\textwidth}
      \centering
          \includegraphics[angle=0,width=0.9\textwidth,clip=true,angle=0]{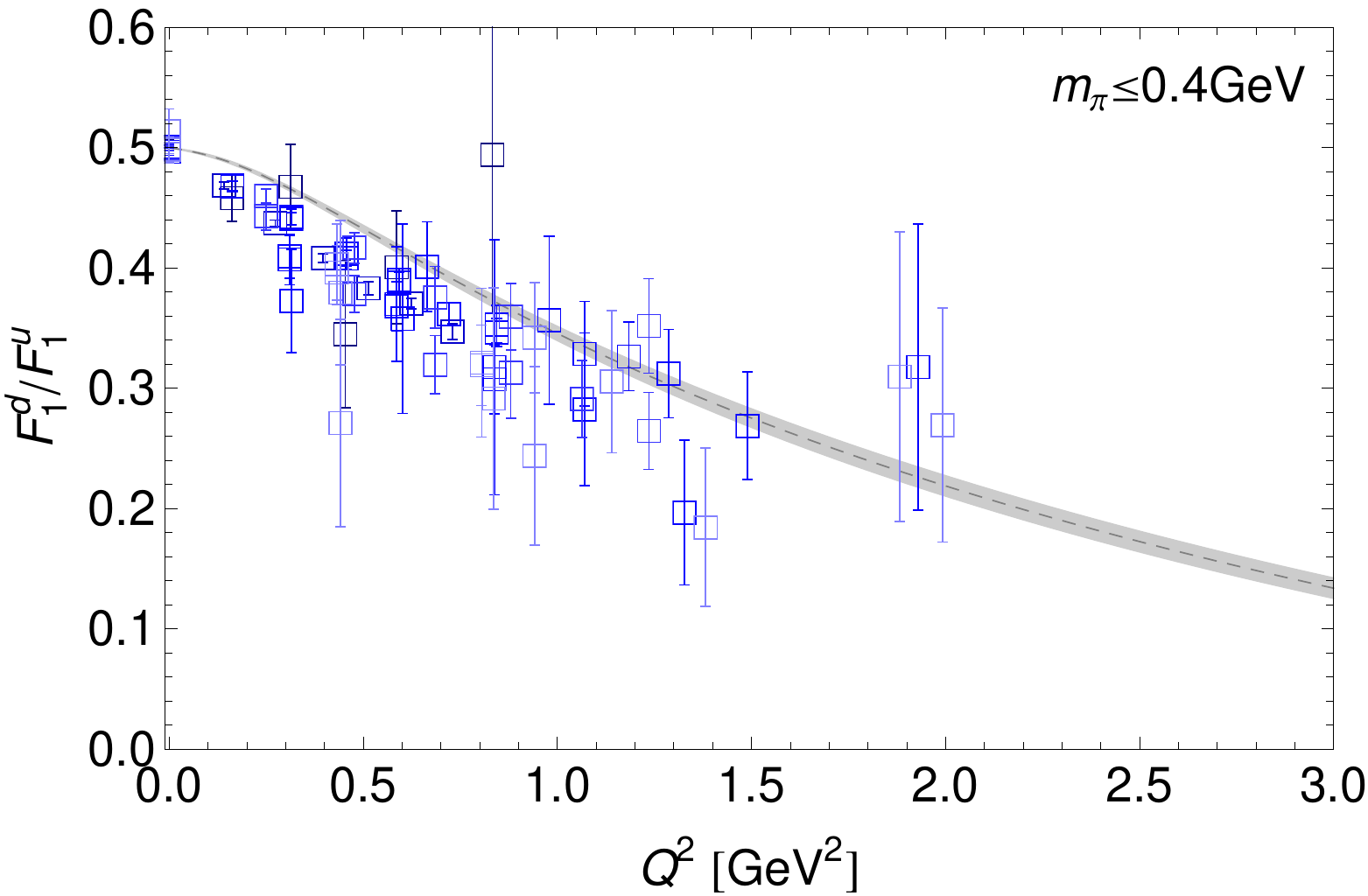}
     \end{minipage}
       \caption{The ratio $F^d_1/F^u_1$ of down to up quark contributions to the Dirac form factor. 
  All ensembles are included. The darker colors correspond to lighter pion masses.     
  The gray shaded band represents the parametrization of Ref.~\cite{Alberico:2008sz} of the experimental data.}
  \label{F1doverF1u}
 \end{figure}

In the cases of $F_1^{u\pm d}$ in Figs.~\ref{F1v} and \ref{F1s}, the normalization at $Q^2=0$ is fixed, and we
clearly see that the slope of the lattice data is significantly smaller than that of the parametrization.
It is interesting to observe, however, that the data points systematically move towards the physical result
as the pion mass decreases.
Concerning $F_2^{u-d}$ in Fig.~\ref{F2v}, it seems at first sight that the lattice data points, which show
little dependence on $m_\pi$, are in rough agreement with experiment over a wide range of $Q^2$.
This can be quite misleading, as we will see in more detail in the following sections: 
Not only is the slope of the lattice data points too small, but the lattice results for $F_2^{u-d}(Q^2=0)=\kappa_{u-d}$ 
(obtained from extrapolations in $Q^2$) are also significantly below the experimental value.
In combination, one naturally finds that the lattice and the experimental results do overlap in 
a certain range of the momentum transfer, however without implying a general agreement for all $Q^2$.
Finally, our results for $F_2^{u+d}$ in Fig.~\ref{F2s} turn out to be compatible with zero within errors
for practically all accessible values of $Q^2$, with the exception of a small number of data points
at lower pion masses and low momentum transfers, in particular the lowest $Q^2\sim0.2\GeV^2$, showing 
a slight trend towards negative values and the experimental error band.

\begin{figure}[t]
    \begin{minipage}{0.48\textwidth}
      \centering
          \includegraphics[angle=0,width=0.9\textwidth,clip=true,angle=0]{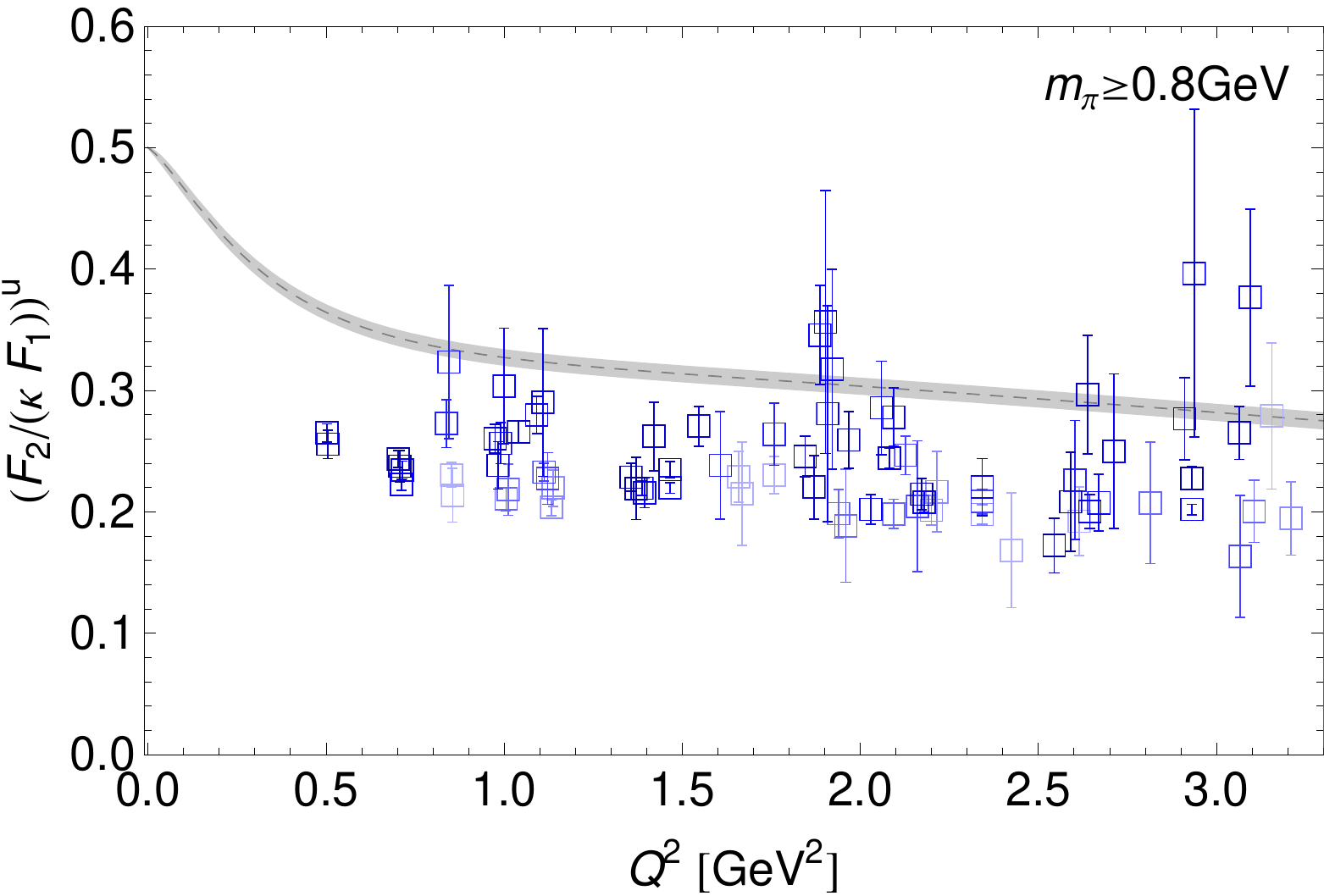}
          \includegraphics[angle=0,width=0.9\textwidth,clip=true,angle=0]{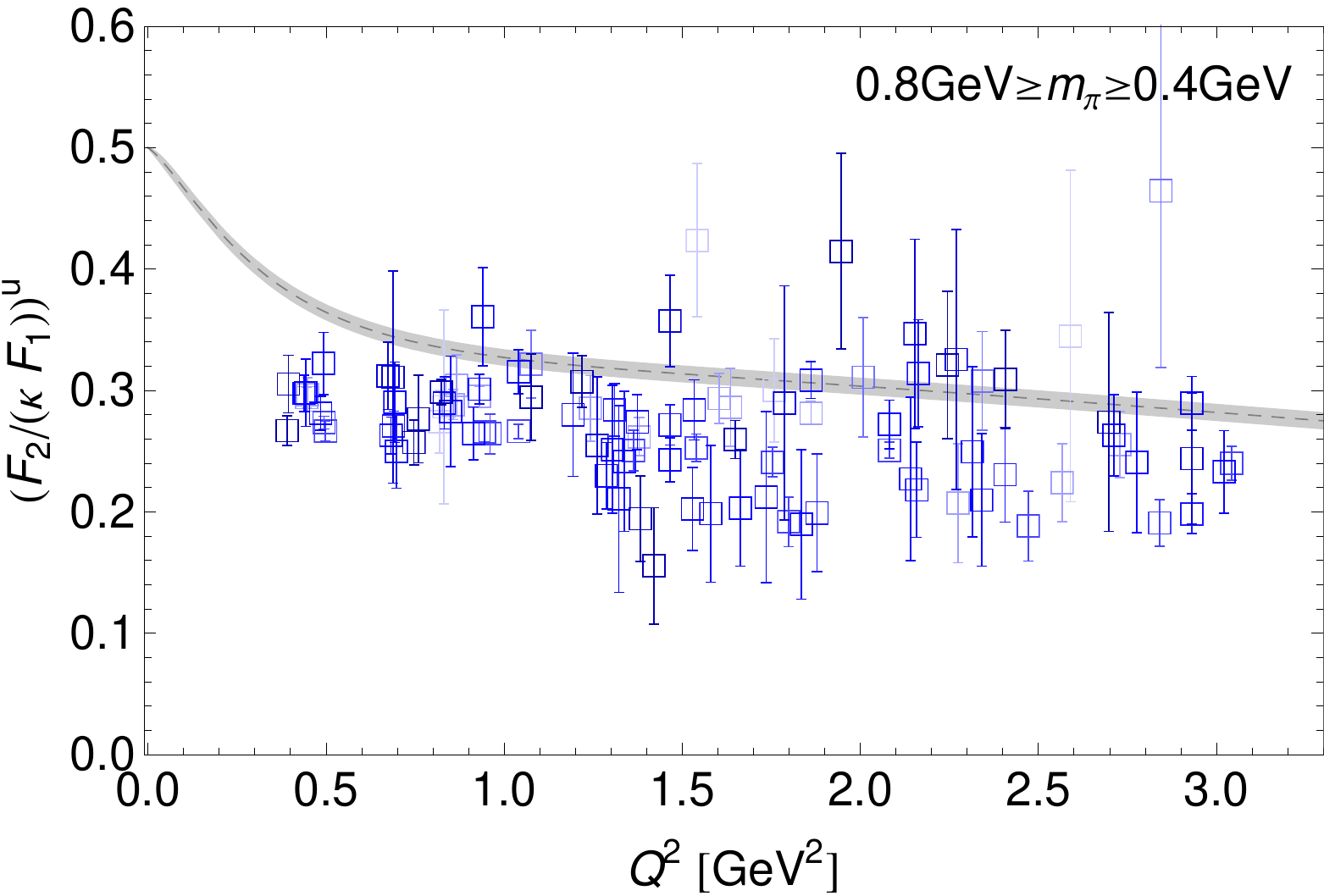}
          \includegraphics[angle=0,width=0.9\textwidth,clip=true,angle=0]{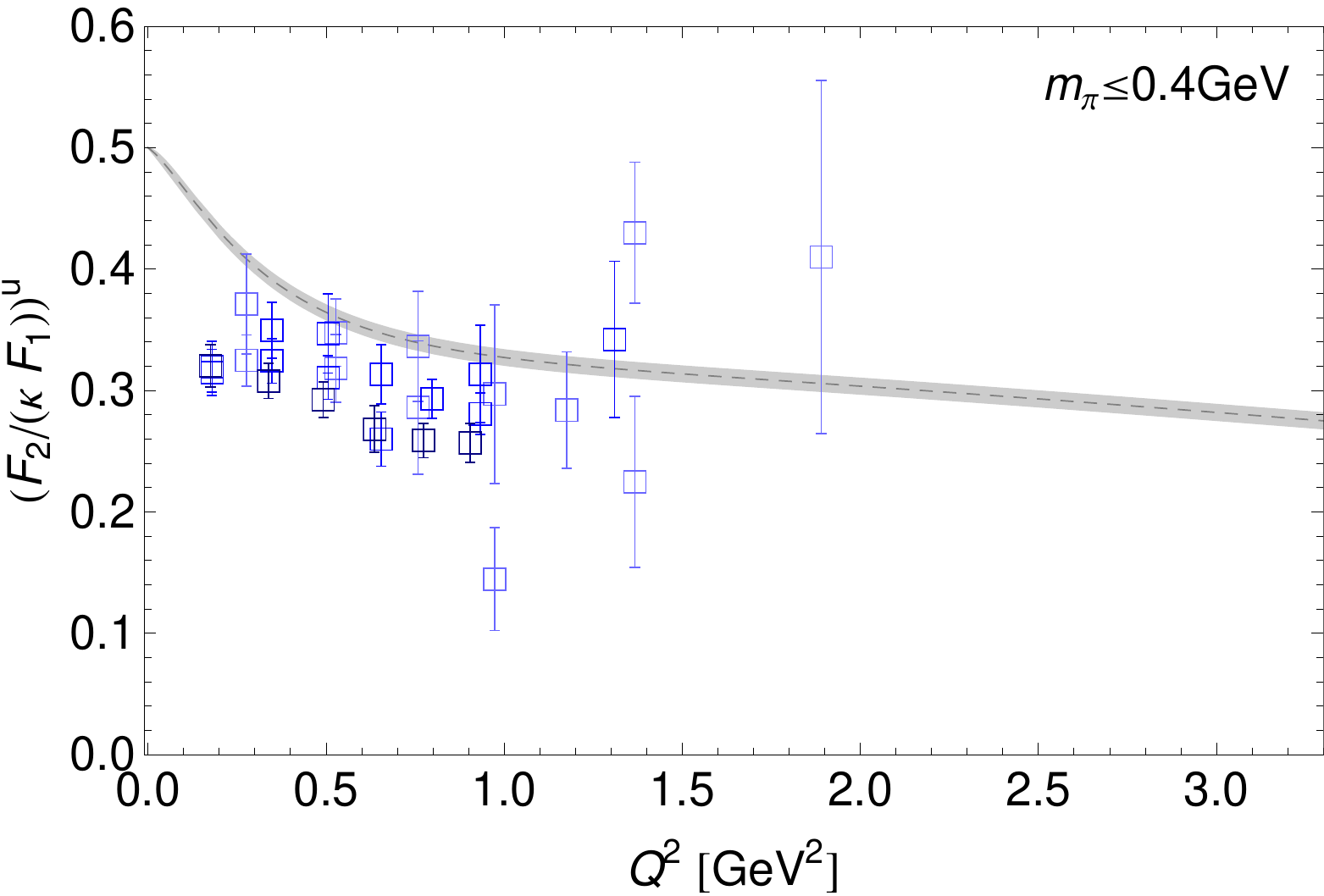}
  \caption{The form factor ratio $F_2/(\kappa F_1)$ for up quarks.  
  All ensembles are included. The darker colors correspond to lighter pion masses.     
  The lattice data points have been obtained using the experimental values for $\kappa_u$ in the ratio.
  The gray shaded band represents the parametrization of Ref.~\cite{Alberico:2008sz} of the experimental data.
  %A legend is provided in Fig.~\ref{F2v}.
  }
  \label{F2overF1u}
     \end{minipage} 
         \hspace{0.2cm}
    \begin{minipage}{0.48\textwidth}
      \centering
      \vspace*{0.15cm}
          \includegraphics[angle=0,width=0.9\textwidth,clip=true,angle=0]{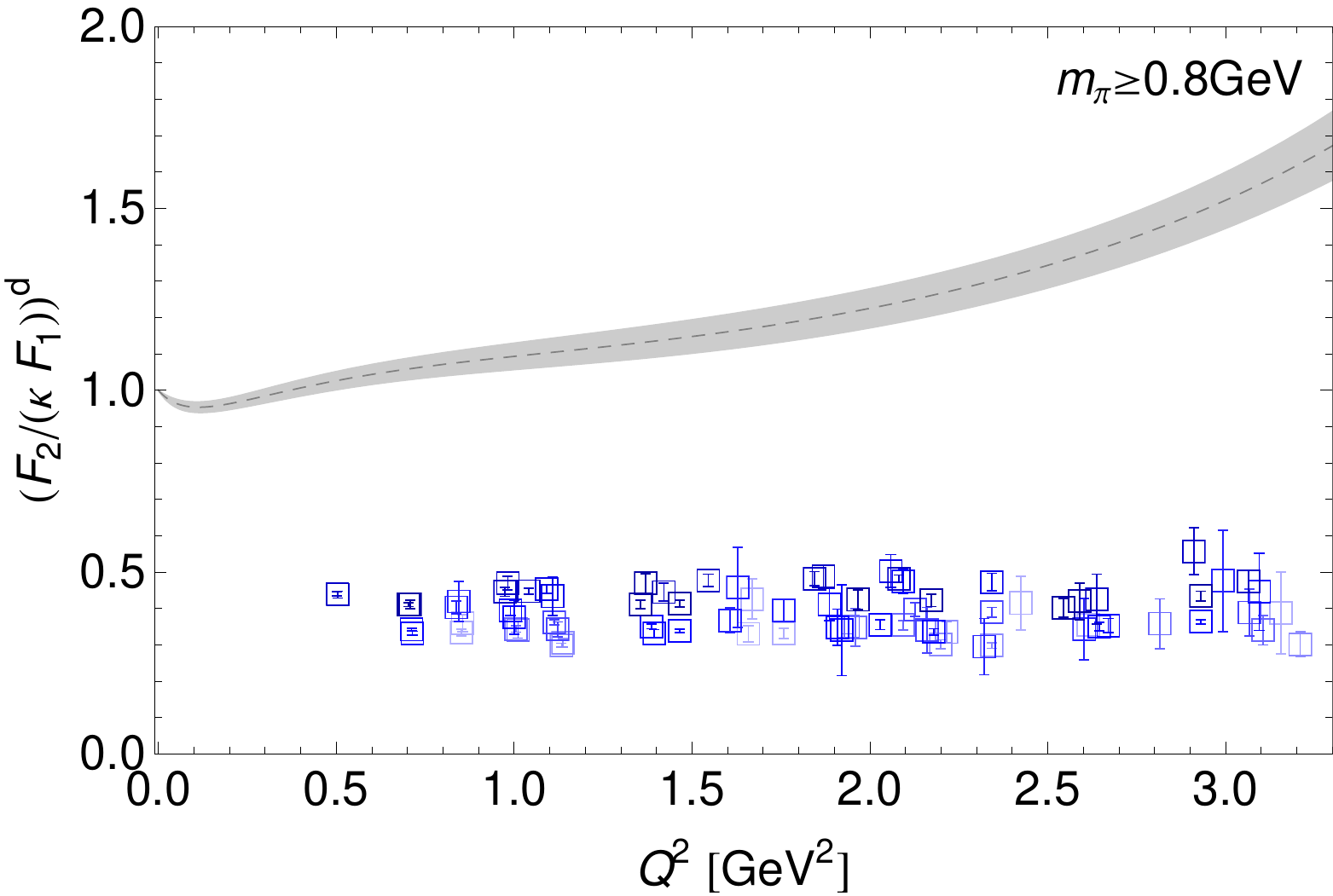}
          \includegraphics[angle=0,width=0.9\textwidth,clip=true,angle=0]{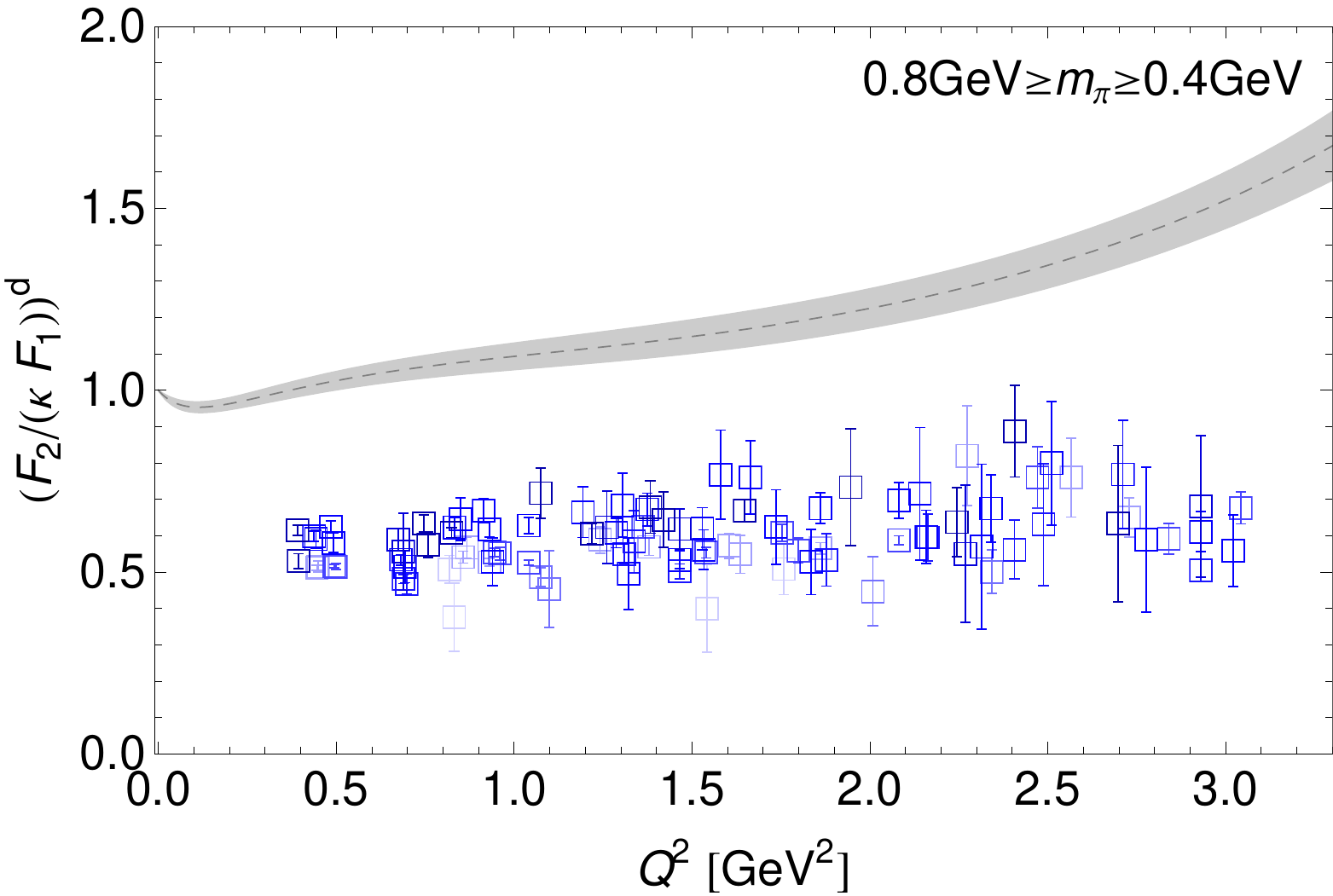}
          \includegraphics[angle=0,width=0.9\textwidth,clip=true,angle=0]{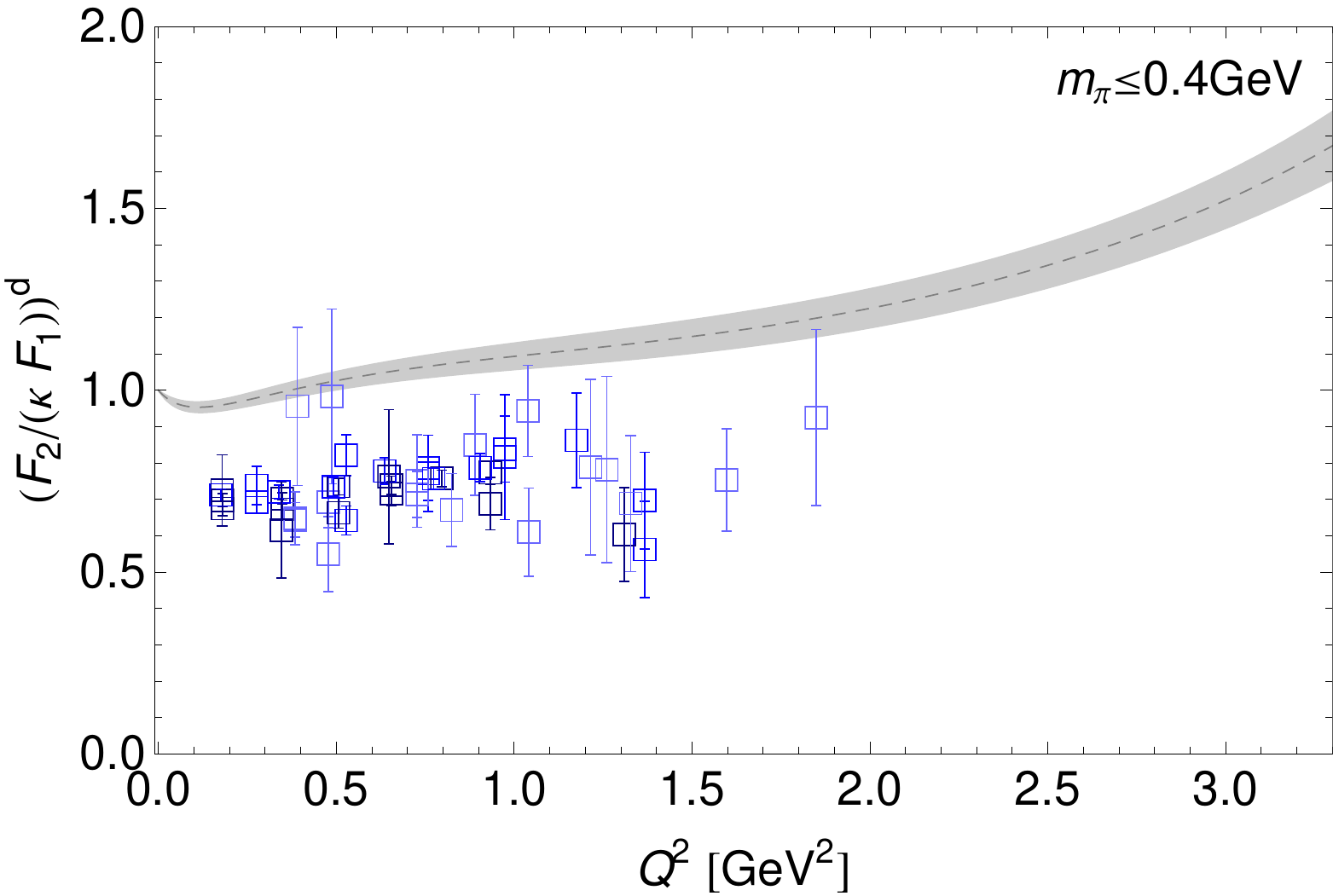}
\caption{The form factor ratio $F_2/(\kappa F_1)$ for down quarks. 
  All ensembles are included. The darker colors correspond to lighter pion masses.     
  The lattice data points have been obtained using the experimental values for $\kappa_d$ in the ratio.
  The gray shaded band represents the parametrization of Ref.~\cite{Alberico:2008sz} of the experimental data.\newline}
  \label{F2overF1d}
     \end{minipage}
 \end{figure}

Before engaging in a more
detailed study of the dependence of $F_{1,2}$ on $Q^2$ and also on the pion mass,
we briefly address two interesting questions that have been discussed before in the literature
and that can be addressed on the basis of ratios of form factors.

The first deals with potentially different $Q^2$-slopes of $F_1$ for the up and the down quarks.
In coordinate/impact parameter space \cite{Burkardt:2000za,Miller:2007uy}, such different slopes would correspond to differently shaped
quark density distributions and therefore provide important information about the inner structure of the nucleon.
Figure \ref{F1doverF1u} gives an overview of our results for the ratio $F_1^d/F_1^u$ as a function of $Q^2$,
where we have included all lattice ensembles, and where darker colors correspond to lighter pions on the lattice.
The parametrization of the experimental data is, as before, illustrated by the error band.
While the data points at the largest pion masses show only a small dependence on $Q^2$, they show a systematic 
downwards trend towards ratios $F_1^d/F_1^u<0.5$ for $Q^2\gtrsim0.5\GeV^2$ as the pions get lighter,
thereby moving closer to the experimental error band.
In the limit $Q^2\rightarrow0$, the experimental result flattens off considerably, which we will
discuss in more detail in section \ref{sec:r1} below on the basis of the separate mean square radii for up and down quarks 
and different parametrizations of the lattice data.

The second question concerns the scaling of $F_2/F_1$ at intermediate to large $Q^2$-values.
Perturbative QCD suggests that $Q^{2}F_2/F_1\sim \text{const.}$ as $Q^2\rightarrow\infty$,
up to logarithmic corrections $\propto \ln{Q^2}$
\cite{Brodsky:1973kr,Lepage:1979za,Belitsky:2002kj}.
Recently, supported by new measurements of $G_M^n$ at JLab Hall A,
it has been noted that $F_2/F_1$ approximately scales as a constant already in an intermediate range of $Q^2=1.5,\ldots,3.5\GeV^2$,
separately for up and for down quarks \cite{Cates:2011pz}.
We show our results together with the parametrization of \cite{Alberico:2008sz} for these ratios in Figs.~\ref{F2overF1u} and \ref{F2overF1d}. 
Clearly, the uncertainties and the scatter of the lattice data, in particular for the up quark case in Fig.~\ref{F2overF1u},
make it difficult to draw any strong conclusions.
Nevertheless, the lattice data points in the different $m_\pi$-ranges are overall compatible with a flat $Q^2$-dependence above $\sim0.5\GeV^2$,
which is most clearly seen for the down quarks in Fig.~\ref{F2overF1d} at the largest pion masses.
While the latter are about a factor of two below the experimental band, we see a clear upwards trend 
as lower pion masses are being approached.
These trends are less clear in the case of the up quarks in Fig.~\ref{F2overF1u}, 
for which the data points are, however, generally closer to experiment.

In any case, more quantitative conclusions with respect to these interesting questions will have to be based 
on precise lattice data at low pion masses that extends up to and beyond squared momentum transfers of $Q^2\sim2\GeV^2$.

To conclude this section, we note that a study of potential systematic uncertainties, as well as the pion mass dependence of the
form factors at fixed $Q^2$, is given further below in section \ref{sec:systematics}.
In short, we do not see any significant, systematic effects due to contributions from excited states
(section \ref{sec:tsnk}), or the finite lattice spacing (section \ref{sec:syseff}), at least within statistical uncertainties.
Although the lattice data points show an approximately linear dependence on $m_\pi$ or $m_\pi^2$ at fixed $Q^2$, we find that
simple linear extrapolations to the physical point would not lead to an agreement with the results from experiment and phenomenology
(section \ref{sec:syseff}). We therefore conclude that a non-trivial pion mass dependence has to set in between
the lowest accessible lattice pion masses of $\sim 180,\ldots,260\MeV$ and $m_\pi^\phys$. 
This will also be studied in greater detail on the basis of the mean square radii and anomalous magnetic moments in section \ref{sec:chiral}.

\subsection{Parametrizations of the $Q^2$-dependence}
\label{sec:para}

We now turn to analytical parametrizations of the $Q^2$-dependence.
These will not only allow us to interpolate between the discrete values of $Q^2$, 
but in particular to extrapolate our results for $F_2$ to the forward limit in order to extract the anomalous magnetic moment. 
Furthermore, well-chosen parametrizations are important to obtain more realistic estimates for the mean
square radii from the slopes of the form factors at $Q^2=0$, 
\bea
\label{eqradii}
\langle r^2\rangle_{i}
  =-\frac{6}{F_i(0)}\left.\frac{dF_i(Q^2)}{dQ^2}\right|_{Q^2=0}\,.
\eea
However, we note that the parametrizations unavoidably introduce some model dependence into the analysis.

\begin{figure}[t]
    \begin{minipage}{0.99\textwidth}
        \centering
          \includegraphics[angle=0,width=0.9\textwidth,clip=true,angle=0]{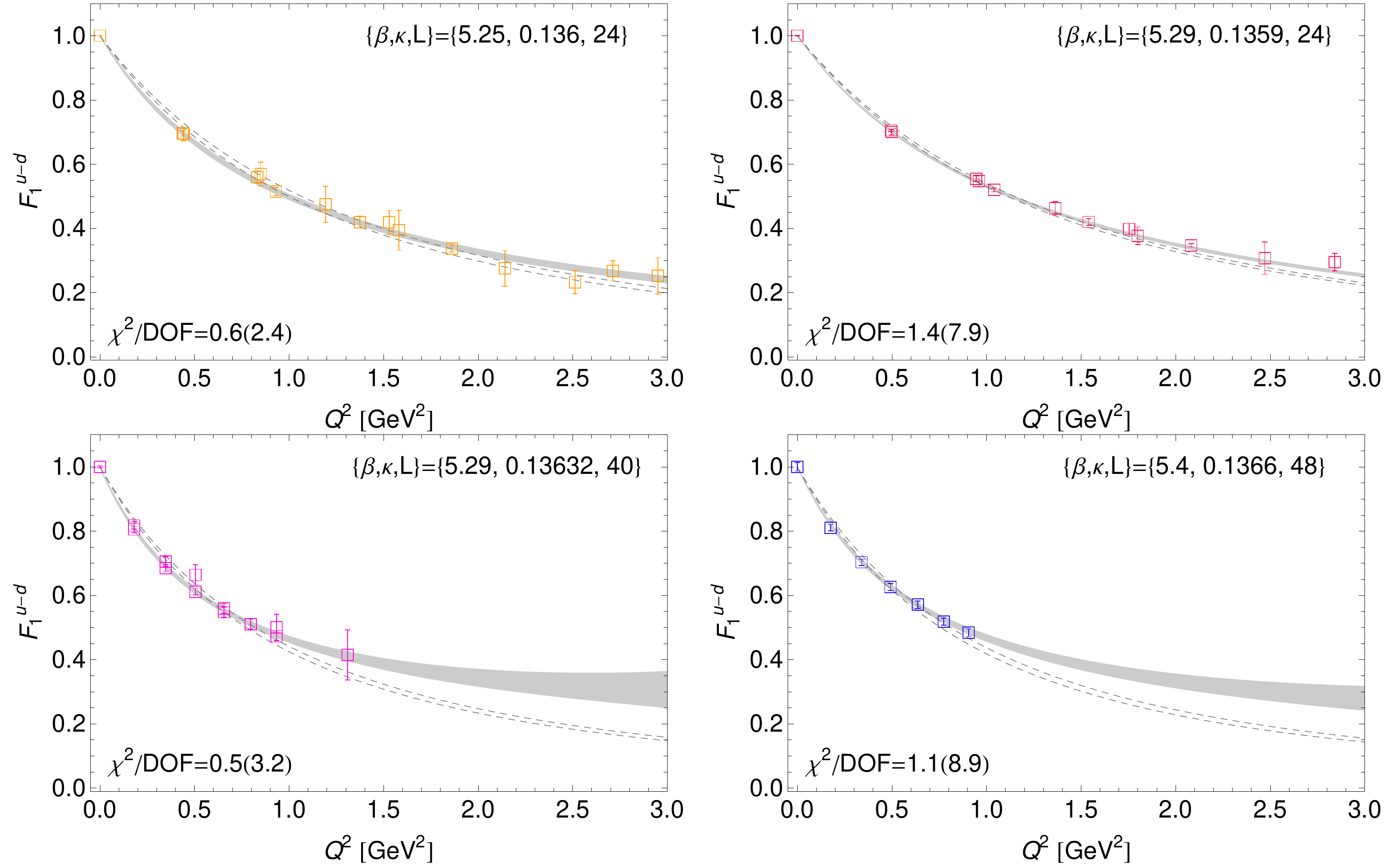}
     \end{minipage}
       \caption{Parametrization of the $Q^2$-dependence of the isovector Dirac form factor lattice data for selected ensembles.
       The shaded bands represent the 2-parameter fits to the lattice data points based on Eq.~(\ref{eqF1poly}). We also 
       show the corresponding values of $\chi^2/DOF$.
       For comparison, the 1-parameter dipole fits based on Eq.~(\ref{eqF1dipole}) are indicated by the dashed lines, 
       with $\chi^2/DOF$ given in parentheses.}
  \label{F1umdDipolePoly}
 \end{figure}

In the following, we compare different ans\"atze for the $Q^2$-dependence of the form factors.
A common ansatz for the Dirac form factor is a dipole,
\be
 F_1(Q^2)=\frac{F_1(0)}{(1+Q^2/m^2_D)^2}\,,
 \label{eqF1dipole}
\ee
with, e.g., $F^{p,u-d}_1(0)=1$, where the dipole mass $m_D$ is a free fit parameter.
The corresponding mean square radius is then given by the squared inverse dipole mass, $\langle r^2\rangle_1=12/m_D^2$.
A more flexible parametrization is obtained with a more general polynomial in the denominator,
\be
 F_1(Q^2)=\frac{F_1(0)}{1+c_{12}Q^2+c_{14}Q^4}\,,
  \label{eqF1poly}
\ee
with $c_{12}$ and $c_{14}$ as free fit parameters. 
Here the mean square radius is obtained from $\langle r^2\rangle_1=6 c_{12}$.
The latter form was already employed in Ref.~\cite{Gockeler:2007hj}, and it 
also allows for a matching to a 
simple vector meson exchange ansatz, as will be discussed below in section \ref{sec:matching}.
Similarly, for the Pauli form factor $F_2$, one could employ a simple dipole or tripole form
\be
 F_2(Q^2)=\frac{F_2(0)}{(1+Q^2/m^2_p)^p}\,,
 \label{eqF2ppole}
\ee
where $p=2$ or $3$, and $F_2(0)$ and the pole mass $m_p$ are the fit parameters.
In this case the Pauli radius is given by $\langle r^2\rangle_2=6p/m_p^2$.
Alternatively, a more general polynomial in the denominator leads to a three-parameter ansatz of the form \cite{Gockeler:2007hj}
\be
 F_2(Q^2)=\frac{F_2(0)}{1+c_{22}Q^2+c_{26}Q^6}\,,
  \label{eqF2poly}
\ee
for which the mean square radius is $\langle r^2\rangle_2=6 c_{22}$.
We note that the choices for the highest powers of $Q^2$ in the denominators of Eqs.~(\ref{eqF1poly}) and (\ref{eqF2poly})
ensure that $F_i(Q^2)\xrightarrow{Q^2\rightarrow\infty}\sim1/(Q^2)^{i+1}$, as expected from perturbative QCD \cite{Brodsky:1973kr}.

Typical results for parametrizations of $F_1^{u-d}(Q^2)$ for selected ensembles are displayed in Fig.~\ref{F1umdDipolePoly}.
We observe that according to the $\chi^2/DOF$, the less restrictive polynomial ansatz seems to describe the data significantly better.
Apart from that, the main difference between the fits based on Eqs.~(\ref{eqF1dipole}) and (\ref{eqF1poly}) is the smaller slope and 
the broadening of the error bands in the case of the 2-parameter polynomial ansatz in the region of larger $Q^2$ where no data points are available. 

\begin{figure}[t]
    \begin{minipage}{0.99\textwidth}
        \centering
          \includegraphics[angle=0,width=0.9\textwidth,clip=true,angle=0]{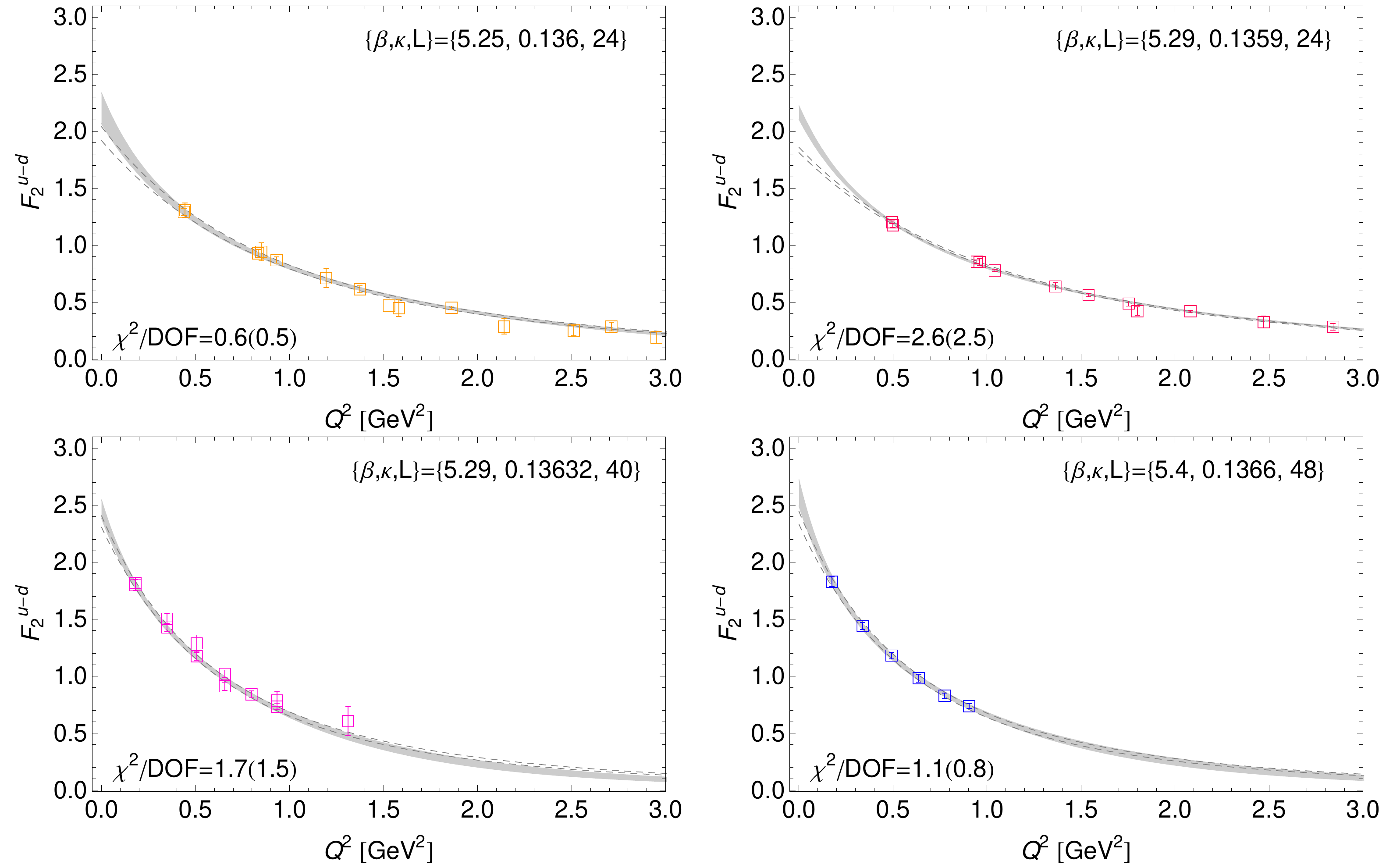}
     \end{minipage}
       \caption{Parametrization of the $Q^2$-dependence of the isovector Pauli form factor lattice data for selected ensembles.
       The shaded bands represent the 3-parameter fits to the lattice data points based on Eq.~(\ref{eqF2poly}). We
       also show the corresponding values of $\chi^2/DOF$.
       For comparison, the 2-parameter tripole fits based on Eq.~(\ref{eqF2ppole}) ($p=3$) are indicated by the dashed lines, 
       with $\chi^2/DOF$ given in parentheses.}
  \label{F2umdTripolePoly}
 \end{figure}

In the case of $F_2^{u-d}$, we are comparing a tripole ($p=3$) ansatz, Eq.~(\ref{eqF2ppole}),
with the more general polynomial parametrization, Eq.~(\ref{eqF2poly}), in Fig.~\ref{F2umdTripolePoly}.
With respect to the $\chi^2/DOF$, the 3-parameter form doesn't have any advantage.
However, due to the somewhat stronger broadening of the error bands at lower and larger values of $Q^2$
in regions where no data points are available, we consider the polynomial ansatz in general to be less biased and 
hence to provide more realistic uncertainties with respect to extrapolations, in particular to $Q^2=0$.
A more quantitative comparison of different parametrizations of $F_2(Q^2)$ will have to be based on precise 
data over a broader range of $Q^2$, e.g., employing partially twisted boundary conditions \cite{Sachrajda:2004mi,Bedaque:2004ax}
to access very small momentum transfers.

In the following sections, we will argue further on the basis of a matching to
a simplistic vector meson exchange ansatz, and the extracted Dirac and Pauli radii,
that the polynomial ans\"atze in Eqs.~(\ref{eqF1poly}) and (\ref{eqF2poly}) provide a more consistent description
of the data. We will therefore consider them in the following as preferred compared to the standard
dipole and tripole forms in Eqs.~(\ref{eqF1dipole}) and (\ref{eqF2ppole}).

A collection of numerical results for the mean square radii and anomalous magnetic moments, obtained from the 
polynomial parametrizations, is provided 
in Appendix \ref{sec:tables}, for all ensembles listed in Table \ref{tab:params}.

\subsection{Dirac radius}
\label{sec:r1}
\begin{figure}[t]
    \begin{minipage}{0.48\textwidth}
        \centering
          \includegraphics[angle=0,width=0.9\textwidth,clip=true,angle=0]{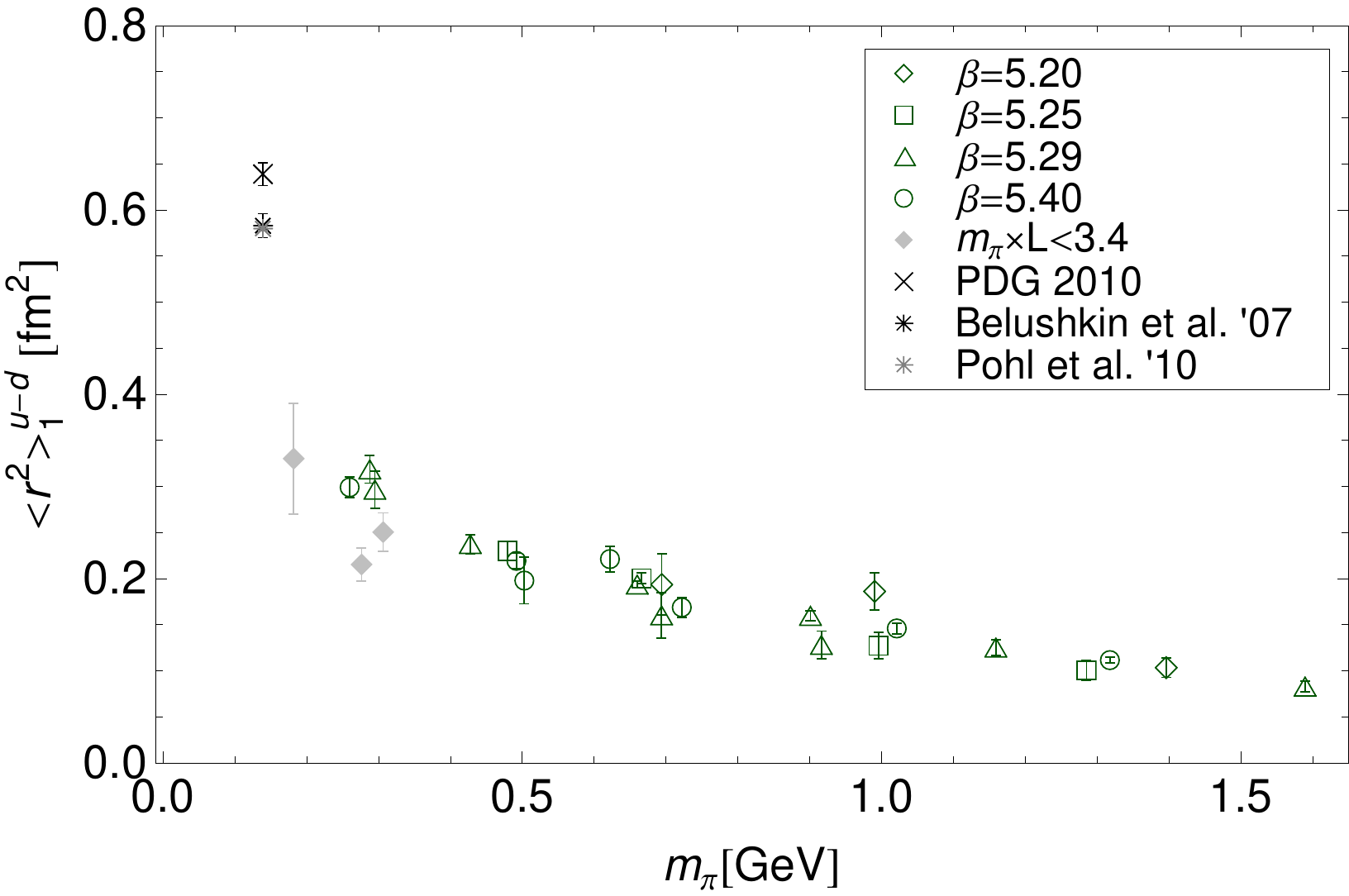}
  \caption{Isovector Dirac radius versus $m_\pi$, as obtained from fits to $F^{u-d}_1$ using the polynomial ansatz in 
  Eq.~(\ref{eqF1poly}) (cf. Fig.~\ref{F1umdDipolePoly}).
  The labels ``PDG 2010" and ``Pohl et al. '10" refer to Refs.~\cite{Nakamura:2010zzi} and \cite{Pohl:2010zz}, respectively.
  Unless specified otherwise, the label ``Belushkin et al. '07" here and below refers to the super-convergence (SC) values
  of Ref.~\cite{Belushkin:2006qa}.}
  \label{r1sqrdPoly}
  \end{minipage} 
         \hspace{0.2cm}
    \begin{minipage}{0.48\textwidth}
      \centering
          \includegraphics[angle=0,width=0.9\textwidth,clip=true,angle=0]{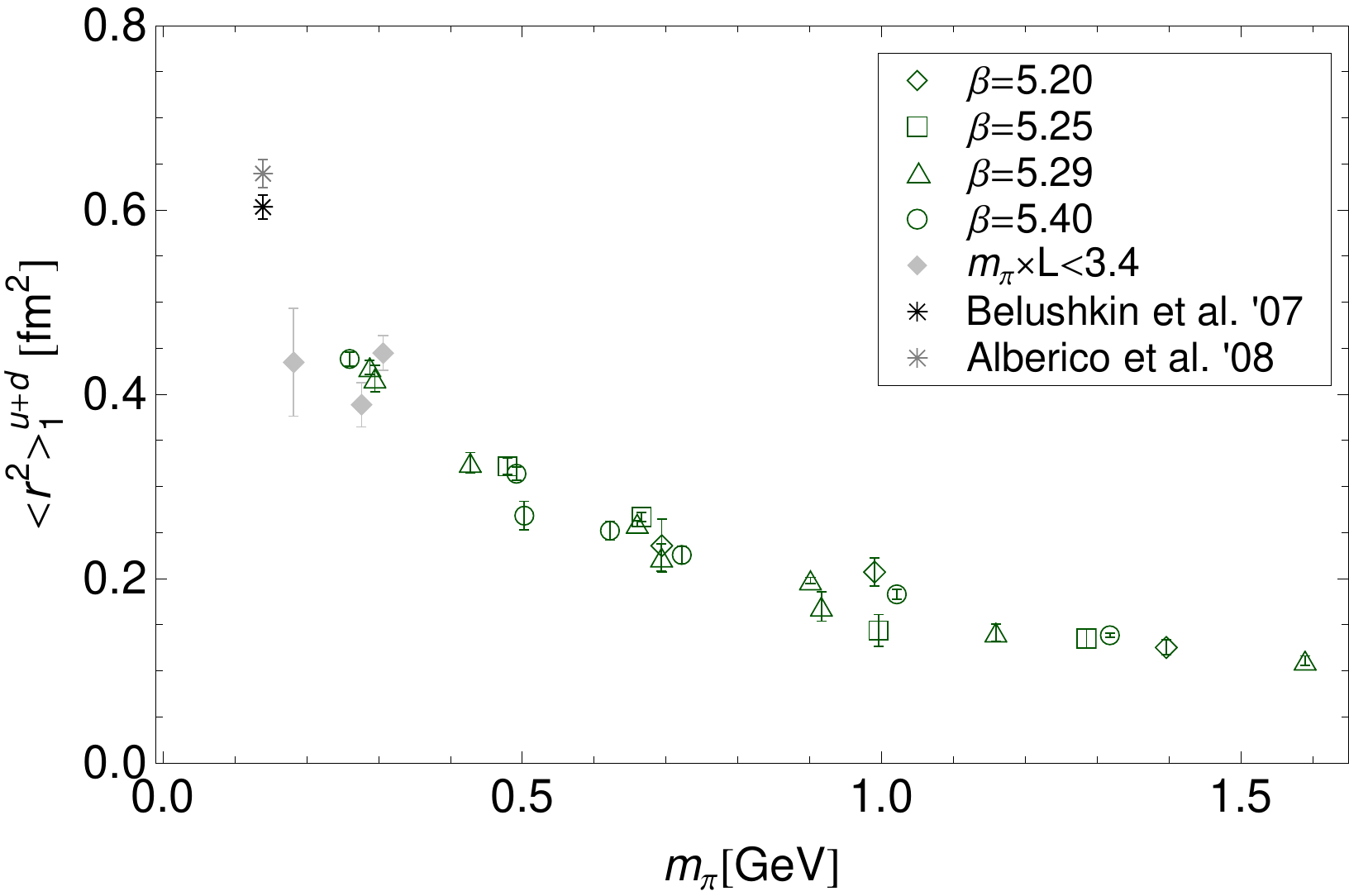}
  \caption{Isosinglet Dirac radius versus $m_\pi$, as obtained from fits to $F^{u+d}_1$ using Eq.~(\ref{eqF1poly}).\newline \newline \newline \newline }
  \label{r1sqrdupdPoly}
     \end{minipage}
 \end{figure}

An overview of our results for the isovector Dirac radius, $\langle r^2\rangle^{u-d}_1$, as a function of $m_\pi$
is provided in Fig.~\ref{r1sqrdPoly}, as obtained from 
the polynomial ansatz in Eq.~(\ref{eqF1poly}).

Although the results from the polynomial ansatz are somewhat larger at lower values of $m_\pi$ relative to the standard dipole fits, 
we find that even at the lowest accessible pion masses of $200-300\MeV$, the lattice data points are about 50\% below the phenomenological
and experimental results. 
It is interesting to point out, however, that we observe an upwards trend for $m_\pi<400\MeV$
that does not seem to follow the otherwise rather linear pion mass dependence of the data points.
In passing, we also note that there is a significant, so far unresolved difference between the values at the physical
point obtained from the recent muonic hydrogen measurements \cite{Pohl:2010zz} and the PDG \cite{Nakamura:2010zzi}. 
Incidentally, the result from an
earlier dispersion relation analysis of experimental form factor data \cite{Belushkin:2006qa} agrees
well with the recent muonic hydrogen study.
The observation that lattice calculations at unphysically large pion masses give mean square radii that are
significantly below experiment has been made already in a number of previous publications, e.g. 
\cite{Gockeler:2003ay,Boinepalli:2006xd,Syritsyn:2009mx,Yamazaki:2009zq,Bratt:2010jn,Alexandrou:2011db}.
In combination with a detailed study of potential discretization and finite volume effects, as well as contaminations from excited states in
section \ref{sec:systematics}, we will come to the conclusion that indeed a strong pion mass dependence has to set in between
the physical pion mass and $m_\pi\sim200\MeV$. This is also in agreement with general predictions from 
chiral perturbation theory, as we will discuss below in section \ref{sec:chiral}.

Apart from the normalization, the most significant difference between 
results from the polynomial and the dipole fits are the relative positions 
of the lattice data points for small values of $m_\pi\times L<3.4$, which are most likely affected
by finite volume effects. They turn out to be residing \emph{above} the data points for larger $m_\pi\times L$ 
in the case of the dipole parametrization, and \emph{below} for the polynomial ansatz in Fig.~\ref{r1sqrdPoly}. 
Since one generically expects the radius of a hadron to decrease as the volume decreases, we find again that
the polynomial fit provides a more physical parametrization of our data.

Corresponding results for the isosinglet Dirac radius are shown in Fig.~\ref{r1sqrdupdPoly}.
Overall, the data points feature small statistical uncertainties and show only little scatter over the full range of pion masses.
Below $m_\pi<700\MeV$, we even find a remarkable upwards tendency, although
the lattice results at $m_\pi\sim250\MeV$ are still $\sim25\%$ below the expected range of values of 
$\langle r^2\rangle^{u+d}_1\sim0.60,\ldots,0.62\fm^2$ at the physical point.

A more detailed discussion of the pion mass dependence of the isovector and isosinglet 
Dirac radii will be given below in section \ref{sec:chiralDirac}.
Although we cannot exclude the presence of some discretization and finite volume effects in $\langle r^2\rangle^{u-d}_1$,
our corresponding analysis in section \ref{sec:r1sys} does not provide any indication that they are larger
than the present statistical uncertainties.

\begin{figure}[t]
    \begin{minipage}{0.48\textwidth}
        \centering
          \includegraphics[angle=0,width=0.9\textwidth,clip=true,angle=0]{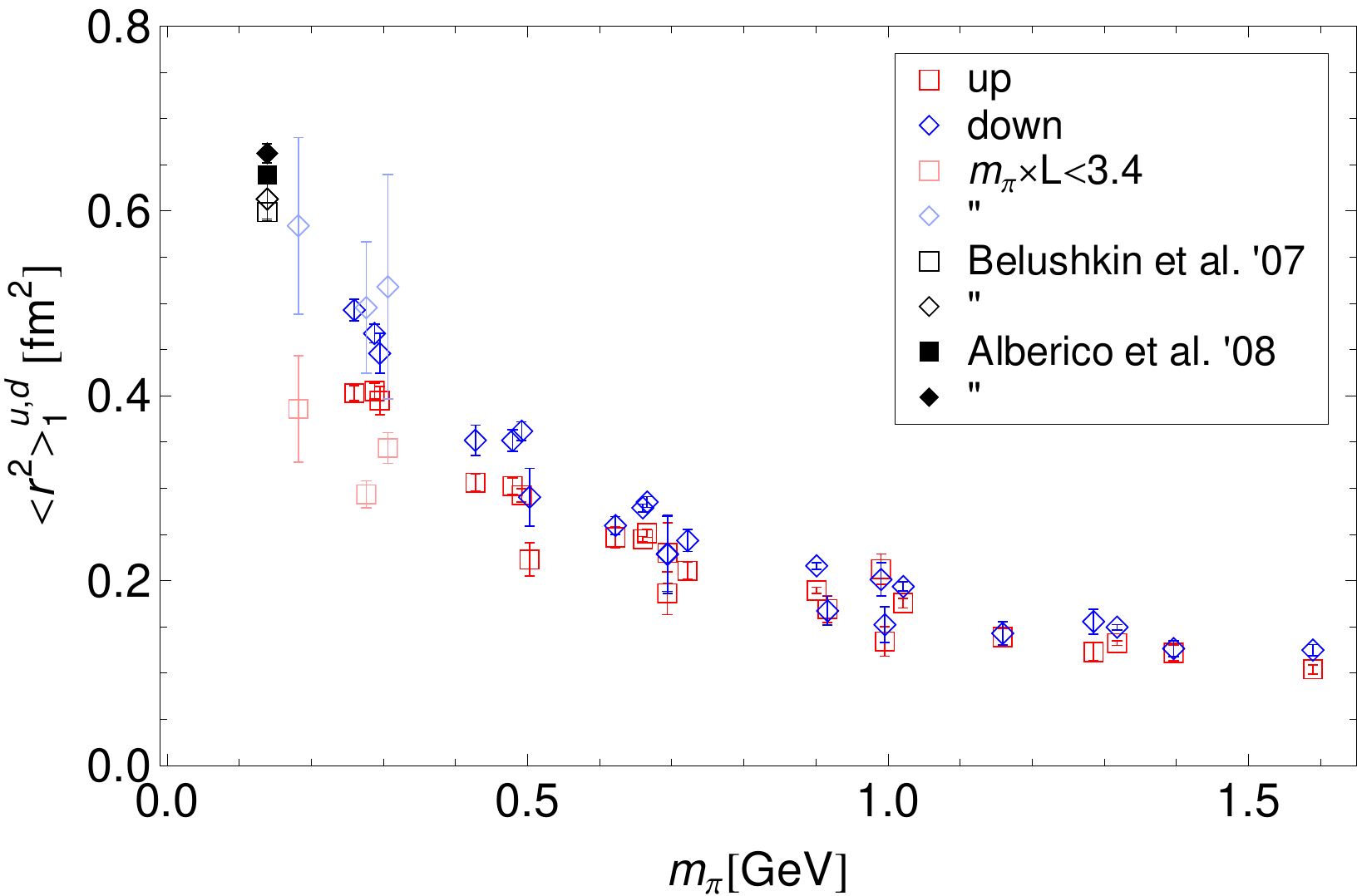}
      \caption{Comparison of Dirac radii for up and down quarks in the proton as a function of the pion mass.}
                \label{r1ud}     
     \end{minipage}
         \hspace{0.2cm}
    \begin{minipage}{0.48\textwidth}
      \centering
          \includegraphics[angle=0,width=0.9\textwidth,clip=true,angle=0]{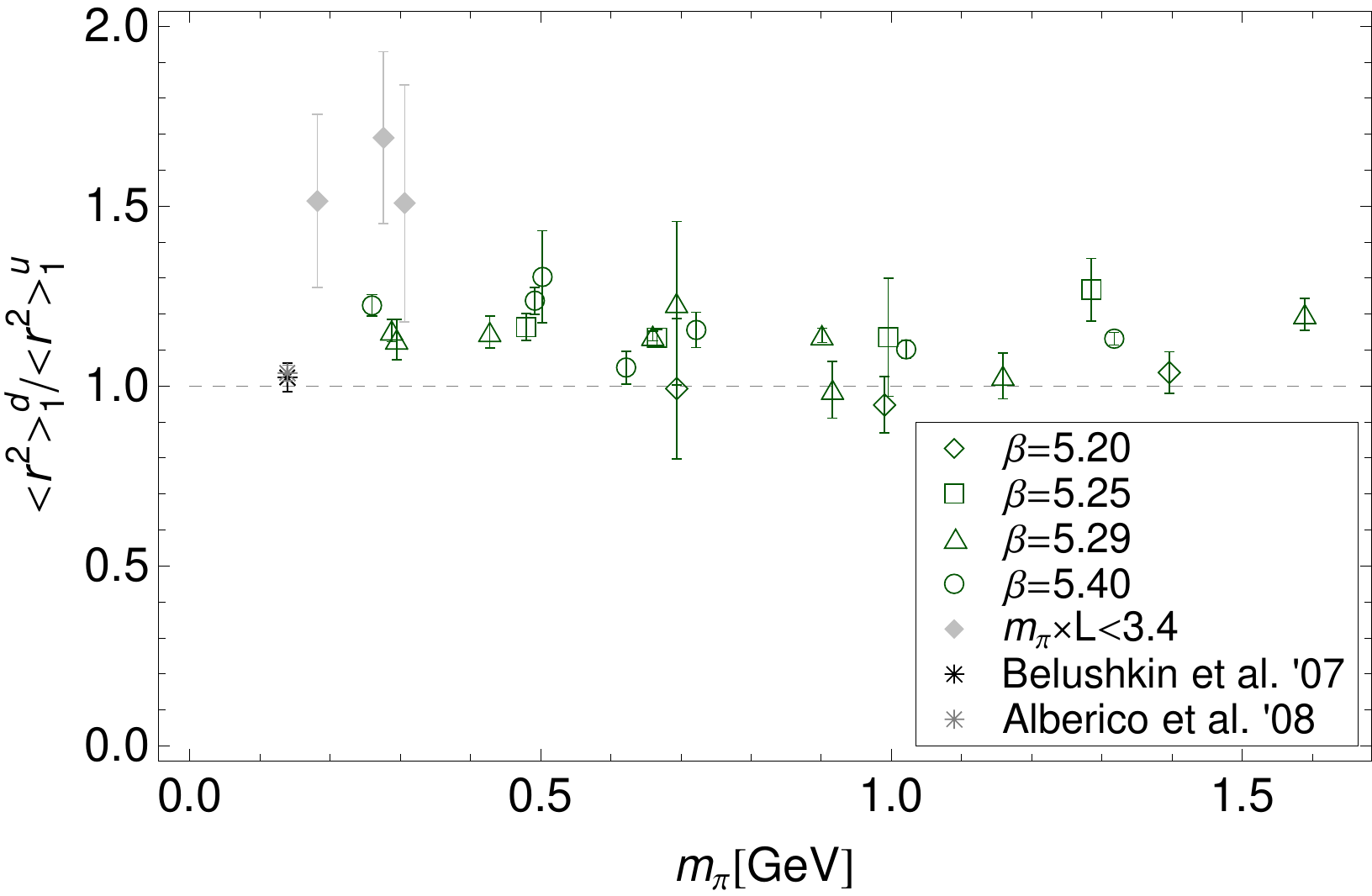}
           \caption{Ratio of the Dirac radius of down to up quarks in the proton as a function of the pion mass.}
                \label{r1uoverd}
       \end{minipage}
 \end{figure}

In conjunction with the comparison of the slopes of $F_1(Q^2)$ for up and for down quarks above 
(see the ratio $F_1^d/F^u_1$ in Fig.~\ref{F1doverF1u}),
we show in Fig.~\ref{r1ud} the corresponding Dirac radii as functions of $m_\pi$.
They were obtained from separate parametrizations of $F^u_1(Q^2)$ and $F^d_1(Q^2)$ using Eq.~(\ref{eqF1poly}).
We find that the mean square radii of the down quarks are systematically larger than those of the up quarks.
The corresponding ratio in Fig.~\ref{r1uoverd} is rather flat over the full range of pion masses,
with an average of $\langle r^2\rangle^{d}_1/\langle r^2\rangle^{u}_1\sim 1.24\pm0.13$.
The observed hierarchy is in agreement with the experimental and phenomenological results,
although the latter show a much smaller difference between the up and the down quark radius of just $\approx2-4\%$.
It will be interesting to study the origin of this feature in more detail in the future.
We also note that the substantial difference between the values obtained from the analysis of Belushkin et al. in \cite{Belushkin:2006qa} 
and the form factor parametrization of Alberico et al. \cite{Alberico:2008sz} in Fig.~\ref{F1doverF1u} is, 
as before, at least to some extent related to
a corresponding difference in the proton charge radius, $\langle r^2\rangle^{p}_E$.

\subsection{Matching to a vector meson exchange ansatz}
\label{sec:matching}

Specific contributions from vector meson exchange, the two-pion continuum, perturbative QCD etc. to nucleon form factors 
have, for example, been investigated in \cite{Belushkin:2006qa} in the framework of a dispersion relation study of experimental data.
Clearly, such a detailed analysis of the $Q^2$-dependence of $F_1$ and $F_2$ is not possible on the basis of the currently available lattice data.
Still, to get some first insight into the physics behind our preferred parametrization in Eq.~(\ref{eqF1poly}), 
we now explore a matching to a simplistic vector meson exchange ansatz of the generic form
\be
 F(Q^2)=\frac{a_1}{M_1^2+Q^2}+\frac{a_2}{M_2^2+Q^2}\,.
  \label{eqVMexchange}
\ee
In the isosinglet channel, one might expect that the lower of the two masses, say $M_1$, corresponds to the $\omega(782)$.
In contrast, in the isovector channel the two-pion continuum contribution plays a
leading role, which also generates a $\rho(770)$-meson exchange contribution.
A comparison with the simple ansatz Eq.~(\ref{eqVMexchange}) might therefore show that 
$M^{u-d}_1\sim m_\rho$.

To facilitate the matching of the two-parameter ansatz in Eq.~(\ref{eqF1poly}) with Eq.~(\ref{eqVMexchange}),
we implement, in addition to charge conservation, i.e. $\sum_ja^{u-d}_j/(M^{u-d}_j)^2=1$ for $F^{u-d}_1$ and similar for 
the isosinglet case, 
also the large-$Q^2$-behavior obtained from perturbative QCD \cite{Brodsky:1973kr},
i.e. $F_i(Q^2)\xrightarrow{Q^2\rightarrow\infty}\sim1/(Q^2)^{i+1}$, by setting
$\sum_ja_j=0$ for the Dirac form factor.
In the case of $F_2$, we would have the additional condition $\sum_ja_jM_j^2=0$.
\begin{figure}[t]
    \begin{minipage}{0.48\textwidth}
        \centering
          \includegraphics[angle=0,width=0.8\textwidth,clip=true,angle=0]{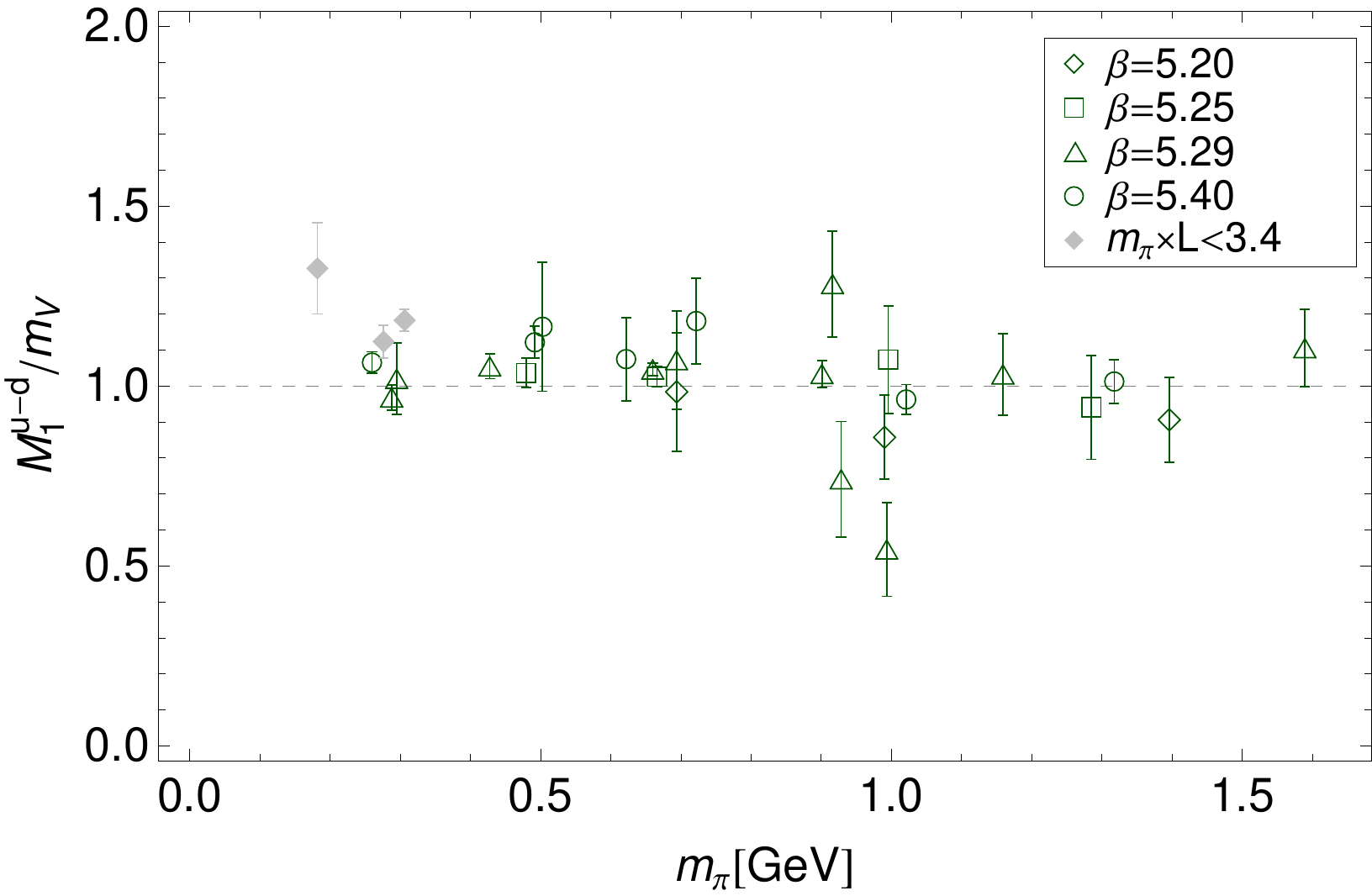}
          \includegraphics[angle=0,width=0.8\textwidth,clip=true,angle=0]{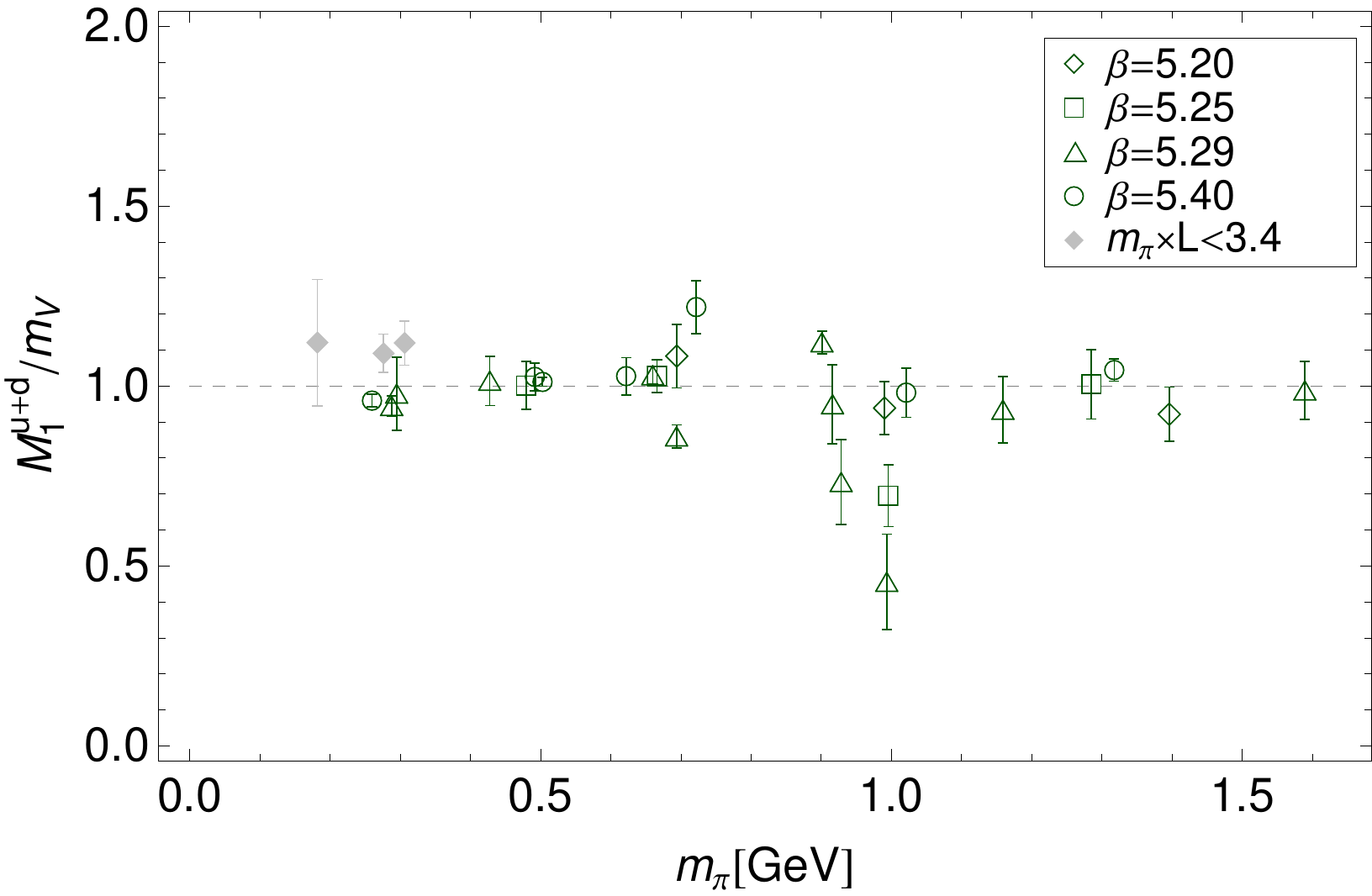}
  \caption{Ratio of the lowest pole mass extracted from $c_{12}$ and $c_{14}$ in the isovector (upper panel) and isosinglet
  (lower panel) channel, cf.~Eq.~(\ref{eqF1poly}), 
  to the lattice vector meson ($\rho$) mass as, a function of $m_\pi$.}
  \label{M1VovermV}
     \end{minipage} 
         \hspace{0.2cm}
    \begin{minipage}{0.48\textwidth}
         \includegraphics[angle=0,width=0.8\textwidth,clip=true,angle=0]{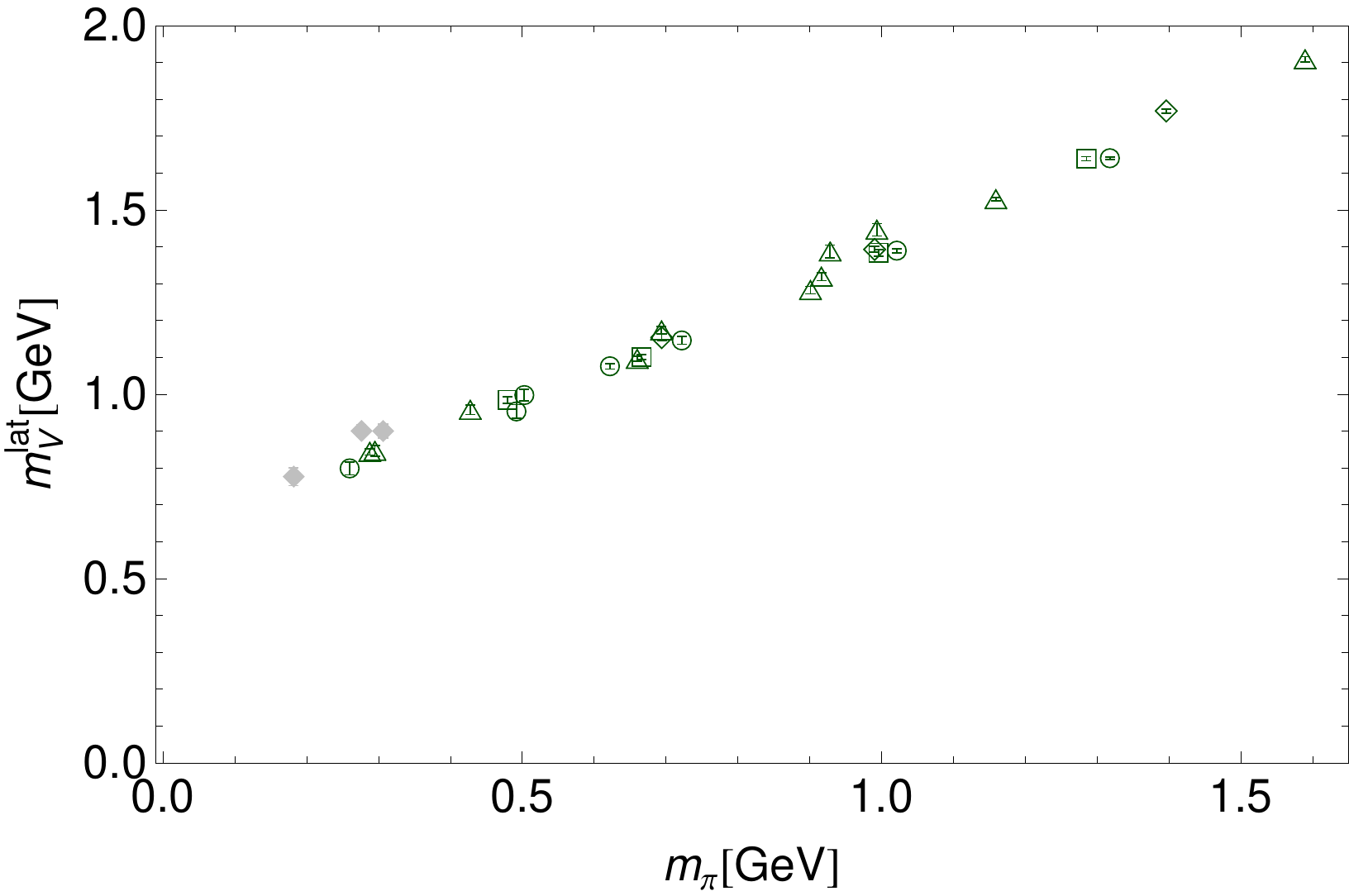}
          \includegraphics[angle=0,width=0.8\textwidth,clip=true,angle=0]{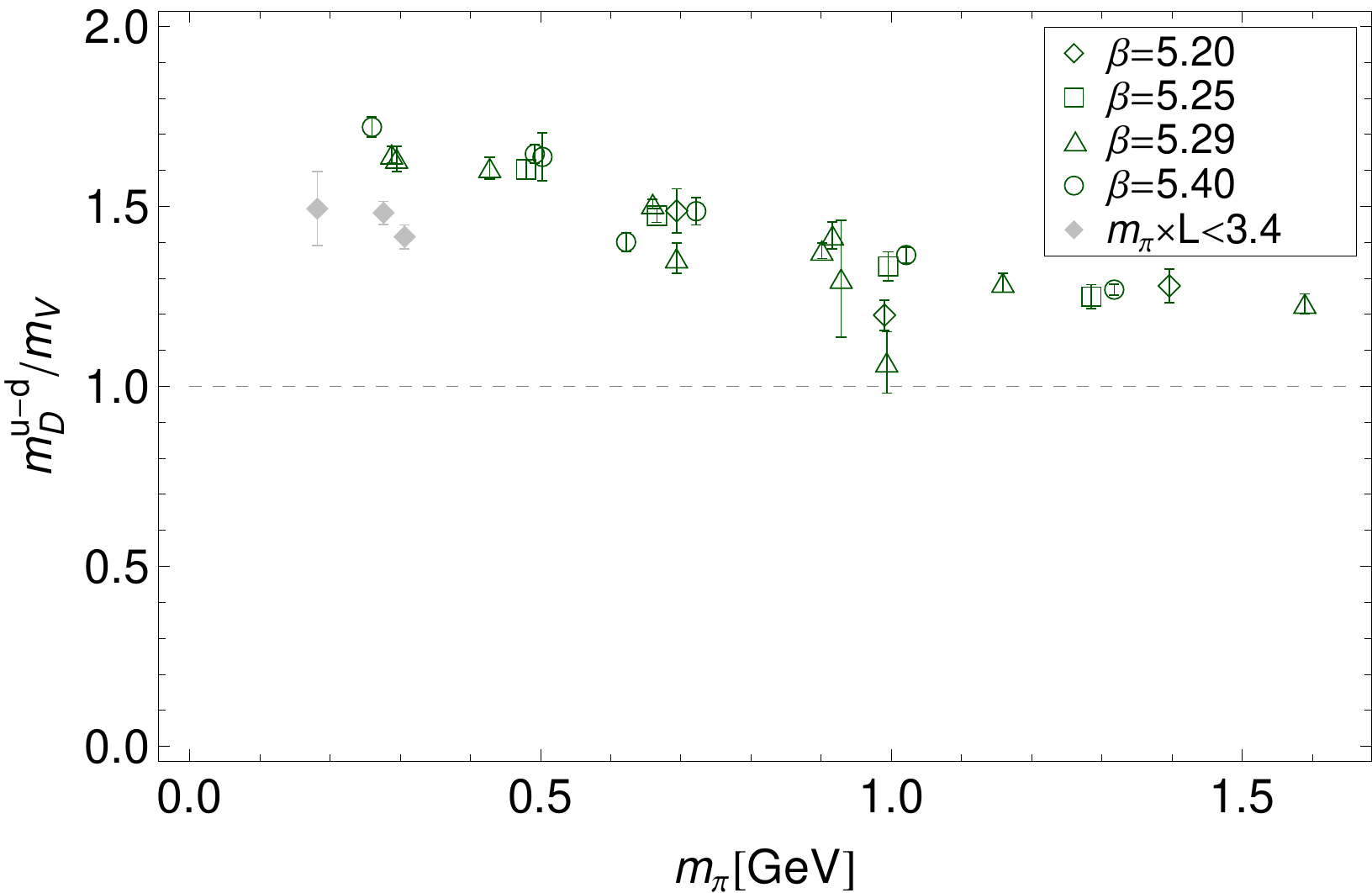}
  \caption{The upper panel shows the pion mass dependence of the lattice vector meson ($\rho$) mass.
  The lower panel shows the dipole mass $m_D$, cf. Eq.~(\ref{eqF1dipole}), over the lattice vector meson ($\rho$) mass as a function of        $m_\pi$, in the isovector channel.}
  \label{MVlatM1DoverMVlat}
     \end{minipage}
 \end{figure}
We then compute the lowest real solution for $M^2_1$ from the parameters $c_{12}$ and $c_{14}$,
which were obtained from the fits discussed above.
The numerical values for the extracted masses are provided in Appendix \ref{sec:tables}.
In Fig.~\ref{M1VovermV},
we display them
in the form of ratios $M^{u-d}_1/m_V$ and $M^{u+d}_1/m_V$ with $m_V\hat=m^{\lat}_\rho(m_\pi)$ as functions of the pion mass.
As we do not have results available for $m^{\lat}_\omega$, we use $m^{\lat}_\rho$
instead also in the isosinglet channel, expecting that $m^{\lat}_\rho\approx m^{\lat}_\omega$ also holds at larger pion masses.
Remarkably, we find that the ratios are very close to, and in most cases within errors fully compatible with, 
unity over the full range of pion masses from $m_\pi\sim1.5\GeV$ down to $m_\pi\sim0.25\GeV$.
That this is a non-trivial observation is supported by the fact that the lattice vector meson ($\rho$) mass,
which has been obtained independently,
shows a strong pion mass dependence, as illustrated in the upper panel of Fig.~\ref{MVlatM1DoverMVlat}.
This pion mass dependence is clearly compensated to a good approximation in the ratios in Fig.~\ref{M1VovermV}.
At the same time, we see from the lower panel in Fig.~\ref{MVlatM1DoverMVlat} that the pion mass dependences do not cancel out
in the ratio $m_D/m_V$ of the dipole mass, obtained from dipole fits to $F_1$, Eq.~(\ref{eqF1dipole}), 
to the lattice vector meson mass.

We interpret these results as providing strong evidence for the assumption that 
the $Q^2$-dependence of $F_1$ (within the accessible ranges) is to a significant extent 
driven by vector meson exchange contributions, in particular from the $\omega$ and $\rho$ mesons.
Furthermore, these findings provide additional support in favor of
our preferred 2-parameter parametrization in Eq.~(\ref{eqF1poly}).   

\subsection{Anomalous magnetic moment}

We now turn to a discussion of the anomalous magnetic moment, $\kappa=F_2(0)$.
As it cannot be extracted directly at $Q^2=0$ from a calculation of the Pauli form 
factor on the lattice with our methods, we have to rely on the $Q^2$-parametrizations discussed in section \ref{sec:para}. 
For the reasons given above, we will focus here on the results from the more flexible 3-parameter parametrization in Eq.~(\ref{eqF2poly}).

\begin{figure}[t]
    \begin{minipage}{0.48\textwidth}
        \centering
          \includegraphics[angle=0,width=0.9\textwidth,clip=true,angle=0]{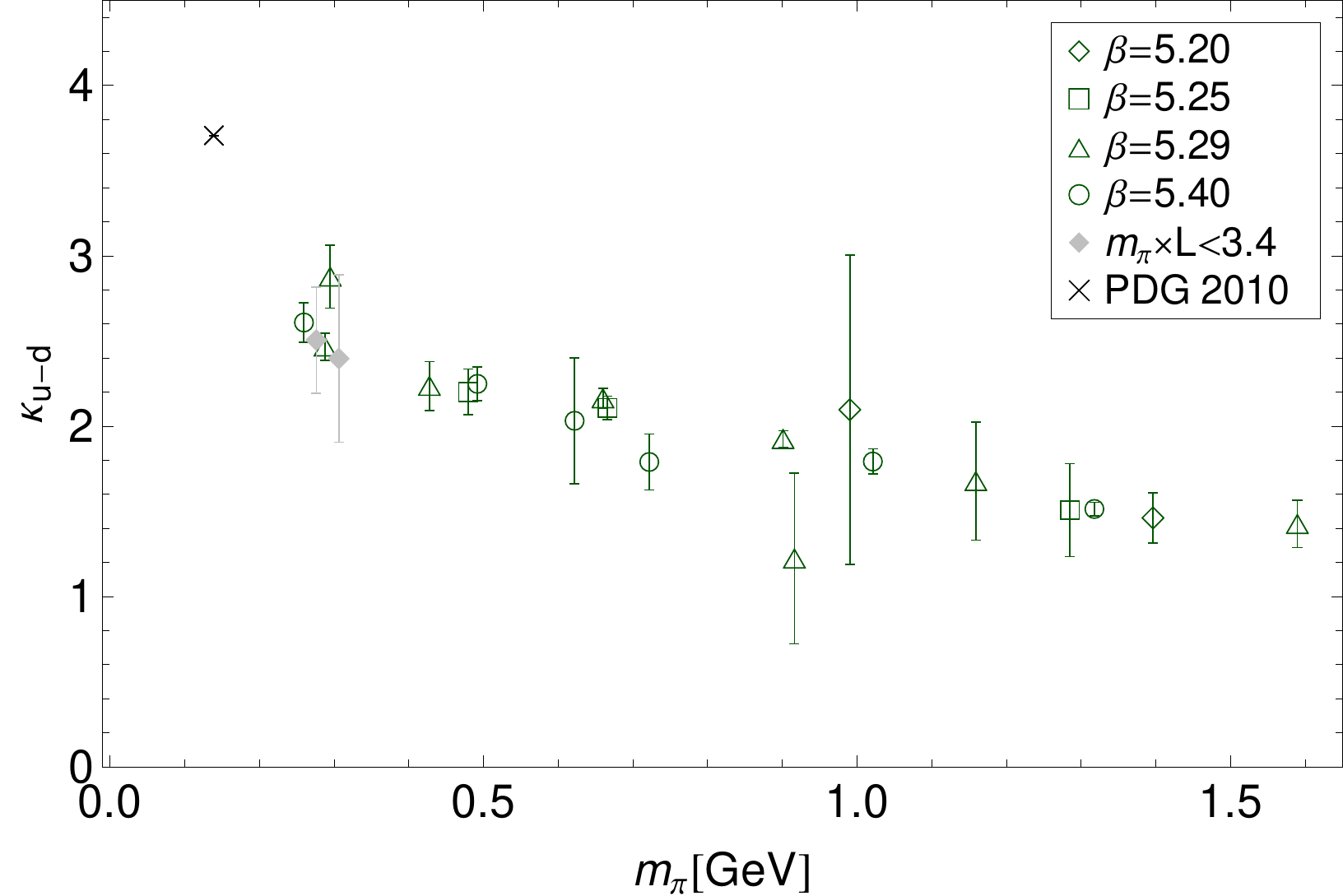}
  \caption{Isovector anomalous magnetic moment versus $m_\pi$, as obtained from fits to $F^{u-d}_2$ using Eq.~(\ref{eqF2poly}).}
  \label{kappaVPoly}
     \end{minipage} 
         \hspace{0.2cm}
    \begin{minipage}{0.48\textwidth}
      \centering
          \includegraphics[angle=0,width=0.9\textwidth,clip=true,angle=0]{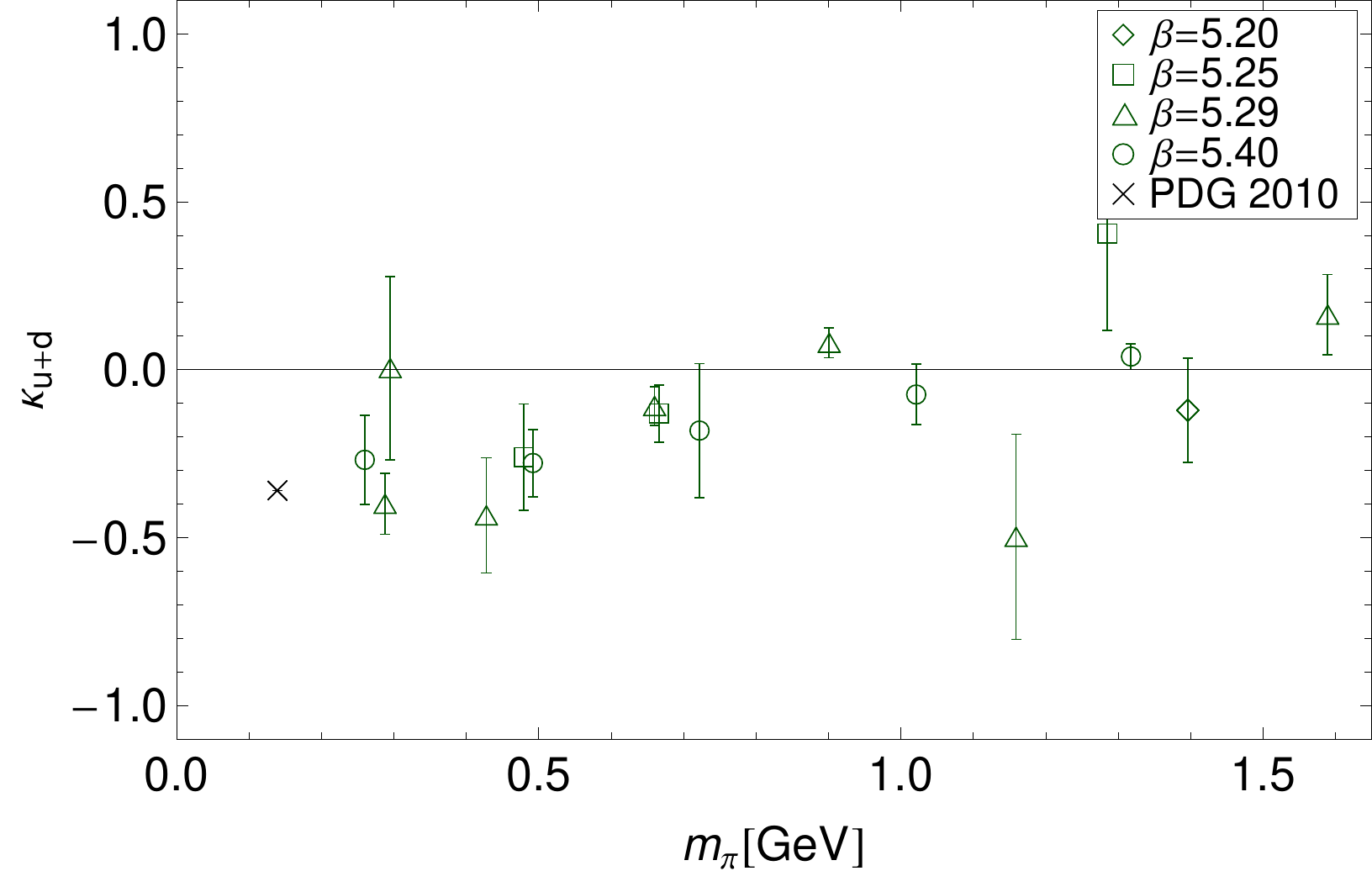}
  \caption{Isosinglet anomalous magnetic moment versus $m_\pi$, as obtained from separate fits to $F^{u}_2$ and $F^{d}_2$ using Eq.~(\ref{eqF2poly}).}
  \label{kappaSPoly}
     \end{minipage}
 \end{figure}

Our results for $\kappa_{u-d}$ as a function of the pion mass are displayed in Fig.~\ref{kappaVPoly}.
While the data points are systematically rising as we approach lower pion masses,
they are still about $25\%$ below the precisely known experimental value of $\kappa_{u-d}=3.7058893$
at the lowest accessible pion masses of $\sim200-300\MeV$.

Corresponding results for the isosinglet ($u+d$) channel are shown in Fig.~\ref{kappaSPoly}.
Since the magnitude of $F^{u+d}_2$ is much smaller than that of $F^{u-d}_2$, the respective lattice data points
are in many cases very close to or even compatible with zero, cf. Fig.~\ref{F2s},
making a reliable extrapolation in $Q^2$ very difficult.
We therefore have fitted the contributions from up and down quarks separately employing the
polynomial ansatz in Eq.~(\ref{eqF2poly}), and subsequently computed $\kappa_{u+d}$ (as well as $\langle r^2\rangle^{u+d}_2$) 
from the individual parts.
While most of the resulting data points in Fig.~\ref{kappaSPoly}
are again compatible with zero within uncertainties, we still can observe
a systematic trend towards negative values at lower pion masses.
In the region $m_\pi<500\MeV$, we even see an overlap with the experimental value within uncertainties.

We will take a closer look at the pion mass dependence of $\kappa$ below in section \ref{sec:kappachiral}. 
As before, we do not find any indications for statistically
significant systematic discretization or finite volume effects for this observable,
as will be discussed in section \ref{sec:r1sys}.

\subsection{Pauli radius}
\label{sec:r2}

\begin{figure}[t]
    \begin{minipage}{0.48\textwidth}
          \includegraphics[angle=0,width=0.9\textwidth,clip=true,angle=0]{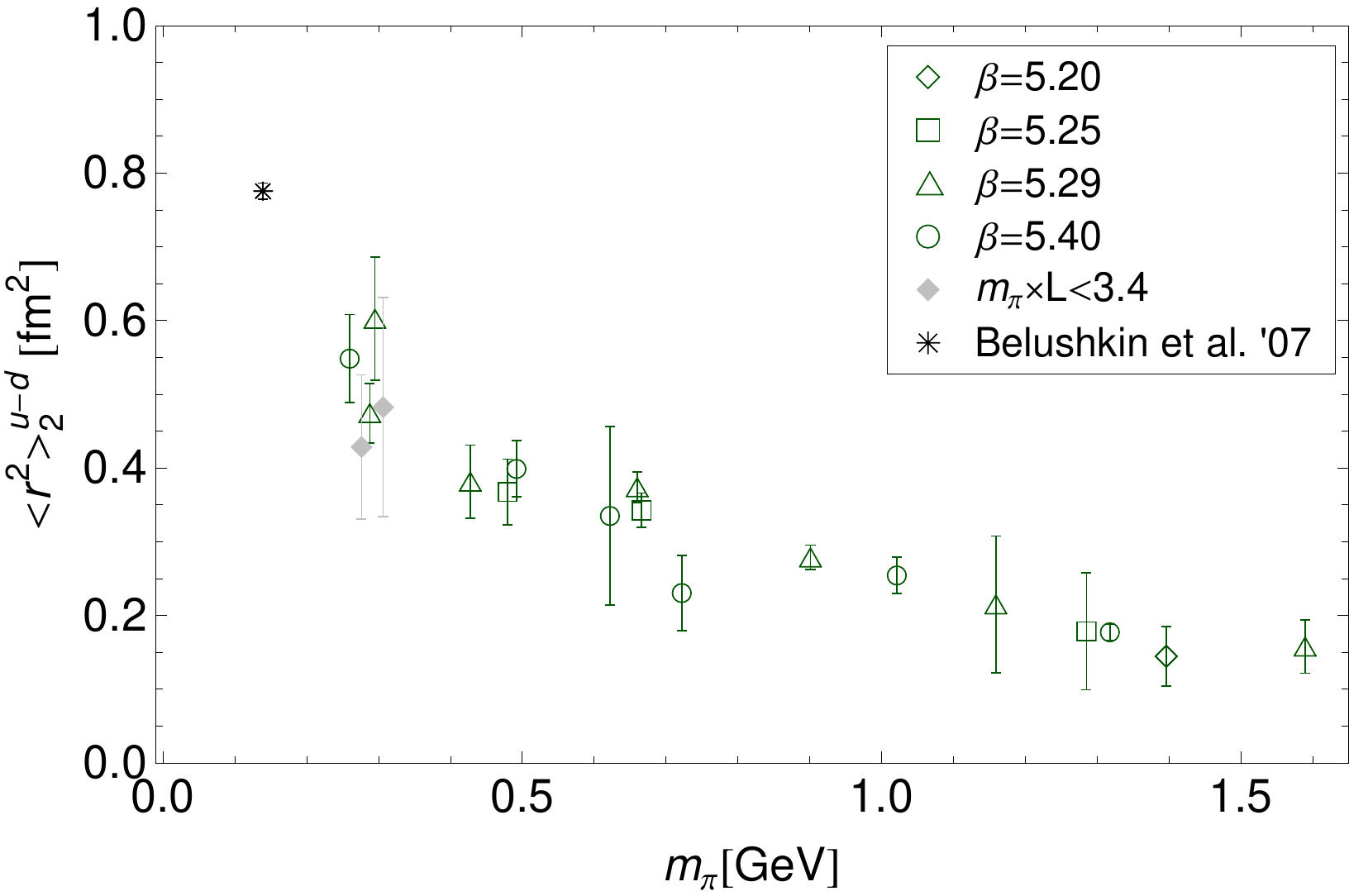}
  \caption{Isovector Pauli radius versus $m_\pi$, as obtained from fits to $F^{u-d}_2$ 
  using Eq.~(\ref{eqF2poly}) (cf. Fig.~\ref{F2umdTripolePoly}).\newline \newline \newline}
  \label{r2sqrdPoly}
  \end{minipage} 
         \hspace{0.2cm}
    \begin{minipage}{0.48\textwidth}
      \centering
          \includegraphics[angle=0,width=0.9\textwidth,clip=true,angle=0]{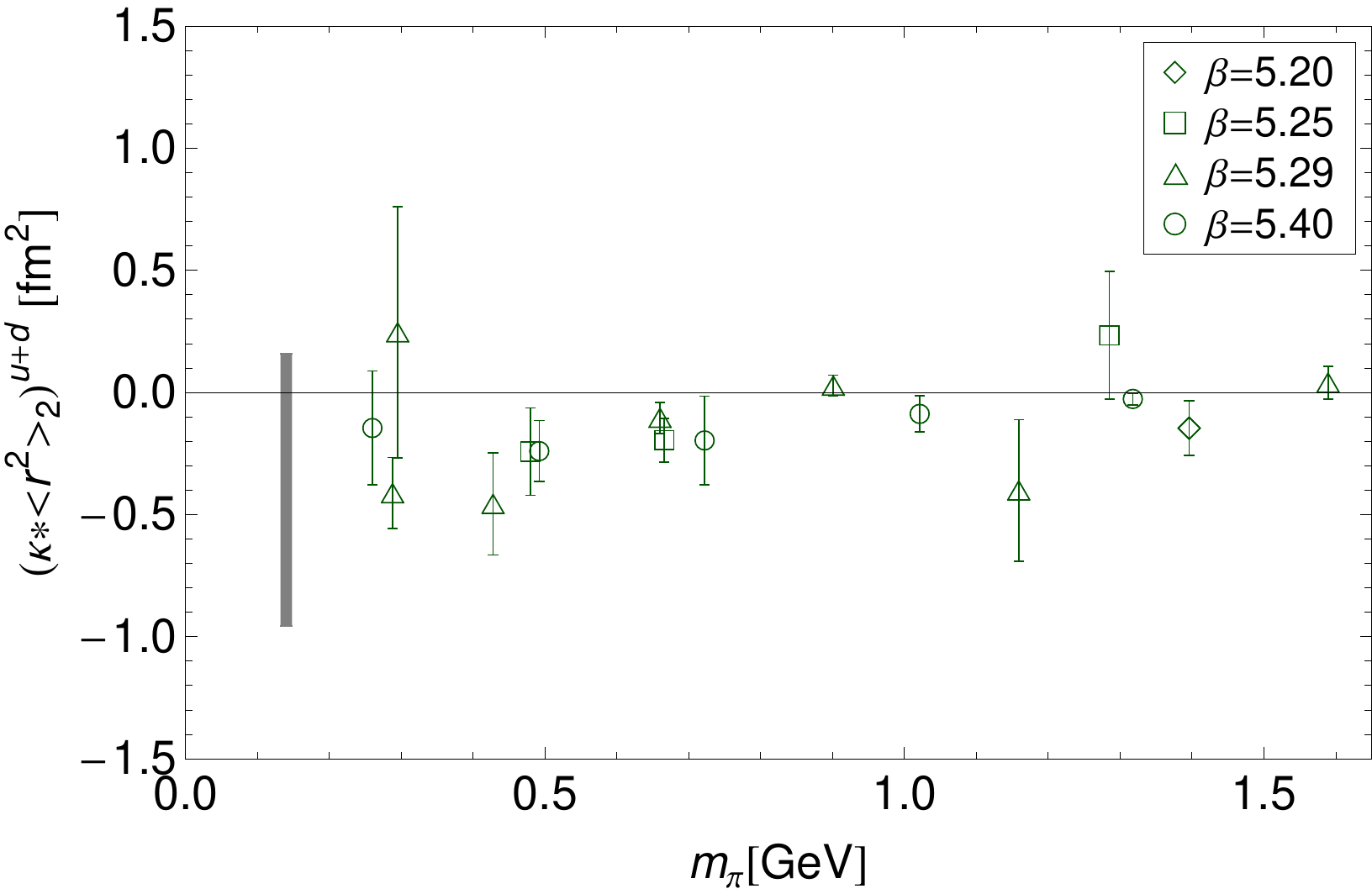}
      \caption{Results for the slope of the Pauli radius for $(u+d)$-quarks in the proton, 
      $(\kappa\times\langle r^2\rangle_2)^{u+d}$, as a function of the pion mass.
      A range of results from experiment and phenomenology is illustrated by the shaded vertical band at the physical pion mass.}
     \label{kappar2upd}  
     \end{minipage}
 \end{figure}

The Pauli radius, $\langle r^2\rangle_2$, is given by the slope of $F_2(Q^2)$ at zero momentum transfer.
Since the lowest values of $Q^2$ for which we can access $F_2$ are in the range of $Q^2\sim0.15,\ldots,0.5\GeV^2$
(for standard periodic boundary conditions in spatial directions, and depending on the lattice parameters),
the computation of the slope heavily relies on the employed parametrization of the $Q^2$-dependence.
The results for our preferred polynomial, Eq.~(\ref{eqF2poly}), ansatz are displayed in Fig.~\ref{r2sqrdPoly} for the isovector case.
Overall, we find that the central values for the polynomial parametrization are higher than for the tripole ansatz.
At the lowest accessible pion masses, the results from the polynomial ansatz in Fig.~\ref{r2sqrdPoly} 
are just about $20\%$ below the phenomenological value.
In contrast, one finds that the corresponding lattice data points from the tripole parametrization
are about $40-50\%$ below the phenomenological number.
Not surprisingly, the uncertainties from the more flexible 3-parameter fits are significantly larger, 
and potentially more realistic, than for the 2-parameter tripole ansatz.
As in section \ref{sec:para} above, we prefer also in this case the more general 
polynomial ansatz over the standard dipole or tripole parametrizations.
With respect to $F_2(Q^2)$, however, a more conclusive assessment probably has to be based on lattice results obtained 
in larger volumes or employing (partially) twisted boundary conditions
in order to get access to lower and more densely spaced values of $Q^2$.

With respect to the isosinglet channel, we first note that $\langle r^2\rangle_2$
can be written as $\langle r^2\rangle^{}_2=-6\rho_2/\kappa$, where 
$\rho_2=dF_2(Q^2)/dQ^2|_{Q^2=0}$ is the slope of the Pauli form factor.
Since $\kappa_{u+d}$ turns out to be small and mostly compatible with zero within errors 
over a wide range of pion masses, cf.\ Fig.~\ref{kappaSPoly},
we will avoid the resulting substantial uncertainties in $\langle r^2\rangle^{u+d}_2$ by considering
instead the slope alone, $-6\rho_2=(\kappa\times\langle r^2\rangle_2)^{u+d}$.

It is also interesting to note that for the isosinglet Pauli radius, or more precisely the slope 
$(\kappa\times\langle r^2\rangle_2)^{u+d}$, one finds a rather widespread range of values from experiment and phenomenology:
The super-convergence approach of Ref.~\cite{Belushkin:2006qa} gives 
(with a Dirac charge radius of $\langle r^2\rangle^p_E\sim0.84\fm^2$ that is close to the recent measurement by Pohl et al. \cite{Pohl:2010zz})
$(\kappa\times\langle r^2\rangle_2)^{u+d}=0.04\pm0.12\fm^2$, where we have obtained the uncertainty
from a standard (uncorrelated) error propagation.
From the same publication \cite{Belushkin:2006qa} the ``Recent determinations" from Table I give
$(\kappa\times\langle r^2\rangle_2)^{u+d}=-0.28\pm0.52\fm^2$, for $\langle r^2\rangle^p_E\sim0.88\fm^2$ 
that is closer to the PDG value \cite{Nakamura:2010zzi}.
An even larger negative value can be obtained from the parametrization of Ref.~\cite{Alberico:2008sz},
$(\kappa\times\langle r^2\rangle_2)^{u+d}=-0.66\pm0.29\fm^2$ (taking into account the error correlation matrix).
In Fig.~\ref{kappar2upd}, we show our lattice results for $(\kappa\times\langle r^2\rangle_2)^{u+d}$
together with the estimated range of phenomenological values.
While the lattice data points at large pion masses are mostly close to, and within uncertainties compatible with, zero,
we observe a trend towards non-zero, negative values below $m_\pi\approx700\MeV$.
Accordingly, our results at the lowest pion masses, with 
$(\kappa\times\langle r^2\rangle_2)^{u+d}\approx0.0,\ldots,-0.7$, 
are fully compatible with the wide range of values from experiment and phenomenology, 
indicated by the shaded band.
We expect the use of (partially) twisted boundary conditions to be of great help
in order to pin down the parametrization of the $Q^2$-dependence of $F^{u+d}_2$ at small $Q^2$, 
which should lead in turn to significantly more precise values for $(\kappa\times\langle r^2\rangle_2)^{u+d}$.

\section{Chiral extrapolations}
\label{sec:chiral}

In the literature, baryon ChPT has often been applied to lattice results with the goal to extrapolate 
them from pion masses $\gtrapprox300\MeV$ downwards to the chiral limit and to obtain in this way
``a priori" or ``a posteriori" predictions at the physical point. 
In this process, less known or previously unknown low energy constants (LECs) are treated as free fit parameters 
and are thereby determined from the available lattice data points and their respective uncertainties.
It is important to keep in mind, however, that in order to provide more than a mere parametrization of the $m_\pi$-dependence of the data, 
such an approach has to rely on the assumptions that (i) the systematic uncertainties of
the lattice calculation are reasonably well under control, and (ii) that the particular ChPT-formula to the
given order is applicable at 
larger pion masses in the first place.
While we are able to study the systematic uncertainties to some extent directly on the basis of our extensive sets of 
lattice ensembles and results, cf. section \ref{sec:systematics},
the latter assumption is generically hard to justify, in particular since for most nucleon observables results from (H)BChPT
are only available at the 1-loop level, and higher order corrections are difficult to quantify.

Therefore, in this work, we follow a somewhat different path: 
\begin{itemize}
\item Well-known constants like $g_A$ of $f_\pi$ will be fixed as usual 
to either their physical or their chiral limit values, see Table \ref{lecs}. 
The resulting uncertainties due to a variation of the constants between these
values will be studied for selected observables. 
\item Low energy constants whose values are at least approximately known will be varied in reasonably wide ranges to assess the related uncertainties.
\item Central physical quantities of interest in this work, in particular the chiral limit value of the anomalous magnetic moment, $\kappa^0$,
as well as regularization scale dependent counter term parameters, will be treated as free fit parameters.
They will be determined by fits preferably only to the experimental and phenomenological values of the 
observable under consideration at the physical point. Only if this turns out to be insufficient, 
we will include our lattice data points for pion masses \emph{below} $260\MeV$ in the fit.
\end{itemize}
In essence, we attempt an ``upwards extrapolation" from $m_\phys$ towards lattice data points at larger pion masses.
In combination with our study of potential systematic uncertainties, 
this approach provides an opportunity to assess the applicability of the different available 
ChPT-schemes in the range in between the physical pion mass and typical lowest lattice pion masses
of $m_\pi\sim200,\ldots,400\MeV$.

\subsection{Dirac radius}
\label{sec:chiralDirac}

In \cite{Hemmert:2002uh,Gockeler:2003ay}, the $m_\pi$ and $Q^2$-dependence of the nucleon vector form factors was studied 
in the small scale expansion (SSE), a heavy baryon scheme with explicit $\Delta$-degrees of freedom, to $\mathcal{O}(\eps^3)$.
The resulting pion mass dependence of $\langle r^2\rangle_{1}^{u-d}$ is given by
\bea
 \langle r^2\rangle_1^{u-d,\text{SSE}}&=&
    -  \frac{1}{(4\pi f_\pi)^2}\left\{1+7 g_{A}^2 +
  \left(10 g_{A}^2 +2\right) \ln\left(\frac{m_\pi}{\lambda}\right)\right\} 
     +  \frac{c_A^2}{54\pi^2 f_{\pi}^2}\Bigg\{
        26+30\ln\left(\frac{m_\pi}{\lambda}\right)  \nonumber\\
     &+& 30\frac{\deltam}{\sqrt{\deltam^2-m_{\pi}^2}}
       \ln\left(\frac{\deltam}{m_\pi}
      + \sqrt{\frac{\deltam^2}{m_{\pi}^2}-1}\right) \Bigg\}
      + \frac{12 B_{10}^{(r)}(\lambda)}{(4\pi f_\pi)^2} \,,
\label{eqr1SSE}
\eea
which depends on four LECs, the pion decay constant $f_\pi$, the isovector axial vector coupling constant $g_A$, 
the axial vector pion-nucleon-$\Delta$ coupling constant $c_A=g_{\pi N\Delta}$, 
and the $\Delta$-nucleon mass difference $\deltam=m_\Delta-m_N$, as well as a
counter term $B_{10}^{(r)}(\lambda)$ that removes the 
regularization scale dependence\footnote{Here and below, an analytic continuation of the form   
$(r^2-1)^{\pm 1/2}\ln\left(r + \sqrt{r^2-1}\right)\rightarrow \mp (1-r^2)^{\pm 1/2}\arccos(r)$ with $r=\deltam/m_\pi$
is regarded as implicit for $m_\pi>\deltam$, i.e. $r<1$.}.
Generically, the LECs are taken in the chiral limit, i.e. $f_\pi=f_\pi^0$ etc., however to the order considered, 
they can as well be taken at the physical point.
Equation (\ref{eqr1SSE}) shows explicitly the well-known logarithmic $\ln m_\pi$ divergence that is expected in the chiral limit of 
$\langle r^2\rangle_1^{u-d}$.

From Eq.~(\ref{eqr1SSE}), the leading 1-loop HBChPT result (see, e.g., \cite{Bernard:1992qa,Bernard:1995dp}) can be easily recovered
by setting $c_A=0$, giving
\bea
 \langle r^2\rangle_1^{u-d,\text{HBChPT}}&=&
    -  \frac{1}{(4\pi f_\pi)^2}\left\{1+7 g_{A}^2 +
  \left(10 g_{A}^2 +2\right) \ln\left(\frac{m_\pi}{\lambda}\right)\right\} 
      + \frac{12 B_{10}^{(r)}(\lambda)}{(4\pi f_\pi)^2} \,.
\label{eqr1HBChPT}
\eea

\begin{figure}[t]
    \begin{minipage}{0.48\textwidth}
        \centering
          \includegraphics[angle=0,width=0.9\textwidth,clip=true,angle=0]{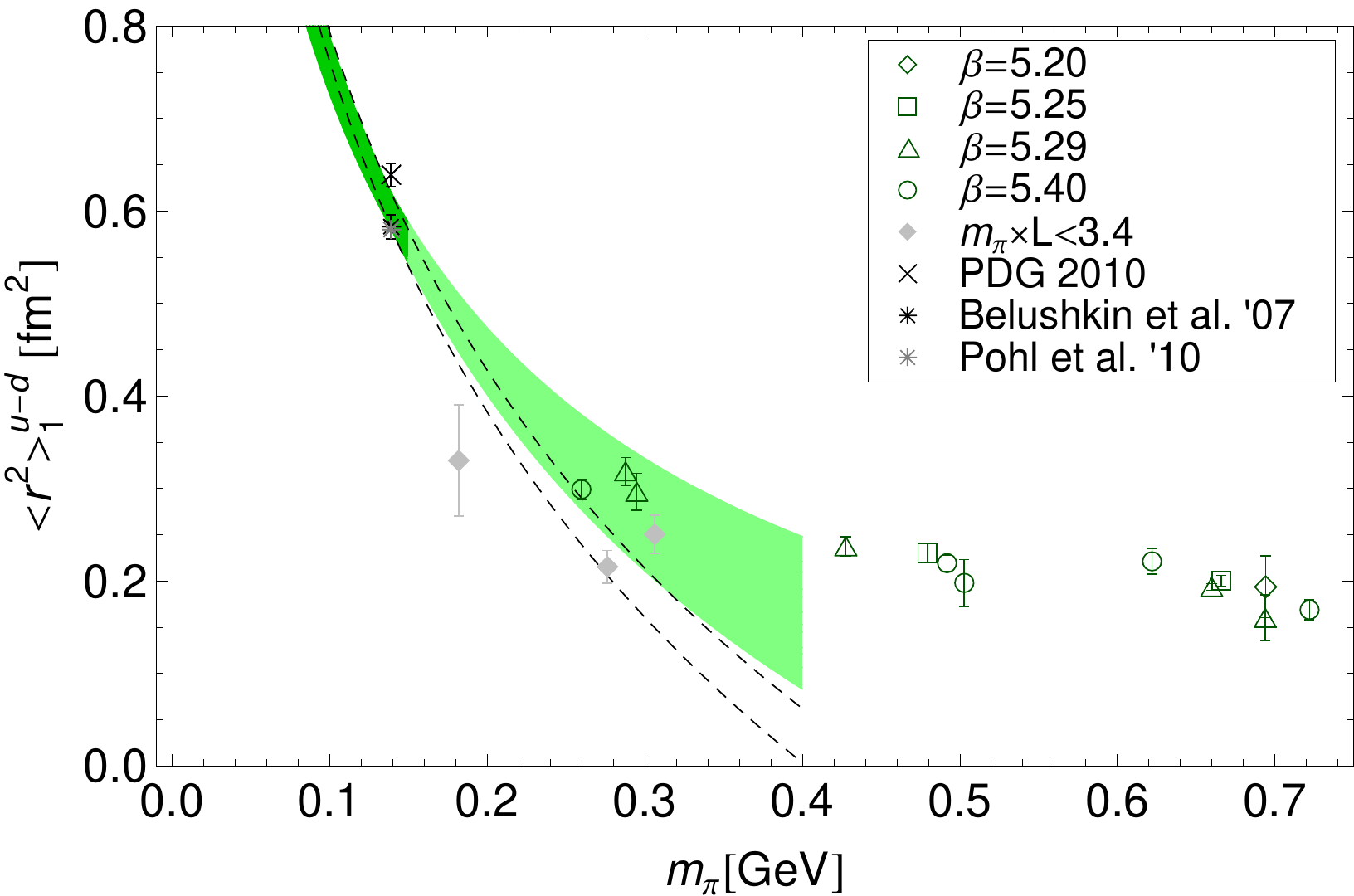}
       \caption{Pion mass dependence of the isovector Dirac radius, as obtained from fits to $F^{u-d}_1$ using Eq.~(\ref{eqF1poly}).
       The band outlined by the dashed curves, and the shaded band represent heavy baryon and SSE chiral extrapolations, respectively.
       For the details, see Eqs.~(\ref{eqr1SSE}) and (\ref{eqr1HBChPT}) and the surrounding text.\newline\newline
       }
  \label{r1sqrdChPT}
    \end{minipage}
             \hspace{0.2cm}
      \begin{minipage}{0.48\textwidth}
        \centering
          \includegraphics[angle=0,width=0.9\textwidth,clip=true,angle=0]{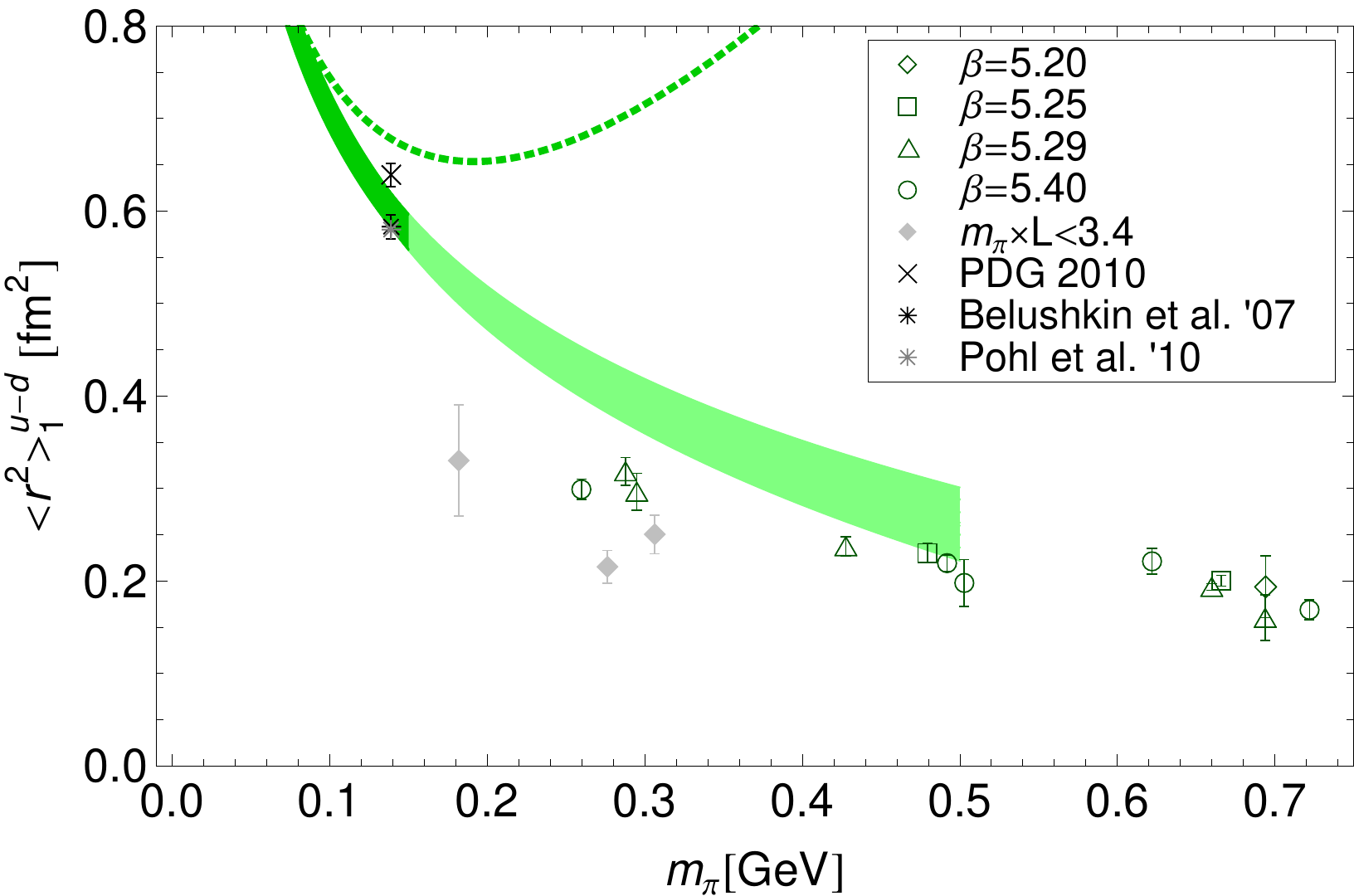}
       \caption{Pion mass dependence of the isovector Dirac radius, as obtained from fits to $F^{u-d}_1$ using Eq.~(\ref{eqF1poly}).
       The shaded band represents a BChPT-fit 
       %of the BChPT result in Eq.~(\ref{eqr1BChPT}) 
       to an average of the 
       phenomenological and experimental values at the physical point with the counter term as the only fit parameter.
   %    Lattice data points were not included in the fit.
       The dotted line represents the heavy-baryon limit of the central covariant fit result.
 %      The error band also includes uncertainties in the coupling $c_6\hat=\kappa^{0,u-d}$, which has been varied in the range
  %     of $4.0\ldots6.0$.
        For the details, see Eq.~(\ref{eqr1BChPT}) and the surrounding text.}
  \label{r1sqrdBChPT}
    \end{minipage}
 \end{figure}

We have employed both the SSE result in Eq.~(\ref{eqr1SSE}) as well as the HBChPT expression in Eq.~(\ref{eqr1HBChPT})
to extrapolate from the physical pion mass upwards in $m_\pi$ towards the lattice data points. 
The counter term $B_{10}^{(r)}$ was in both cases fitted to the
average phenomenological value at the physical point. 
This was done for a range of values of the low energy constants 
$f_\pi$, $g_A$, $m_N$ and $\deltam$, which have been varied in between their physical
and chiral limit values, cf. Table \ref{lecs}.
The coupling $c_A$ in the SSE approach has been varied at the same time in the range of $c_A=1,\ldots,1.5$.
The outcome of this procedure is shown in Fig.~\ref{r1sqrdChPT}, where
the dashed lines outline the uncertainty band from the heavy baryon fits, and the shaded band represents the SSE approach.
Here and below, a lighter shading is used for the ChPT-extrapolation band 
for pion masses larger than those included in the fit.
It is interesting to see that both approaches show a rapidly decreasing isovector Dirac radius
as the pion mass increases, even leading to an overlap with the lattice data points at $m_\pi\approx250,\ldots,300\MeV$. 
We find that the adjusted counter term parameter $B_{10}^{(r)}$ varies significantly for
the different combinations of parameters and ChPT-approaches: 
In the SSE-approach, $B_{10}^{(r)}\approx-1.28,\ldots,-0.37$,
while for the HBChPT case, $B_{10}^{(r)}\approx-0.16,\ldots,0.0$, for a regularization scale of $\lambda=0.89\GeV$.
As has already been noted in \cite{Gockeler:2003ay}, it seems doubtful that these two ChPT approaches 
to the given orders are quantitatively applicable at or above the physical pion mass.
The overlap with the lattice data points should therefore be interpreted with some care, as it might be
accidental and not the result of a physically meaningful chiral extrapolation.

\begin{table}
\begin{center}
\begin{tabular}{c|c|c|c|c}
\hline
$m_\pi[\MeV]$&$f_\pi[\MeV]$&$g_A$&$m_N[\MeV]$&$\deltam[\MeV]$\\
\hline
$0$&$86$&$1.2$&$890$& $330$\\
$139$&$92$&$1.269$&$938$& $271$\\
\hline
\end{tabular}
\caption{\label{lecs}Standard low energy constants at the physical point and in the chiral limit (estimated).
We denote the $\Delta$-nucleon mass difference by $\deltam=m_\Delta-m_N$. 
}
\end{center}
\end{table}

Results for the pion mass dependence of $\langle r^2\rangle_1^{u-d}$ in the covariant BChPT-scheme of Ref.~\cite{Dorati:2007bk} 
(without explicit $\Delta$-DOFs) have been obtained in \cite{Gail:2007}.
To $\mathcal{O}(p^4)$, it reads
\bea
 \langle r^2\rangle_1^{u-d,\text{BChPT}}&=&
  B_{c1} + (r^2_1)^{u-d,(3)} +
  (r^2_1)^{u-d,(4)} \,,
\label{eqr1BChPT}
\eea
where the individual higher order contributions $(r^2_1)^{u-d,(3,4)}$ are given in Appendix \ref{app} in 
Eqs.~(\ref{eqr1BChPT2a0}) to (\ref{eqr1BChPT2b}).
It is interesting to note that 
in contrast to the SSE-expansion to 
$\mathcal{O}(\eps^3)$, the $\mathcal{O}(p^4)$-contribution in Eq.~(\ref{eqr1BChPT2b}) introduces a dependence
on the coupling $c_6$, which determines the isovector anomalous magnetic moment in the chiral limit, 
$\kappa^{0,u-d}\hat=c_6$, as we will see below.
We also note that the regularization scale in Eqs.~(\ref{eqr1BChPT2a0}) to (\ref{eqr1BChPT2b}) has been set equal to
the nucleon mass in the chiral limit, $m^0_N\approx0.89\GeV$.

For the covariant BChPT extrapolation, we have varied the LEC $c_6$ in a range of $4,\ldots,6$ to account for the related
systematic uncertainties. As before, for a given value of $c_6$, we have fitted the counter term $B_{c1}$ to the average 
of the phenomenological values at $m_\pi^\phys$, giving a relatively stable $B_{c1}(m^0_N)\approx-1.5\GeV^{-2}$. 
The result is represented by the error band in Fig.~\ref{r1sqrdBChPT},
which in addition includes a variation of the standard LECs as described above. 
Compared to Fig.~\ref{r1sqrdChPT}, the extrapolation band falls off more slowly, and lies about $20\%$ above the lattice data points
at $m_\pi\sim260\MeV$.
Notably, a heavy baryon expansion of the covariant BChPT result leads to a curve that quickly bends upwards
above the physical pion mass, as illustrated by the dotted line in Fig.~\ref{r1sqrdBChPT}.
Assuming that the BChPT formula to the given order is applicable at the physical pion mass, this would indicate
that the corresponding HBChPT expansion has a much smaller radius of convergence, and starts to break down already above $m_\pi\sim100\MeV$.
In comparing HBChPT, SSE and BChPT expansions it is interesting to note that progress with respect to the isovector nucleon form factors 
has been reported very recently in BChPT including the $\Delta$-resonance in the so-called $\delta$-power counting scheme,
see \cite{Ledwig:2011iw} and references therein.

At the one-loop level in HBChPT, with or without explicit $\Delta$-DOFs, the isosinglet Dirac radius
turns out to be independent of $m_\pi$, $\langle r^2\rangle_1^{u+d,\text{HBChPT,SSE}}=\text{const.}$ \cite{Bernard:1995dp,Bernard:1998gv}.
Furthermore, while the analytical expression of the BChPT result of Ref.~\cite{Gail:2007} for $\langle r^2\rangle_1^{u+d}$ 
shows at first sight a rather non-trivial pion mass dependence, it turns out to be nearly flat 
in practice\footnote{Assuming that the anomalous magnetic moment in the chiral limit fulfills 
$0>\kappa^0_{u+d}\approx\kappa^\phys_{u+d}\sim -0.36$, which is confirmed
by the observed pion mass dependence of $\kappa_{u+d}$ and the corresponding chiral extrapolations, see, 
e.g., Ref.~\cite{Syritsyn:2009mx} and our discussion below in section \ref{sec:kappachiral}.}.
To the contrary, a rather strong pion mass dependence for this observable is observed on the lattice down 
to $m_\pi\sim 230\MeV$, cf. Fig.~\ref{r1sqrdupdPoly}.
Lacking any reason to assume that the pion mass dependence suddenly flattens off at the physical pion mass, 
we conclude that the available ChPT results for $\langle r^2\rangle_1^{u+d}$ are most likely not even qualitatively applicable
at or above $m^\phys_\pi$.
We therefore refrain from extracting the relevant LECs from fits to the phenomenological values at the physical point.
It is interesting to note that, although lacking a theoretical foundation, a 
naive linear extrapolation in $m_\pi$ of the lattice data points below $m_\pi=500\MeV$ would get reasonably close to the experimental and
phenomenological values at the physical point in Fig.~\ref{r1sqrdupdPoly}. 

\subsection{Anomalous magnetic moment}
\label{sec:kappachiral}

\begin{figure}[t]
    \begin{minipage}{0.48\textwidth}
        \centering
          \includegraphics[angle=0,width=0.9\textwidth,clip=true,angle=0]{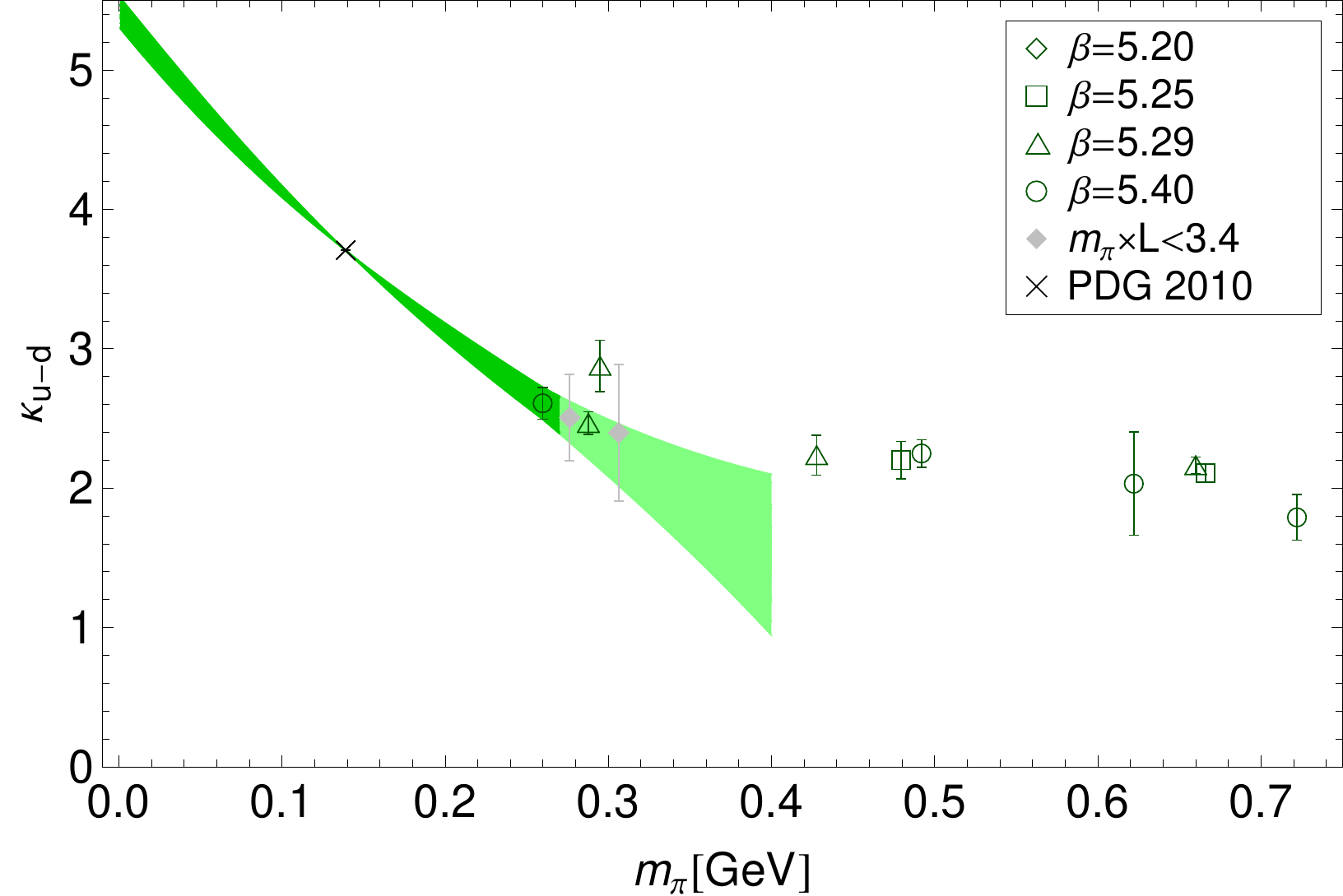}
      \caption{SSE chiral extrapolation of the isovector anomalous magnetic moment.
      The shaded error band represents the fit of the SSE Eq.~(\ref{kappaSSE}) with two free parameters ($\kappa^0$ and a counter-term) to
      the experimental value and the lattice data for $m_\pi\le 260\MeV$.
      %The width of the band also shows the uncertainty due to the value of the isovector nucleon-$\Delta$-coupling $c_V$,
      %which was varied in the range $c_V=-1.5\ldots -3.5\GeV^{-1}$, and the corresponding axial-vector coupling $c_A$,
      %varied in the range $c_A=1.0\ldots 1.5$.
      For the details, see Eq.~(\ref{kappaSSE}) and the surrounding text.\newline}
                \label{fig:kappaSSE}     
     \end{minipage}
         \hspace{0.2cm}
    \begin{minipage}{0.48\textwidth}
      \centering
          \includegraphics[angle=0,width=0.9\textwidth,clip=true,angle=0]{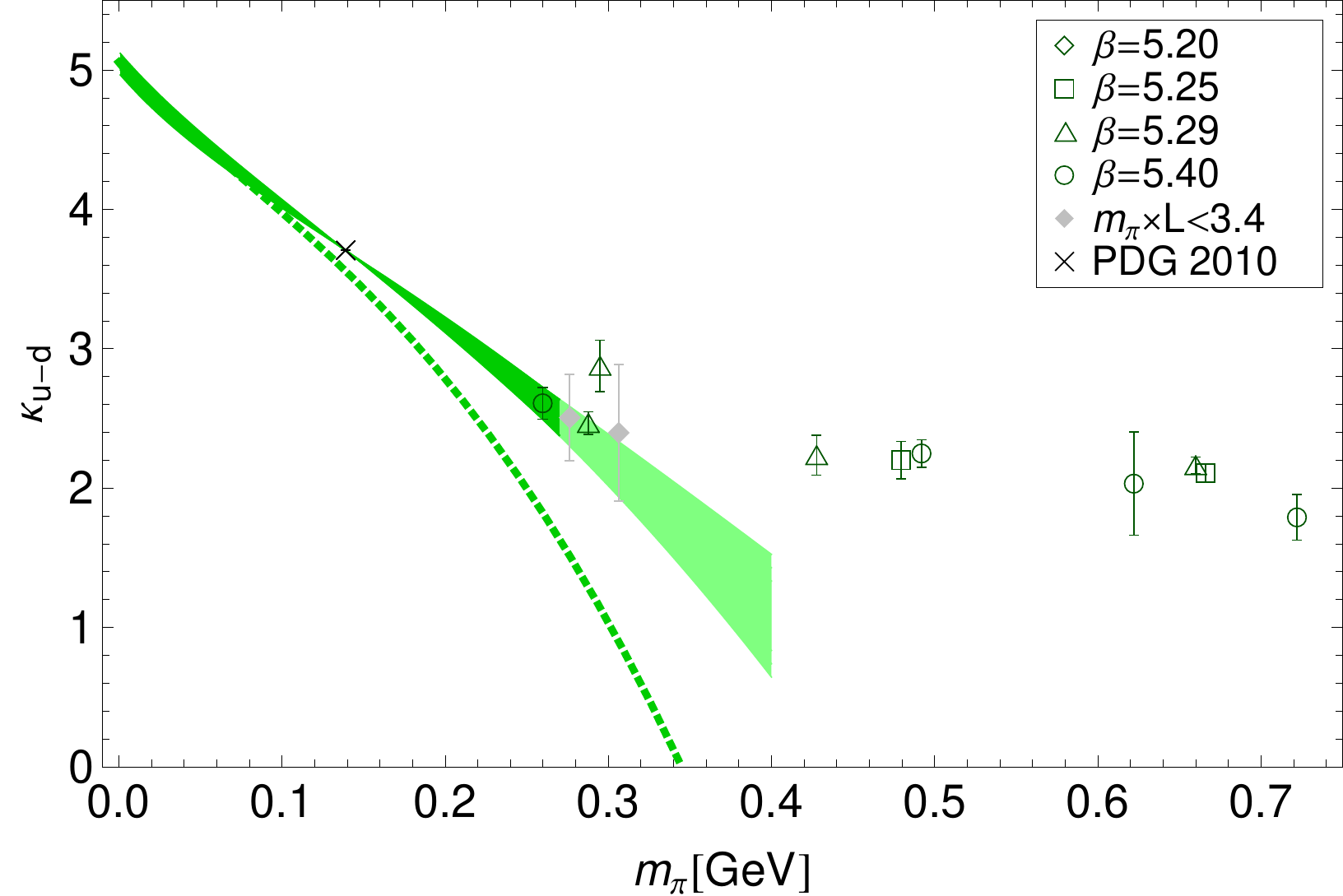}
           \caption{BChPT extrapolation of the isovector anomalous magnetic moment.
      The shaded error band represents fits of Eq.~(\ref{kappaBChPT}) with two free parameters ($c_6\hat=\kappa^0$ and a counter-term) to
      the experimental value and the lattice data for $m_\pi\le 260\MeV$.
      The heavy-baryon limit of the central covariant fit result is indicated by the dotted line.
      %The width of the band also shows the uncertainty due to the value of the coupling $c_4$,
      %which was varied in the range $c_4=3.2\ldots 4.0\GeV^{-1}$.
      For the details, see Eq.~(\ref{kappaBChPT}) and the surrounding text.}
                \label{kappaCovariant}
       \end{minipage}
 \end{figure}

The pion mass dependence of $\kappa_{u-d}$ in the small scale expansion to $\mathcal{O}(\eps^3)$
can be written as \cite{Hemmert:2002uh,Gockeler:2003ay}
\bea
 \kappa_{u-d}^{\text{SSE}}&=&
          \kappa_{u-d}^0
					+\Kappa_{u-d}(m_\pi)
          -   8 E_1^{(r)} (\lambda) m_N m_\pi^2\,,
\label{kappaSSE}
\eea
where we provide the explicit expression for $\Kappa_{u-d}(m_\pi)$ in Appendix \ref{app} in Eq.~(\ref{kappaSSE2}).
In addition to the LECs that were already discussed above, $\kappa_{u-d}^{\text{SSE}}$ depends on 
the isovector anomalous magnetic moment $\kappa^{0}_{u-d}$ and 
the isovector nucleon-$\Delta$ coupling constant $c_V=c^0_V$ in the chiral limit.
The counter-term parameter $E_1^{(r)}(\lambda)$ removes the regularization scale dependence to the given order.
Neither $c_V$ nor $c_A$ are known to great precision. In order to reduce the number of fit parameters, 
we will keep them fixed but perform various fits for $c_V=-1.5,\ldots,-3.5\GeV^{-1}$ and $c_A=1.0,\ldots,1.5$
in order to assess the related uncertainties.
The two remaining unknowns, $\kappa^{0}_{u-d}$ and $E_1^{(r)}(\lambda)$, are treated as free fit parameters and
can be obtained from fits to the experimental value at the physical point and the lattice data point at $m_\pi\sim260\MeV$.
We show the results of the SSE-fits to $\kappa_{u-d}$  in Fig.~\ref{fig:kappaSSE}.
In the chiral limit, we obtain $\kappa^{0}_{u-d}\sim5.3,\ldots,5.5$, which is remarkable $40\%$ above the
precisely known value at the physical point.
For the counter-term parameter, we find values of $E_1^{(r)}\sim-2.9,\ldots,-5.2\GeV^{-3}$ for $\lambda=0.89\GeV$.
Since the extrapolation band continues to fall off above $m_\pi\sim300\MeV$,
it misses the lattice data points to the right, which show little pion mass dependence between $m_\pi\sim300\MeV$ and $m_\pi\sim700\MeV$.

The corresponding expression in the covariant BChPT-approach of Ref.~\cite{Gail:2007} reads
\bea
  \kappa_{u-d}^\text{BChPT} &=& \frac{m_N^{(n)}}{m_N^0}
  \bigg\{c_6 - 16 m_N^0m_\pi^2 e_{106}^r(\lambda) + (\kappa_{u-d})^{(3)} +
    (\kappa_{u-d})^{(4)}\bigg\}\,,
   \label{kappaBChPT}
\eea
where $m_N^{(n)}$ denotes the nucleon mass used in front of $F_2$ in the parametrization of the current in Eq.~(\ref{eq:em-me}).
In our case, $m_N^{(n)}=m_N^\phys=0.938\GeV$, and we will explicitly replace $m_N^{(n)}$ by $m_N^\phys$ in the following.
The contributions at $\mathcal{O}(p^3)$ and $\mathcal{O}(p^4)$, i.e.\ $(\kappa_{u-d})^{(3,4)}$, are given 
in Appendix \ref{app} in Eqs.~(\ref{kappaBChPT2}) and (\ref{kappaBChPT3}), respectively.

These expressions depend on $c_6\hat=\kappa^0$ as well as the additional LEC $c_4$, plus a counter-term parameter $e_{106}^r$.
Varying $c_4$ in the range of $3.2,\ldots,4.0\GeV^{-1}$, we have determined $c_6$ and the counter-term
from fits to the experimental value and the lattice data at $m_\pi\sim260\MeV$.
The result is shown in Fig.~\ref{kappaCovariant}.
Again, we find a rather large value for $\kappa_{u-d}$ in the chiral limit,
$\kappa^{0}_{u-d}\sim4.8,\ldots,5.1$, somewhat below the values of the SSE extrapolation.
For the counter-term parameter, we obtain $e_{106}^r\sim0.5,\ldots,1.0\GeV^{-3}$ for $\lambda=0.89\GeV$.
The error band is close to the one in Fig.~\ref{fig:kappaSSE} for the SSE case up to pion masses of $\sim300\MeV$, but
then falls off more strongly, already lying a factor of about two below the data point at $\sim400\MeV$.
More interesting is the observation that the non-relativistic limit, $m_N\rightarrow\infty$,
of the covariant fit, indicated by the dotted line in  Fig.~\ref{kappaCovariant},
drops off even more strongly and starts to deviate from the full result already at the physical point.
In the case that the covariant approach is at all quantitatively applicable in these ranges of the pion mass,
this would suggest in turn that the range of applicability of the corresponding heavy baryon expansion is much more limited.

\begin{figure}[t]
    \begin{minipage}{0.48\textwidth}
        \centering
          \includegraphics[angle=0,width=0.9\textwidth,clip=true,angle=0]{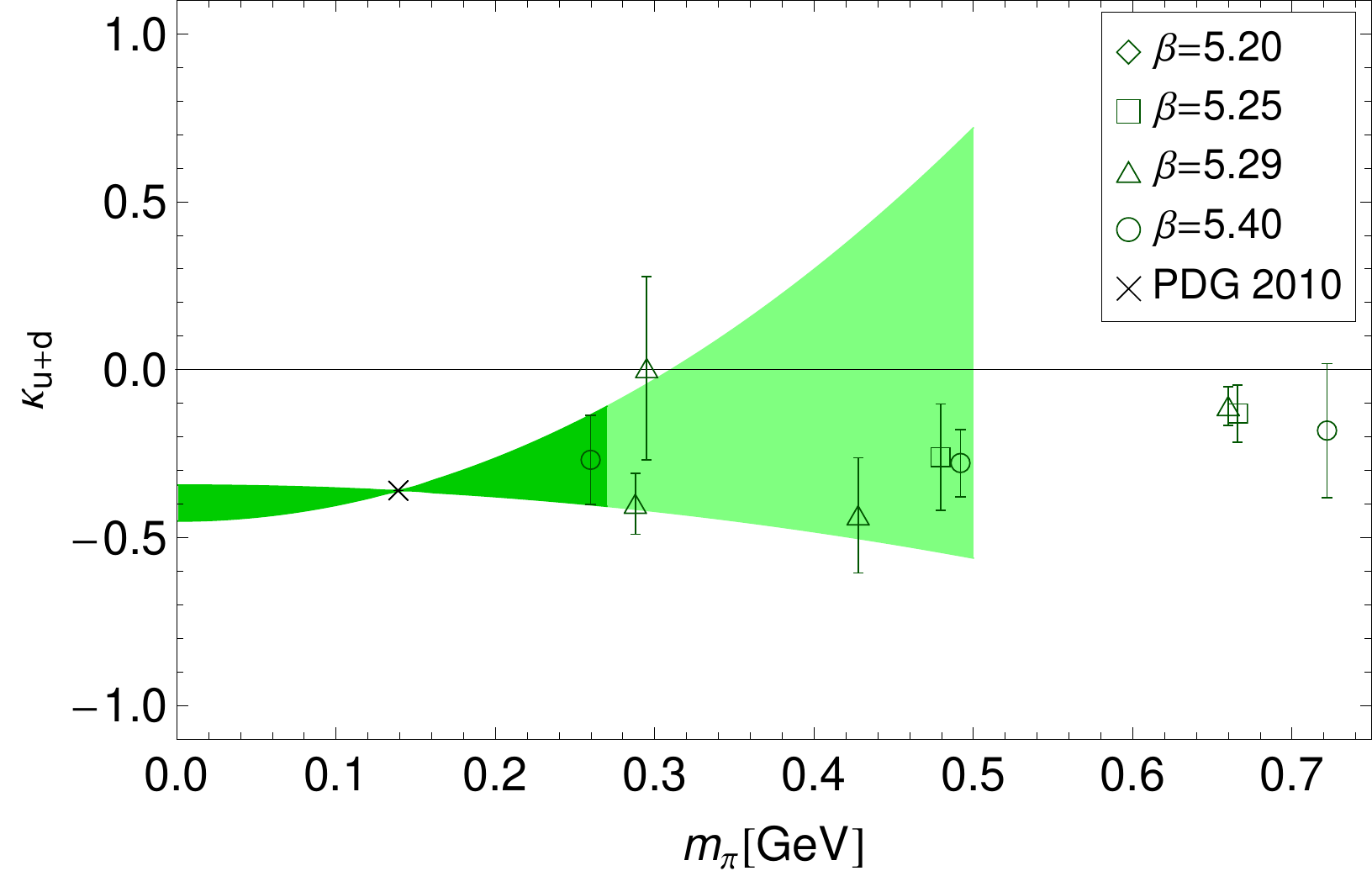}
      \caption{SSE chiral extrapolation of the isosinglet anomalous magnetic moment.
      The shaded error band represents the fit of the SSE Eq.~(\ref{kappaSSEupd}) with two free parameters ($\kappa^0$ and a counter-term) to
      the experimental value and the lattice data for $m_\pi\le 260\MeV$. \newline      }
                \label{kappaupdSSSE}     
     \end{minipage}
         \hspace{0.2cm}
    \begin{minipage}{0.48\textwidth}
      \centering
          \includegraphics[angle=0,width=0.9\textwidth,clip=true,angle=0]{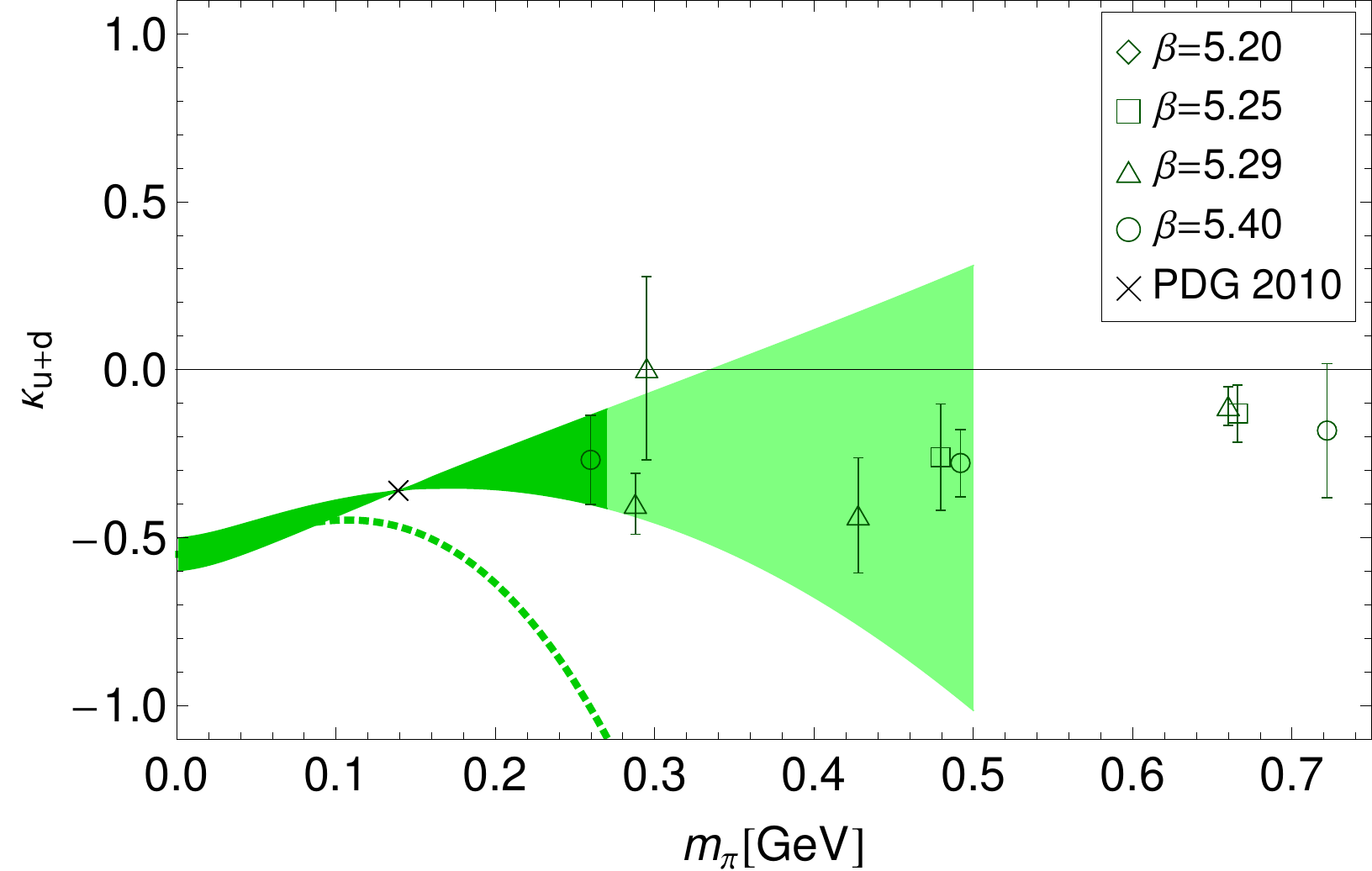}
           \caption{BChPT extrapolation of the isosinglet anomalous magnetic moment.
      The shaded error band represents fits of Eq.~(\ref{kappaBChPTupd}) with two free parameters ($\kappa^0_s\hat=\kappa_{u+d}^0/3$ and a counter-term) to
      the experimental value and the lattice data for $m_\pi\le 260\MeV$.
      The dotted line illustrates the heavy-baryon limit of the central covariant fit result.}
                \label{kappaupdCovariant}
       \end{minipage}
 \end{figure}

Turning our attention to the isosinglet channel, we first note that in HBChPT at one-loop level, 
a pion mass dependence is only observed in the modified SSE counting scheme of Ref.~\cite{Hemmert:2002uh} (denoted by 'scheme C').
It is given by
\bea
 \kappa_{u+d}^{\text{SSE}}&=&
          \kappa_{u+d}^0 -24 E_2 m_N m_\pi^2\,,
\label{kappaSSEupd}
\eea
where $E_2$ is a counter-term parameter.
The result of a fit to the experimental value at the physical point and the lattice data point at $m_\pi\sim260\MeV$
is illustrated by the shaded error band in Fig.~\ref{kappaupdSSSE}.
We find a non-zero, negative value for the isosinglet anomalous magnetic moment in the chiral limit, $\kappa_{u+d}^0=-0.40\pm0.05$,
while the counter-term parameter turns out to be small and compatible with zero within errors, $E_2=-0.08\pm0.12\GeV^{-3}$.

The pion mass dependence in the BChPT calculation of Ref.~\cite{Gail:2007} reads
\bea
  \kappa_{u+d}^\text{BChPT} &=&\frac{m_N^\phys}{{m_N^0}} \bigg\{ \kappa_{u+d}^0 - 48{m_N^0} m_\pi^2 e_{105}^r(\lambda) 
    + (\kappa_{u+d})^{(3)} + (\kappa_{u+d})^{(4)}\bigg\}\,,
   \label{kappaBChPTupd}
\eea
where the contributions $(\kappa_{u+d})^{(3,4)}$ of $\mathcal{O}(p^3)$ and $\mathcal{O}(p^4)$ 
are given in Eqs.~(\ref{kappaBChPTupd2a}) and (\ref{kappaBChPTupd2b}) in Appendix \ref{app}.
One finds that $\kappa_{u+d}$ depends on the chiral limit value $\kappa_{u+d}^0=3\kappa_{s}^0=\kappa_{p}^0+\kappa_{n}^0$, 
and a counter-term parameter $e_{105}^r(\lambda)$. As before, the regularization scale has been set to $\lambda=0.89\GeV$.
From a fit to the experimental value and the lattice data point at $m_\pi\sim260\MeV$, we obtain
a negative $\kappa_{u+d}^0\sim-0.6,\ldots,-0.49$, with a small value for the counter-term parameter of $e_{105}^r\sim0.58,\ldots,0.79\GeV^{-3}$.
The result of the fit is shown by the shaded error band in Fig.~\ref{kappaupdCovariant}.
With just two data points constraining the fit, the band quickly broadens at larger pion masses, thereby prohibiting
a quantitative assessment.
We note, however, that the center of the band provides a good description of the lattice data points up to $m_\pi\sim500\MeV$.
In the heavy-baryon limit, we find a strongly downwards bending curve directly above the physical pion mass,
illustrated by the dotted line. 
This indicates once more that the radius of convergence of the heavy-baryon approach at one-loop level is limited to the region
below $m^\phys_\pi$ (assuming that the BChPT result is applicable up to $m_\pi\sim260\MeV$ in our fit in the first place).

\subsection{Pauli radius}
\label{sec:r2chiral}

To separate the pion mass dependence of the slope of $F_2(Q^2)$ from that of $F_2(0)=\kappa$,
we focus here on the product $\kappa\times\langle r^2\rangle_2$ instead of $\langle r^2\rangle_2$ (see also \cite{Syritsyn:2009mx}).
This avoids in particular a potential issue related to the expression for $\kappa(m_\pi)$ that is used
in the denominator of the chiral expansion of $\langle r^2\rangle^{}_2=-6\rho_2(m_\pi)/\kappa(m_\pi)$, where 
$\rho_2=dF_2(Q^2)/dQ^2|_{Q^2=0}$ is the slope. Depending on the order of the ChPT-calculation, 
one is in general allowed to employ ChPT-expressions of different orders for $\kappa(m_\pi)$ in the denominator
without affecting the overall consistency of the chiral expansion of $\langle r^2\rangle_2$ to the given order. 
Since the pion mass dependence of $\kappa$ can be rather strong (as discussed in the previous section,
where $\kappa_{u-d}(m_\pi^{\phys})$ increases by as much as $40\%$ as $m_\pi\rightarrow0$), 
the ambiguity in the choice of $\kappa(m_\pi)$ could have a significant impact on 
the uncertainty of the chiral extrapolation of $\langle r^2\rangle_2$.

\begin{figure}[t]
    \begin{minipage}{0.48\textwidth}
        \centering
          \includegraphics[angle=0,width=0.9\textwidth,clip=true,angle=0]{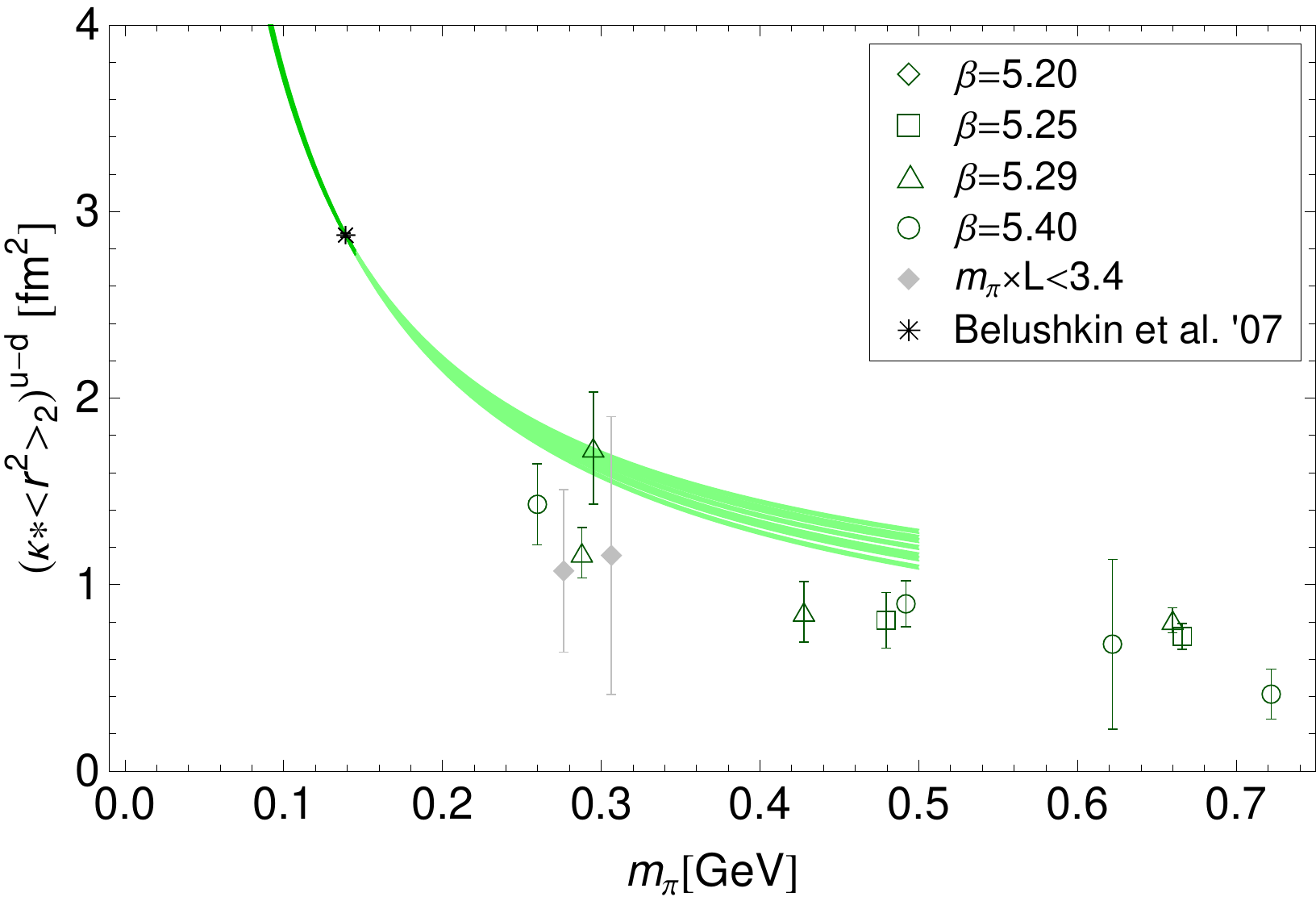}
      \caption{SSE chiral extrapolation of $\kappa\times \langle r^2\rangle_2$ in the isovector channel.
      The shaded error bands represent fits of the SSE results in Eq.~(\ref{kappaXr2SSE}) with one free parameter (a counter-term) to
      the experimental value.
      %, where the axial vector $\Delta$-nucleon coupling has been varied in the range of $c_A=1.0,\ldots, 1.5$.
      }
                \label{kappar2SSE}     
     \end{minipage}
         \hspace{0.2cm}
    \begin{minipage}{0.48\textwidth}
      \centering
          \includegraphics[angle=0,width=0.9\textwidth,clip=true,angle=0]{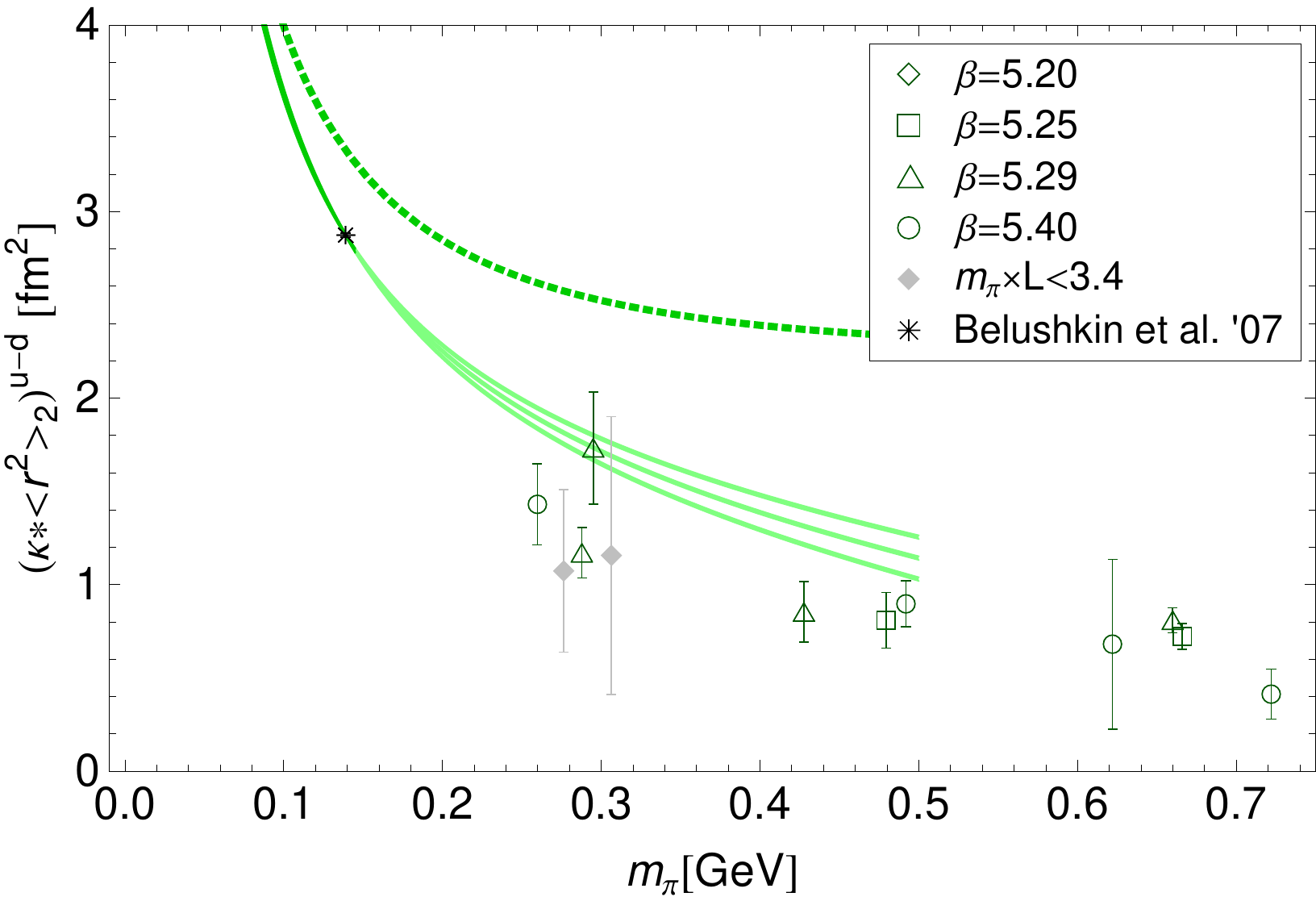}
           \caption{BChPT extrapolation of $\kappa\times \langle r^2\rangle_2$ in the isovector channel.
      The shaded error bands represent fits of the BChPT results in Eq.~(\ref{kappaXr2BChPT}) with one free parameter (a counter-term) to
      the experimental value.}
                \label{kappar2Covariant}
       \end{minipage}
 \end{figure}

We therefore consider in the following the SSE and BChPT expansions of $\kappa\times\langle r^2\rangle_2$.
The result in the SSE \cite{Hemmert:2002uh,Gockeler:2003ay} to $\mathcal{O}(\eps^3)$ is
\begin{equation} 
 \Big(\kappa\times \langle r^2\rangle_{2}\Big)^{u-d,\text{SSE}} = 
   %\Bigg\{
   \frac{g_{A}^2m_N}{8 \pi f_{\pi}^2  m_\pi}
  +\frac{c_A^2m_N}{9 \pi^2 f_{\pi}^2 \sqrt{\deltam^2-m_{\pi}^2}} 
    \ln\left(\frac{\deltam}{m_\pi} + \sqrt{\frac{\deltam^2}{m_{\pi}^2}-1}\right)
  + 24 m_N B_{c2}\,, 
\label{kappaXr2SSE}
\end{equation}
showing explicitly the well-known linear divergence in $m_\pi$ expected in the chiral limit.
It depends again on the coupling $c_A$, and a counter-term parameter $B_{c2}$.
As in the case of $\kappa_{u-d}$, we have varied $c_A=1.0,\ldots,1.5$, and determined $B_{c2}$ as the only free parameter
from a fit to the phenomenological value at the physical point. We find $B_{c2}=0.32,\ldots,0.82\GeV^{-3}$.
The result is illustrated in Fig.~\ref{kappar2SSE} by the shaded bands, in comparison to 
the lattice data points obtained from the polynomial ansatz Eq.~(\ref{eqF2poly}) for the $Q^2$-dependence of $F_2$.
While the extrapolation band quickly decreases above $m_\pi^\phys$,
it overshoots the lattice results by about $20-40\%$ in the region of $m_\pi\sim260,\ldots,500\MeV$.

A significantly more involved expression for the $m_\pi$-dependence has been obtained in the BChPT-scheme 
of Ref.~\cite{Dorati:2007bk} to $\mathcal{O}(p^4)$, which can be written as \cite{Gail:2007}
\bea
  \Big(\kappa\times\langle r^2\rangle_2\Big)^{u-d,\text{BChPT}} & = & \frac{m_N^\phys}{m_N^0} \left(
   24m_N^0e_{74}^r(\lambda) + (\kappa r_2^2)^{u-d,(3)} + (\kappa r_2^2)^{u-d,(4)}\right) \,,
      \label{kappaXr2BChPT}
\eea
where the individual terms, $(\kappa r_2^2)^{u-d,(3)}$ and $(\kappa r_2^2)^{u-d,(4)}$, are provided in Appendix \ref{app}
in Eqs.~(\ref{kappaXr2BChPT2}) and (\ref{kappaXr2BChPT3}).
This result depends, apart from the counter-term parameter $e_{74}^r(\lambda)$, also on the couplings $c_4$ and $c_6\hat=\kappa_{u-d}^0$.
The regularization scale has been fixed to $\lambda=0.89\GeV$.
To study the predicted $m_\pi$-dependence, we have varied, as before, $c_4=3.2,\ldots,4.0\GeV^{-1}$, 
and used the corresponding values obtained for $c_6=\kappa^{0}_{u-d}$ from the BChPT analysis of $\kappa_{u-d}$ above.
The unknown parameter $e_{74}^r$ has been fitted to the phenomenological value at the physical point,
giving $e_{74}^r\sim1.6,\ldots,2.0\GeV^{-3}$.
We compare this approach with the $m_\pi$-dependence of our results obtained from the polynomial ansatz for $F^{u-d}_2(Q^2)$
in Fig.~\ref{kappar2Covariant}. 
Similarly to the SSE-extrapolation discussed before, 
the extrapolation curves lie somewhat above the lattice data points for $m_\pi\sim260\MeV$ to $500\MeV$.
Taking the heavy-baryon limit of the central band, we obtain the dotted curve in Fig.~\ref{kappar2Covariant}.
Under the assumption that the BChPT result is applicable at the physical pion mass, 
we find that the contributions of $\mathcal{O}(1/(m_N)^n)$, which are included in the covariant BChPT approach, 
start to play an important role already before $m_\pi^\phys$ is reached.
Hence also for this observable, our results indicate that the range of applicability of the leading order heavy-baryon ChPT result  is 
restricted to pion masses $\lesssim100\MeV$.

With respect to the isosinglet channel, we note that at the one-loop level in HBChPT, $\langle r^2\rangle_2^{u+d}$ 
is predicted to vanish, i.e.\ the 
form factor is independent of $Q^2$, $F^{u+d}_2(Q^2)=\text{const.}$ \cite{Bernard:1998gv}.
Similar to the case of the isosinglet Dirac radius, the BChPT calculation gives at first sight
a rather non-trivial $m_\pi$-dependence \cite{Gail:2007}, but in practice it turns out to be flat over the full range
of relevant pion masses, $(\kappa\times\langle r^2\rangle_2)^{u+d,\text{BChPT}}\approx \text{const.}\,$
Given the poorly determined value at the physical point, we conclude that it is currently difficult
to provide even a semi-quantitative chiral extrapolation of $(\kappa\times\langle r^2\rangle_2)^{u+d}$ from
$m^\phys_\pi$ to the chiral limit and to larger pion masses.

\section{Systematic uncertainties}
\label{sec:systematics}

\subsection{Contaminations from excited states}
\label{sec:tsnk}

A potentially important source of systematic uncertainties is given by contributions from excited states in the nucleon
correlation functions.
For too small distances between the operator insertion time $\tau$ and the source and sink times
they could adversely affect the ratios of three- to two-point functions in Eq.~(\ref{eq:ratio}),
and thereby the plateau values from which we extract the form factors.
In turn, if the sink time $t_\snk$ and accordingly $\tau$ are chosen too large, the signal-to-noise ratio begins to deteriorate, 
and the data points start to fluctuate more strongly\footnote{Also the two-point functions taken at
$t_\snk$ in the ratio might fluctuate around zero within errors, thereby leading to an unreliable result for the plateau.}.
This is mostly an issue at larger hadron momenta required for the analysis of the form factors at $Q^2>0$.
Hence we have to seek a compromise between potential contaminations from excited states on the one hand, and 
noisy/fluctuating correlation functions and plateaus on the other.

\begin{figure}[t]
    \begin{minipage}{0.48\textwidth}
        \centering
          \includegraphics[angle=0,width=0.9\textwidth,clip=true,angle=0]{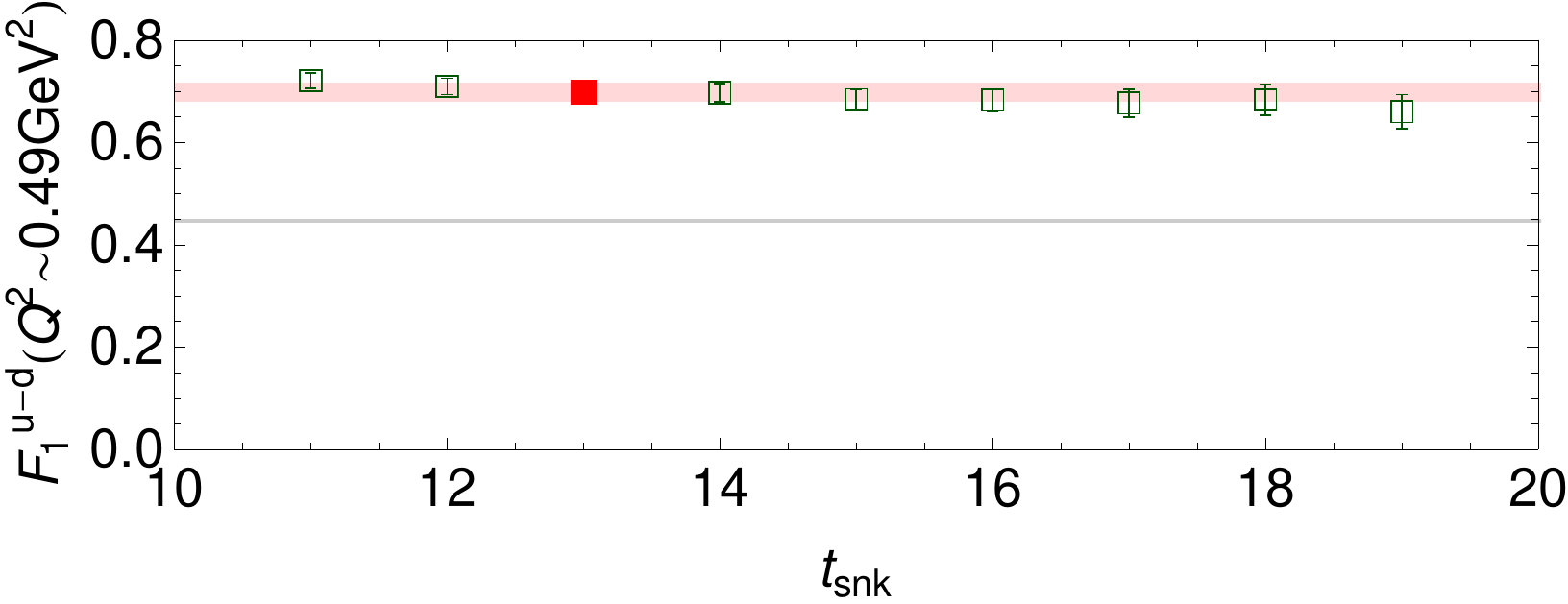}
          \includegraphics[angle=0,width=0.9\textwidth,clip=true,angle=0]{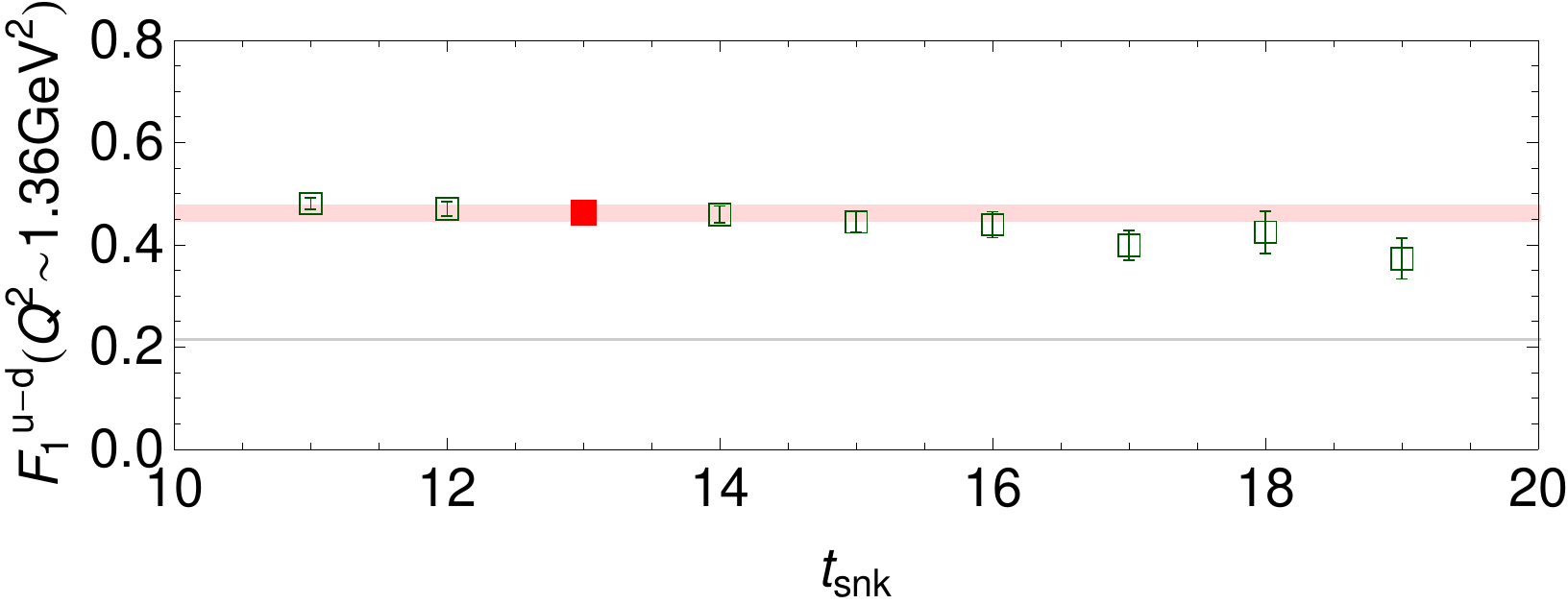}
     \end{minipage} 
         \hspace{0.2cm}
    \begin{minipage}{0.48\textwidth}
      \centering
          \includegraphics[angle=0,width=0.9\textwidth,clip=true,angle=0]{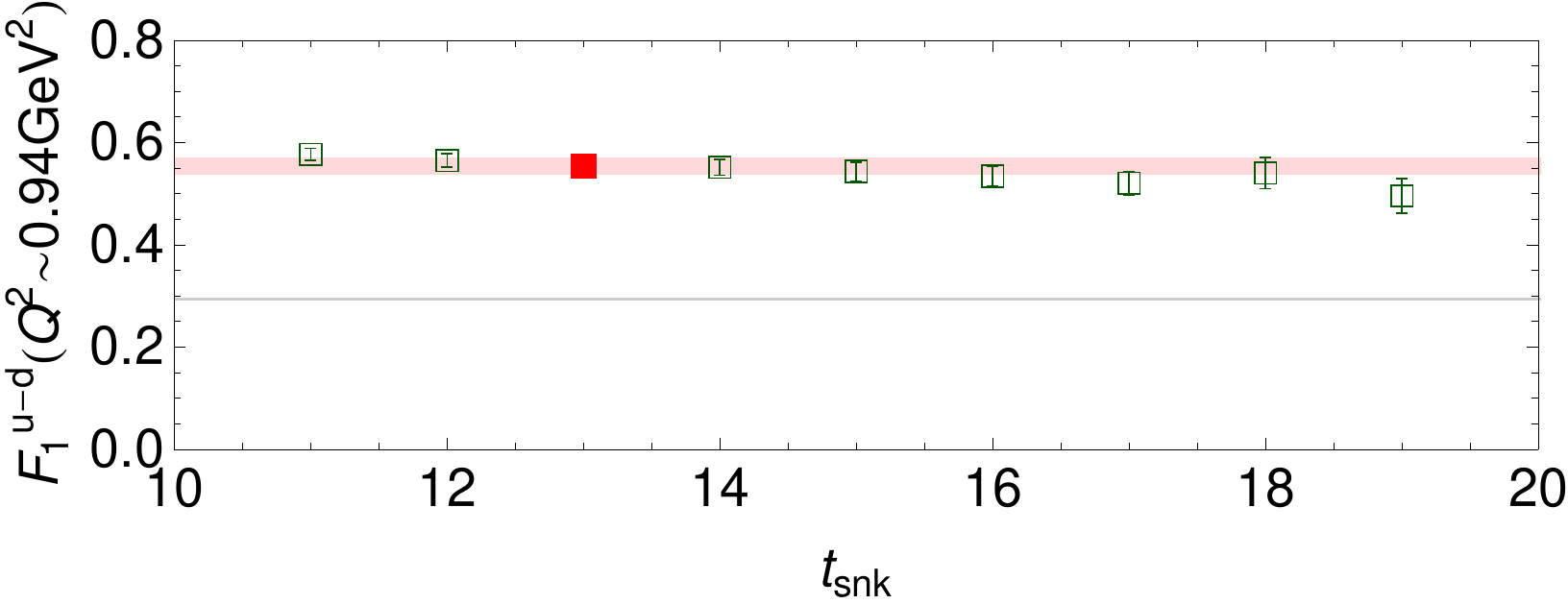}
          \includegraphics[angle=0,width=0.9\textwidth,clip=true,angle=0]{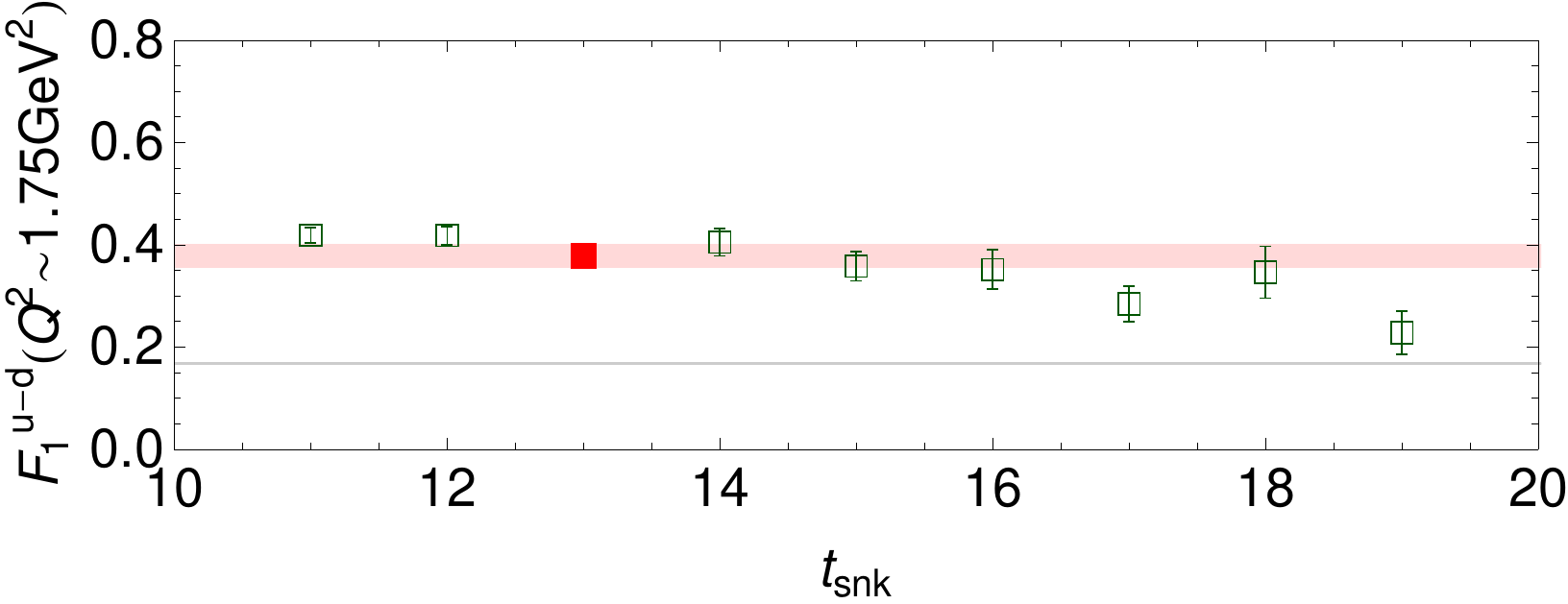}
     \end{minipage}
       \caption{Dependence of $F^{u-d}_1$ at fixed values of $Q^2$ on the sink time $t_\snk$ in the three point function,
       for $\beta=5.29,\kappa=0.13590$.
       Results for our choice $t_\snk=13$ are indicated by the filled (red) points and corresponding bands.
       The thin gray shaded band represents the parametrization of Ref.~\cite{Alberico:2008sz} of the experimental data.}
         \label{F1tsnk}
 \end{figure}

In this work, we have chosen primarily a fixed 
distance between source and sink of about $0.95\fm$.
Instead of studying the excited states contributions directly by performing, e.g., multi-exponential fits 
(which are notoriously unstable) of the correlations functions, we have analyzed the form factors
for a range of different sink times $t_\snk=11,\ldots,19$, for a single ensemble with $\beta=5.29,\kappa=0.13590$.
The dependence of, e.g., the data points for $F_1$ at fixed $Q^2$ on $t_\snk$ is then a direct indicator
for the possible influence of excited state contributions on our results.
The results of this study are displayed in Fig.~\ref{F1tsnk}, showing $F^{u-d}_1$ as a function of $t_\snk$
for the four values $Q^2\sim0.49,0.94,1.36,1.75\GeV^2$.
We note that the broader band (corresponding to our primary choice $t_\snk=13$) is compatible
with all data points at sink times up to and including $t_\snk=16$ within errors.
While the central values decrease on average by a small amount as $t_\snk$ increases from $13$ to $16$, 
no clear systematic trend can be established when the uncertainties are taken into account.
At large $t_\snk\ge17$, we find that the data points start to fluctuate more strongly as the momentum transfer increases.
This indicates that the plateaus indeed become unstable due to deteriorating signal-to-noise ratios 
of the correlation functions in the ratio at large times and momenta.
At larger $Q^2>1.8\GeV^2$ (not shown), we even find that the extracted values for $F_1$ quickly approach zero as $t_\snk\rightarrow19$.

In summary, for the given ensemble, we cannot identify a systematic dependence of our results for $F_1$
on the sink time within errors, excluding large $t_\snk$ values where strong fluctuations and 
low signal-to-noise ratios make a quantitative analysis impossible. 
This indicates that the uncertainty due to excited state contaminations is not larger than
the statistical errors in our study.

\subsection{Discretization effects and pion mass dependence at fixed $Q^2$}
\label{sec:syseff}

Studies of discretization effects and the pion mass dependence of the lattice results
are usually directly performed for the fundamental observables of interest.
In our case, these are the radii of the Dirac and Pauli form factor, as well
as the anomalous magnetic moments. However, due to the discrete values of the momentum transfer
that can be accessed in a finite volume, in particular the still rather large, lowest non-zero
$Q^2$ of $\approx0.2\GeV^2$ in our case, the extraction of these
observables requires non-trivial inter- and extrapolations of the form factor data in $Q^2$.
To avoid an intermixture of the primary lattice artifacts with 
uncertainties due to the required $Q^2$-parametrization, we now attempt to investigate the $a$- and $m_\pi$-dependences
of the form factor data directly for fixed values of $Q^2>0$.
In this regard, one has to keep in mind that changes in the lattice volume, spacing, and the quark mass (and thereby the nucleon mass),
lead in general to different sets of values of $Q^2$ for the different ensembles. 
To study the lattice spacing dependence, we have therefore scanned our data sets for
narrow ranges in $Q^2$ and $m_\pi$ (with maximum relative widths of 8\%) 
for which data points for three or more couplings $\beta$ are available.
The residual $Q^2$- and $m_\pi$-dependences within these narrow windows were taken into account
by interpolations and subsequent relative shifts of the data points to the central
values of  $Q^2$ and $m_\pi$ in the respective ranges.
As a test, we have monitored the relative shifts and found that their absolute values are
about the same size as the statistical errors of the shifted data points.

\begin{figure}[t]
    \begin{minipage}{0.99\textwidth}
        \centering
          \includegraphics[angle=0,width=0.9\textwidth,clip=true,angle=0]{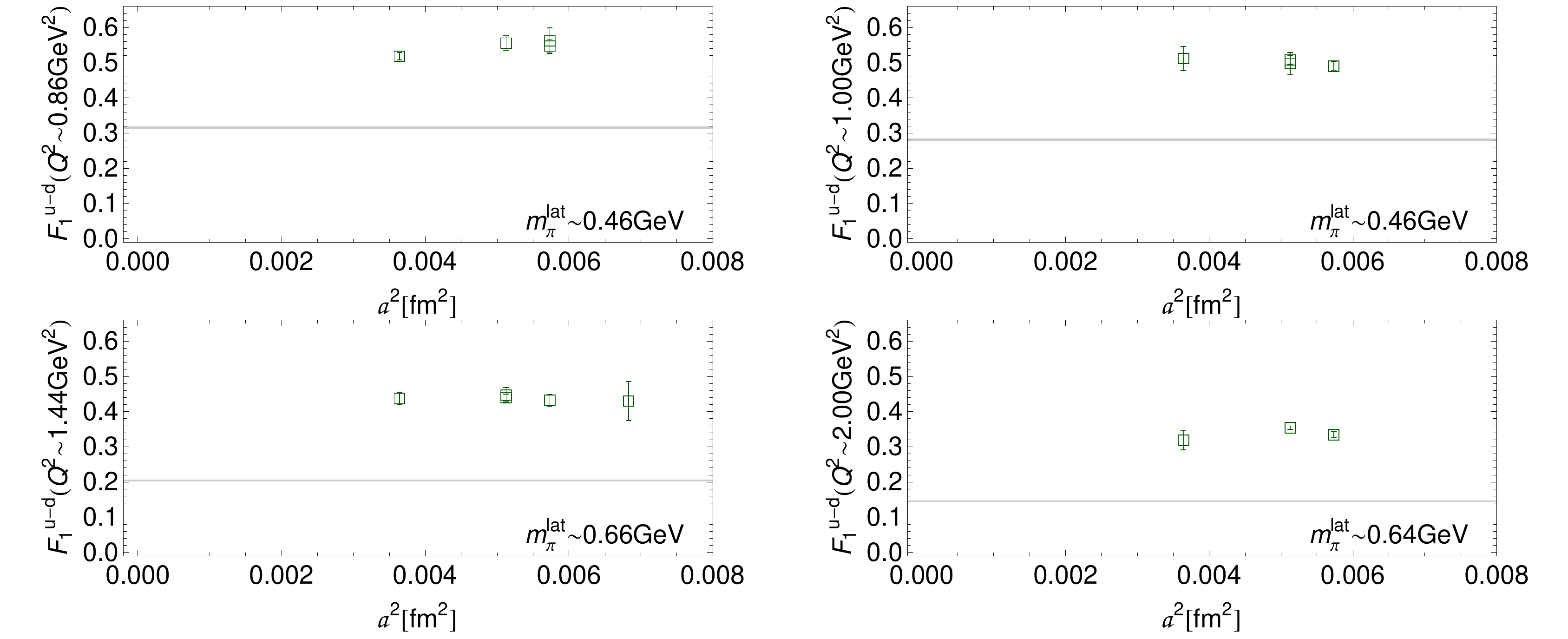}
     \end{minipage}
       \caption{Lattice spacing dependence of the isovector Dirac form factor for fixed narrow ranges in $Q^2$ and $m_\pi$.
  The gray shaded bands represent the parametrization of Ref.~\cite{Alberico:2008sz} of the experimental data.}
  \label{F1vsa}
 \end{figure}

\begin{figure}[t]
    \begin{minipage}{0.99\textwidth}
        \centering
          \includegraphics[angle=0,width=0.9\textwidth,clip=true,angle=0]{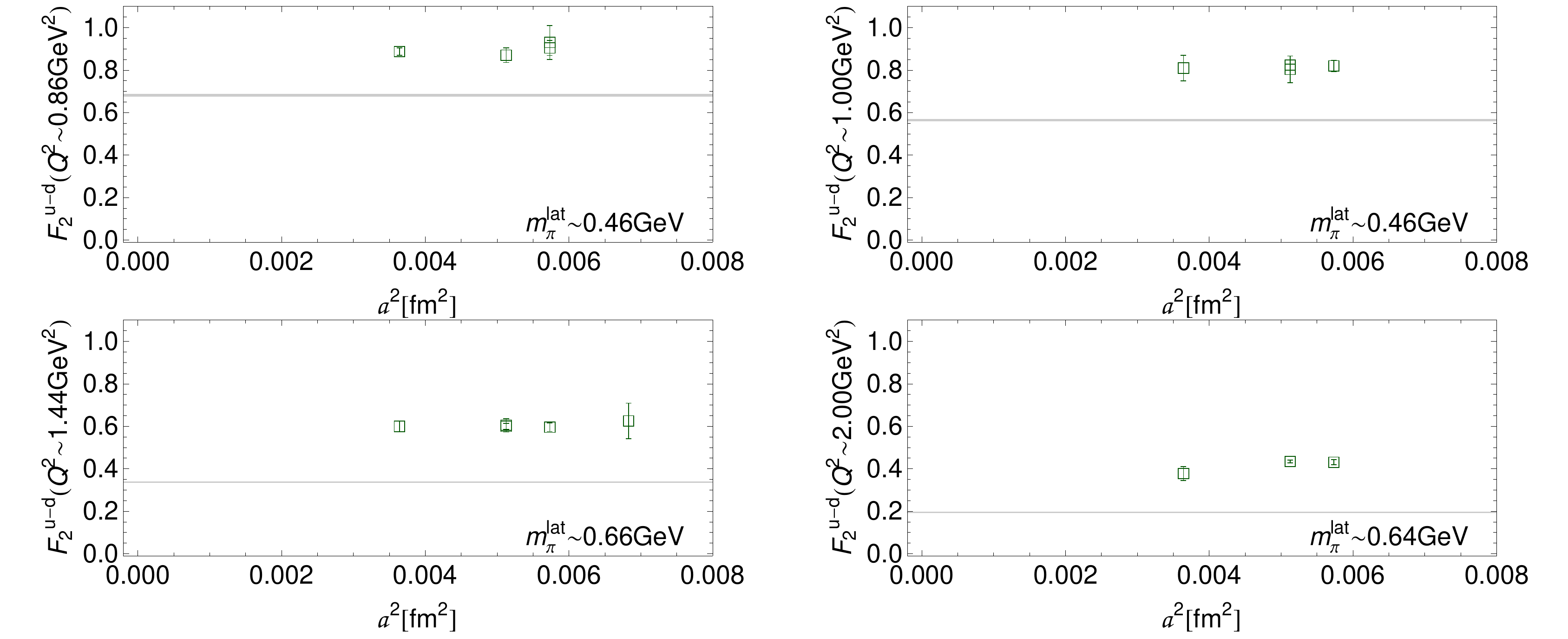}
     \end{minipage}
       \caption{Lattice spacing dependence of the isovector Pauli form factor for fixed narrow ranges in $Q^2$ and $m_\pi$.
  The gray shaded bands represent the parametrization of Ref.~\cite{Alberico:2008sz} of the experimental data.}
  \label{F2vsa}
 \end{figure}

The results for the $a^2$-dependence are displayed in Figs.~\ref{F1vsa} and \ref{F2vsa} for $F^{u-d}_1$ and $F^{u-d}_2$, respectively.
While some fluctuations of the central values as functions of $a^2$ are visible, they do not seem to follow a systematic pattern.
Overall, the data points are compatible with a constant behavior within statistical errors.
In combination, the uncertainties and the fluctuations of the data for the given $Q^2$-values 
are however too large to allow for a consistent, quantitative continuum extrapolation.
Still, although we cannot exclude the presence of some discretization effects, we do not
see any evidence that they could significantly reduce the large gap between the lattice data points and
the experimental result illustrated by the shaded bands in Figs.~\ref{F1vsa} and \ref{F2vsa}.

\begin{figure}[t]
    \begin{minipage}{0.99\textwidth}
        \centering
          \includegraphics[angle=0,width=0.9\textwidth,clip=true,angle=0]{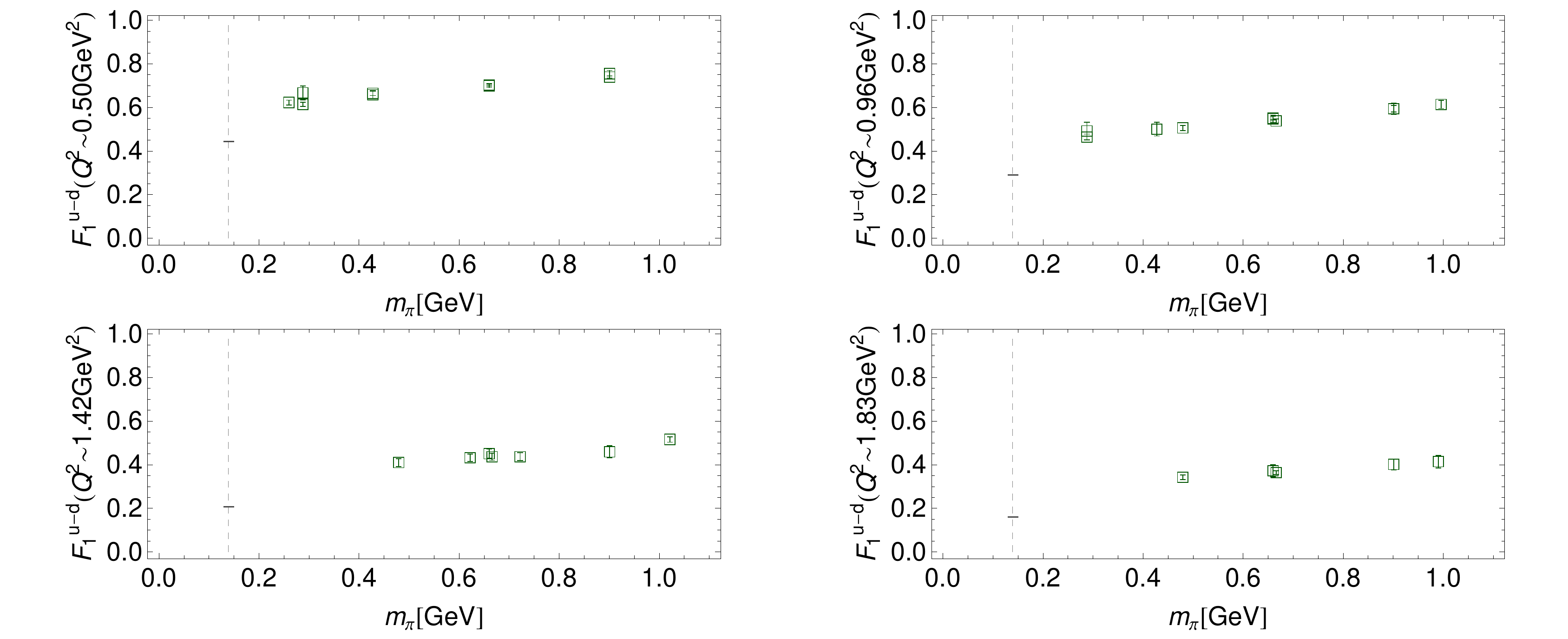}
     \end{minipage}
       \caption{Pion mass dependence of the isovector Dirac form factor for selected ranges in $Q^2$ and $m_\pi$.
  The gray bars represent the parametrization of Ref.~\cite{Alberico:2008sz} of the experimental data at the physical point.}
  \label{F1vsmPi}
 \end{figure}

\begin{figure}[t]
    \begin{minipage}{0.99\textwidth}
        \centering
          \includegraphics[angle=0,width=0.9\textwidth,clip=true,angle=0]{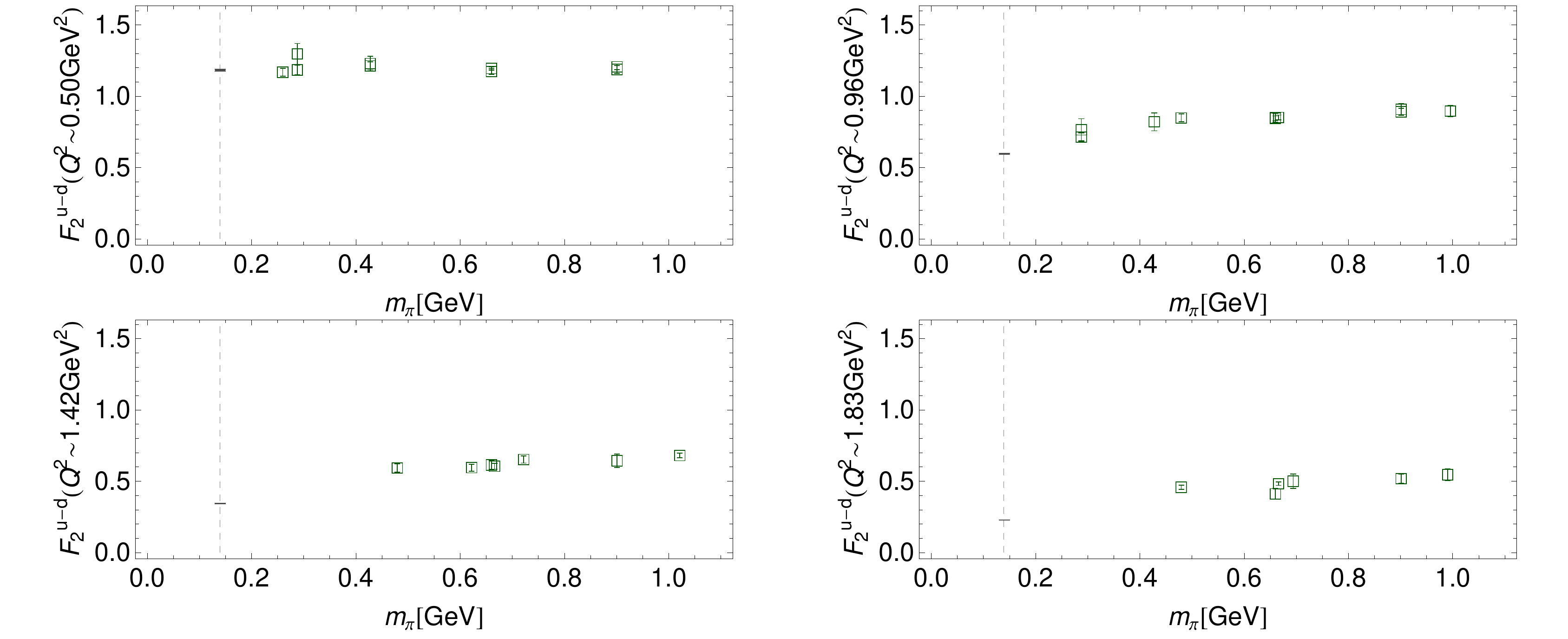}
     \end{minipage}
       \caption{Pion mass dependence of the isovector Pauli form factor for selected ranges in $Q^2$ and $m_\pi$.
  The gray bars represent the parametrization of Ref.~\cite{Alberico:2008sz} of the experimental data at the physical point.}
  \label{F2vsmPi}
 \end{figure}

A similar approach to the pion mass dependence at fixed $Q^2$ leads to the results displayed in 
Figs.~\ref{F1vsmPi} and \ref{F2vsmPi}.
Following the above findings on the $a^2$-dependence, we have in this case included all $\beta$ on an equal footing.
For $F_1^{u-d}$ in Fig.~\ref{F1vsmPi}, we observe an approximately linear dependence on $m_\pi$ 
over a wide range of $Q^2$ from $\sim0.5\GeV^2$ up to $\sim1.8\GeV^2$.
While the data points do show a slight downwards trend in the right direction, simple linear extrapolations would clearly miss 
the experimental values by about 20\% to 40\% at the physical pion mass.
Keeping in mind that chiral perturbation theory predicts a logarithmically diverging slope of $F_1^{u-d}$ at $Q^2=0$ as $m_\pi\rightarrow0$,
it is not surprising that also for $F_1^{u-d}(Q^2\not=0)$, 
a non-linear $m_\pi^2$-dependence has to set in at low pion masses.

In the case of $F_2^{u-d}$ in Fig.~\ref{F2vsmPi}, we find again that the lattice results are, to a good approximation, linear in $m_\pi$. 
While the data points at lower $Q^2$ are constant within the uncertainties, a slight downwards slope
seems to develop as we approach larger momentum transfers.
In contrast to $F_1^{u-d}$, a naive linear extrapolation in $m_\pi$ would even lead to an overlap with the experimental
values at the physical pion mass, at least for the lowest value of $Q^2\sim0.50\GeV^2$ 
in Fig.~\ref{F2vsmPi}.
This does not imply, however, that the lattice results for $m_\pi>m_\pi^\phys$ provide a good description of the experimental data 
over a wider range of $Q^2$. 
At large $Q^2$, linear extrapolations in $m_\pi$ would lead to values for $F_2^{u-d}$ that are systematically larger
than in experiment. 
We therefore find again that the $Q^2$-slope of the lattice data (even when naively extrapolated to $m_\pi^\phys$ at fixed $Q^2$) 
is too small, and a typical dipole or tripole extrapolation (see section \ref{sec:para}) to $Q^2=0$ would then lead 
to a $F_2^{u-d,\lat}(Q^2\!=\!0)=\kappa_{u-d}^{\lat}<\kappa_{u-d}^{\phys}$. 
The apparently good agreement of the lattice data points with the experimental values in the top row of Fig.~\ref{F2vsmPi} has
to be interpreted as the result of a too small slope and, at the same time, a too low normalization (at $Q^2=0$) 
at unphysically large lattice pion masses.
This is studied explicitly in sections \ref{sec:kappachiral} and \ref{sec:r2chiral} above, 
where we discuss the pion mass dependence and chiral extrapolations
of $\kappa_{u-d}$ and the slope of $F_2^{u-d}(Q^2)$, respectively.

\subsection{Discretization and finite volume effects in the Dirac radius and the anomalous magnetic moment}
\label{sec:r1sys}

\begin{figure}[t]
    \begin{minipage}{0.48\textwidth}
        \centering
         \includegraphics[angle=0,width=0.85\textwidth,clip=true,angle=0]{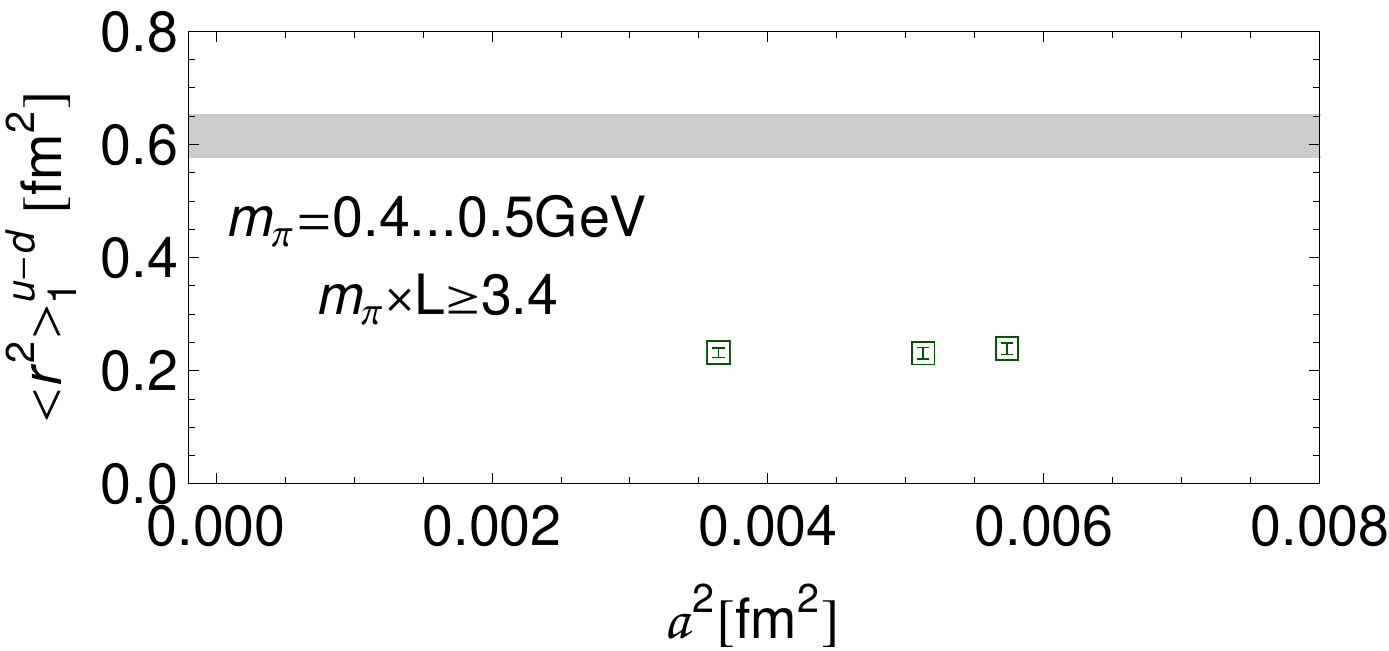}
     \end{minipage}
         \begin{minipage}{0.48\textwidth}
        \centering
          \includegraphics[angle=0,width=0.85\textwidth,clip=true,angle=0]{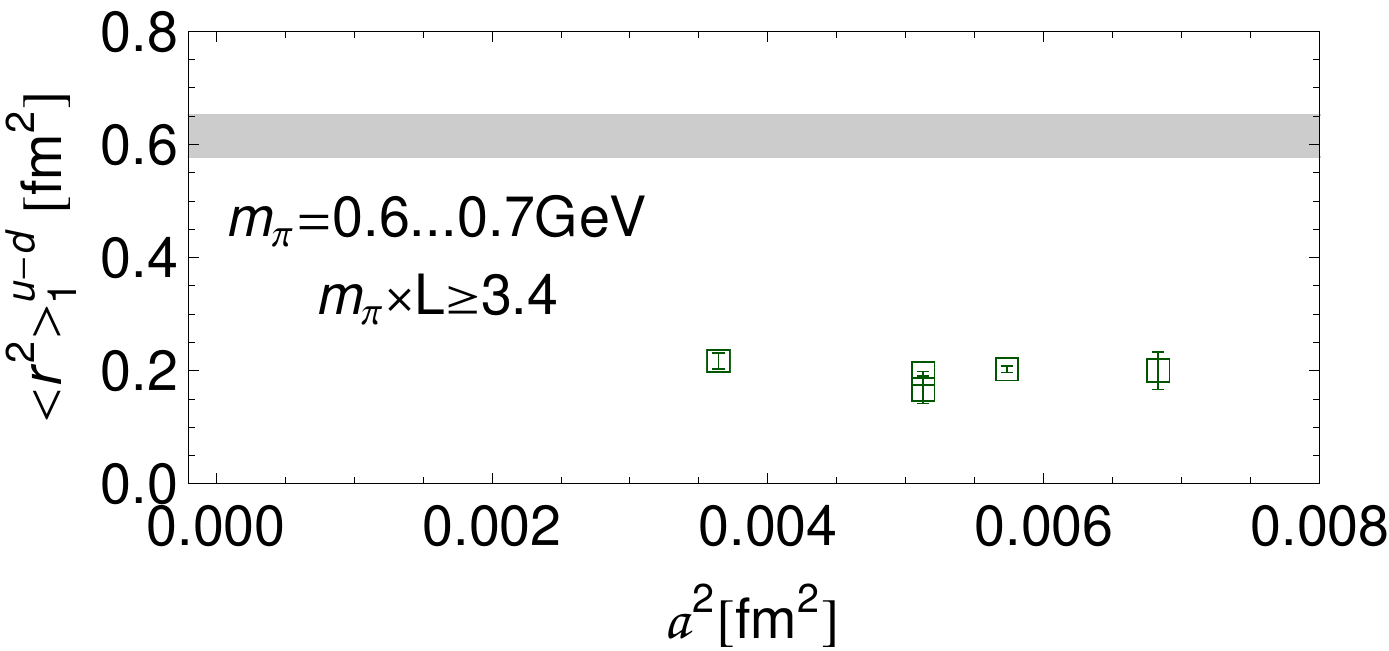}
     \end{minipage} 
         \hspace{0.2cm}
    \begin{minipage}{0.48\textwidth}
      \centering
          \includegraphics[angle=0,width=0.85\textwidth,clip=true,angle=0]{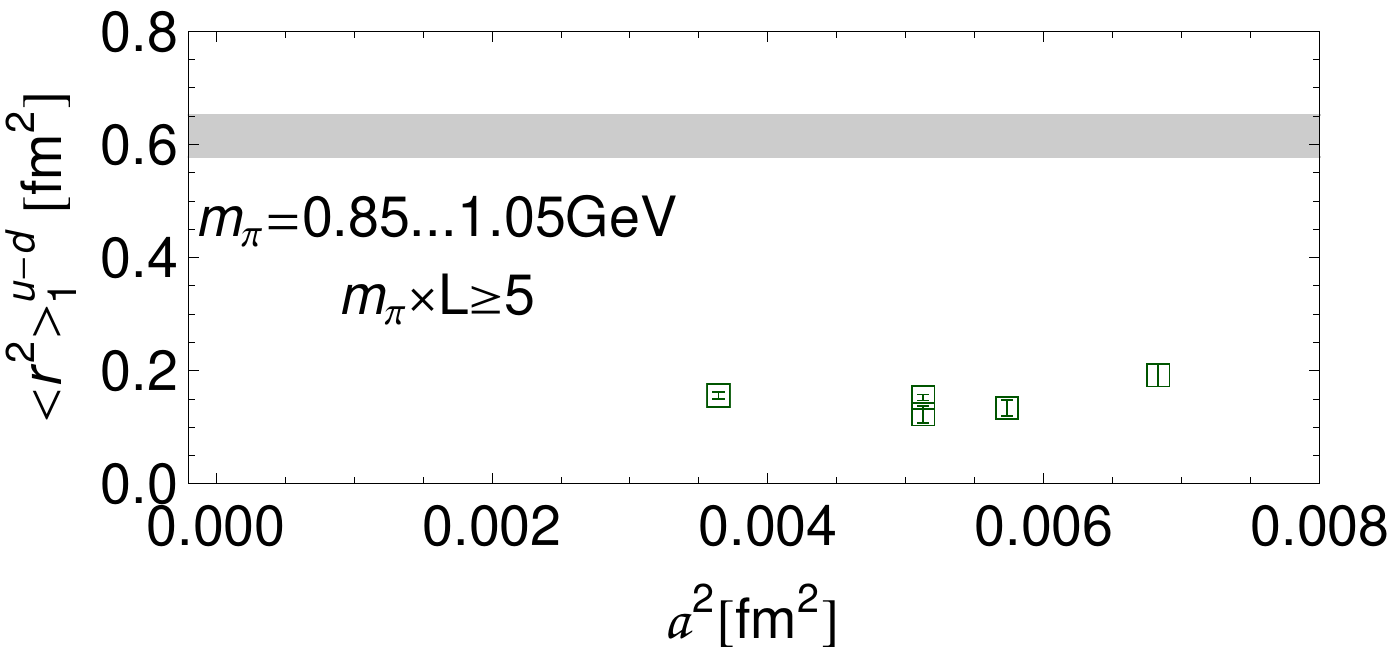}
     \end{minipage}
     \caption{Lattice spacing dependence of the isovector Dirac radius for fixed ranges of $m_\pi$, as obtained from fits to $F^{u-d}_1$ using Eq.~(\ref{eqF1poly}) (cf. Fig.~\ref{F1umdDipolePoly}). The residual pion mass dependence in the given $m_\pi$-windows has been taken into account by restricted linear fits, and the data points have been shifted accordingly to the central $m_\pi$-values.
       The gray bands represent the range of values obtained from experiment and phenomenology at the physical point.}
         \label{r1vsa}
 \end{figure}

Our study of the $a$-dependence of $F_1$ at fixed $m_\pi$ and $Q^2$ above in section \ref{sec:syseff} already indicated
that the discretization errors are small.
Here, we perform a similar analysis for the isovector Dirac radius obtained from the $Q^2$-parametrization
based on Eq.~(\ref{eqF1poly}).
In Fig.~\ref{r1vsa} we show our results for $\langle r^2\rangle^{u-d}_1$ for narrow ranges of $m_\pi$ as a function of $a^2$. 
For direct comparison, we also show an average of the experimental results at the physical point as a gray error band.
The residual pion mass dependence of the lattice data within the $m_\pi$-ranges has been accounted
for by linear fits to the pion mass dependence and subsequent relative shifts of the data points
to the central $m_\pi$-values.
As expected, we find that the results are compatible within statistical uncertainties for the three or four different available values of $a^2$. 
Apart from some small fluctuations, which, however, do not show a systematic trend,
we find that even the central values of the data points are in good agreement.
Overall, in the accessible parameter ranges, and for $a^2\sim0.0035,\ldots,0.007\fm^2$, we therefore
do not observe any significant, systematic lattice spacing dependence of our results.
Most importantly, a naive extrapolation of our data in $a^2$ to the continuum limit would not
bring us any closer to the experimental value indicated by the gray band.

\begin{figure}[t]
    \begin{minipage}{0.48\textwidth}
        \centering
          \includegraphics[angle=0,width=0.85\textwidth,clip=true,angle=0]{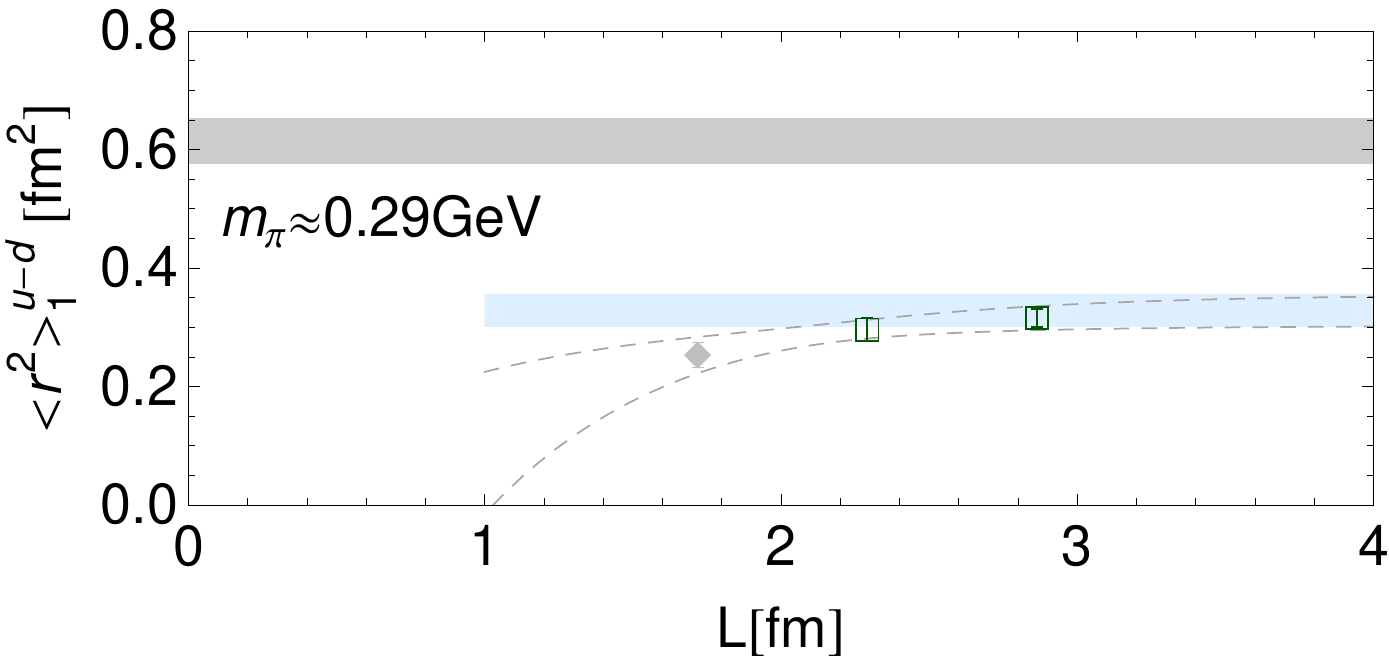}
          \includegraphics[angle=0,width=0.85\textwidth,clip=true,angle=0]{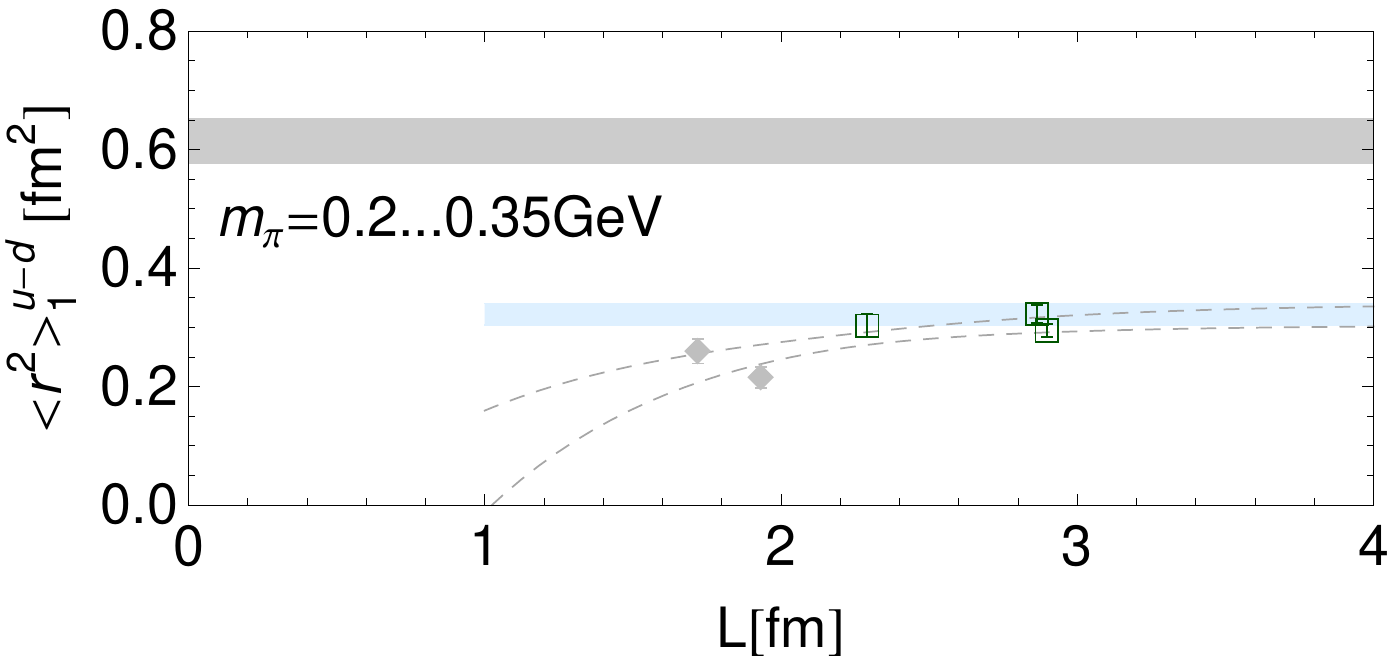}
     \end{minipage} 
         \hspace{-0.8cm}
    \begin{minipage}{0.48\textwidth}
      \centering
          \includegraphics[angle=0,width=0.85\textwidth,clip=true,angle=0]{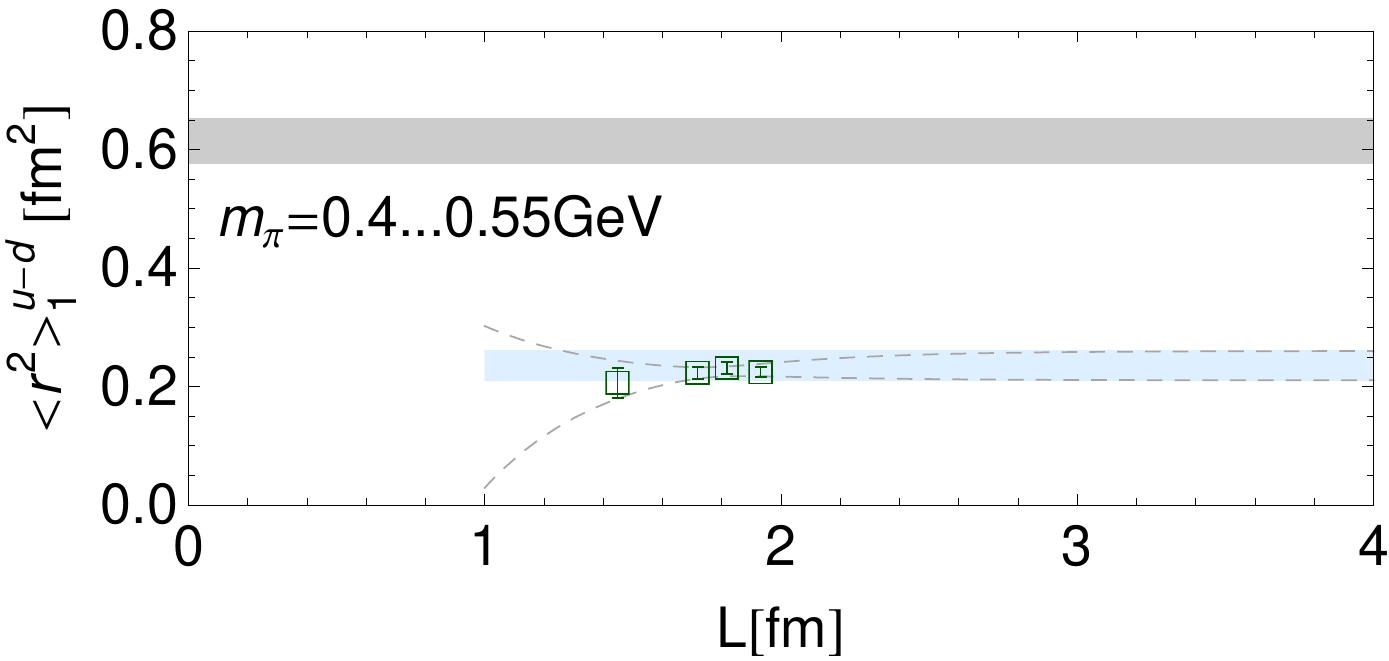}
          \includegraphics[angle=0,width=0.85\textwidth,clip=true,angle=0]{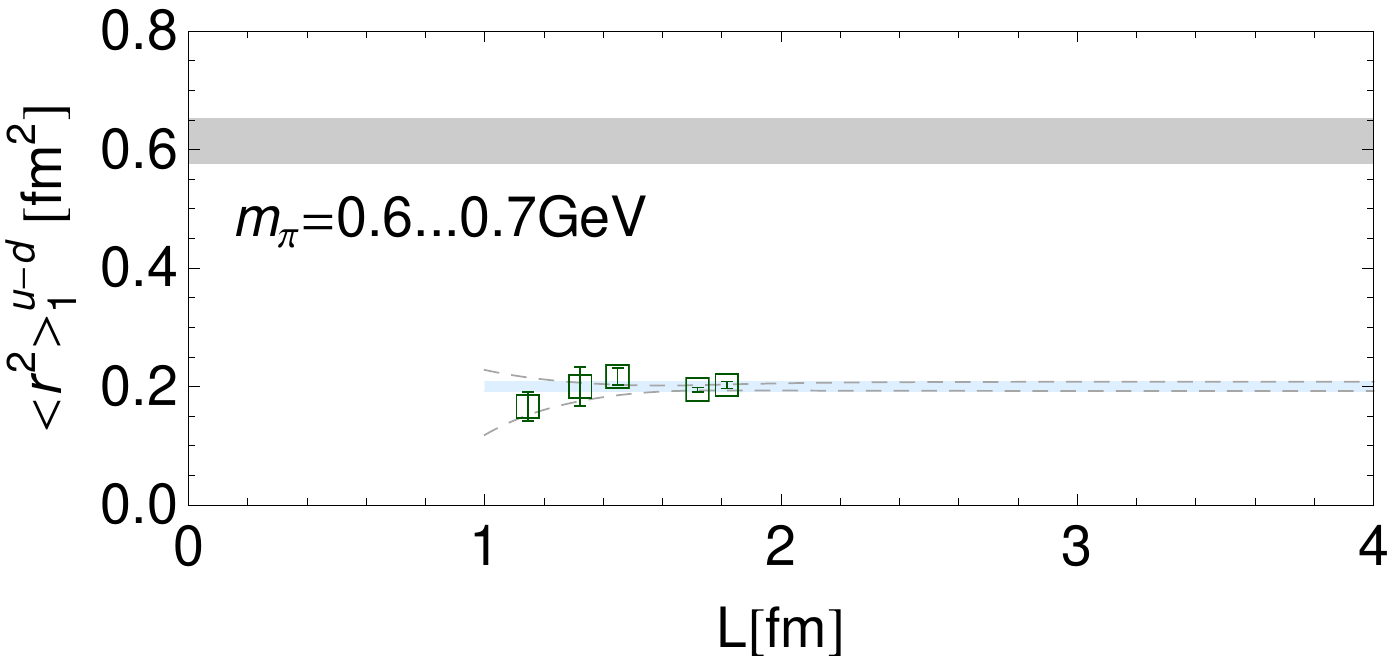}
     \end{minipage}
       \caption{Volume dependence of the isovector Dirac radius for fixed values/ranges of $m_\pi$, as obtained from fits to $F^{u-d}_1$ using Eq.~(\ref{eqF1poly}) (cf. Fig.~\ref{F1umdDipolePoly}). The residual pion mass dependence in the given $m_\pi$-windows has been taken into account by restricted linear fits, and the data points have been shifted accordingly to the central $m_\pi$-values.
       Fits to the $L$-dependence of the data points are indicated by the dashed lines, and the light shaded bands represent 
       the resulting value in the infinite volume limit.
       The upper gray bands indicate the range of values obtained from experiment and phenomenology at the physical point.
      }
         \label{r1vsL}
 \end{figure}

With respect to finite volume effects, we display in Fig.~\ref{r1vsL} our results for $\langle r^2\rangle_1$ as a function
of the box length $L$.
As before, the lattice data points were shifted to the central values of the indicated
narrow ranges in $m_\pi$ employing linear interpolations in order to account for the residual pion mass dependence.
In contrast to the absence of any $a^2$-dependence discussed before, we observe a slight, systematic upwards trend
of the data points as $L$ increases. 
In a first attempt to quantify this observation, we have fitted the $L$-dependence of the data points in each pion mass range 
with a simple exponential ansatz inspired by predictions from chiral perturbation theory: $a+b\exp(-m_\pi L)$.
The results of the fits are indicated by the dashed lines, and the corresponding estimated values in the infinite volume
limit are shown as light shaded bands.
We note that all data points, apart from the ones with $m_\pi\times L<3.4$ (filled diamonds),
show at least a small overlap with the infinite volume band within uncertainties.
The rightmost points at larger volumes are in all cases fully compatible with the estimated results at $L=\infty$
and hence can be regarded as corresponding to the infinite volume limit.

\begin{figure}[t]
    \begin{minipage}{0.48\textwidth}
        \centering
          \includegraphics[angle=0,width=0.85\textwidth,clip=true,angle=0]{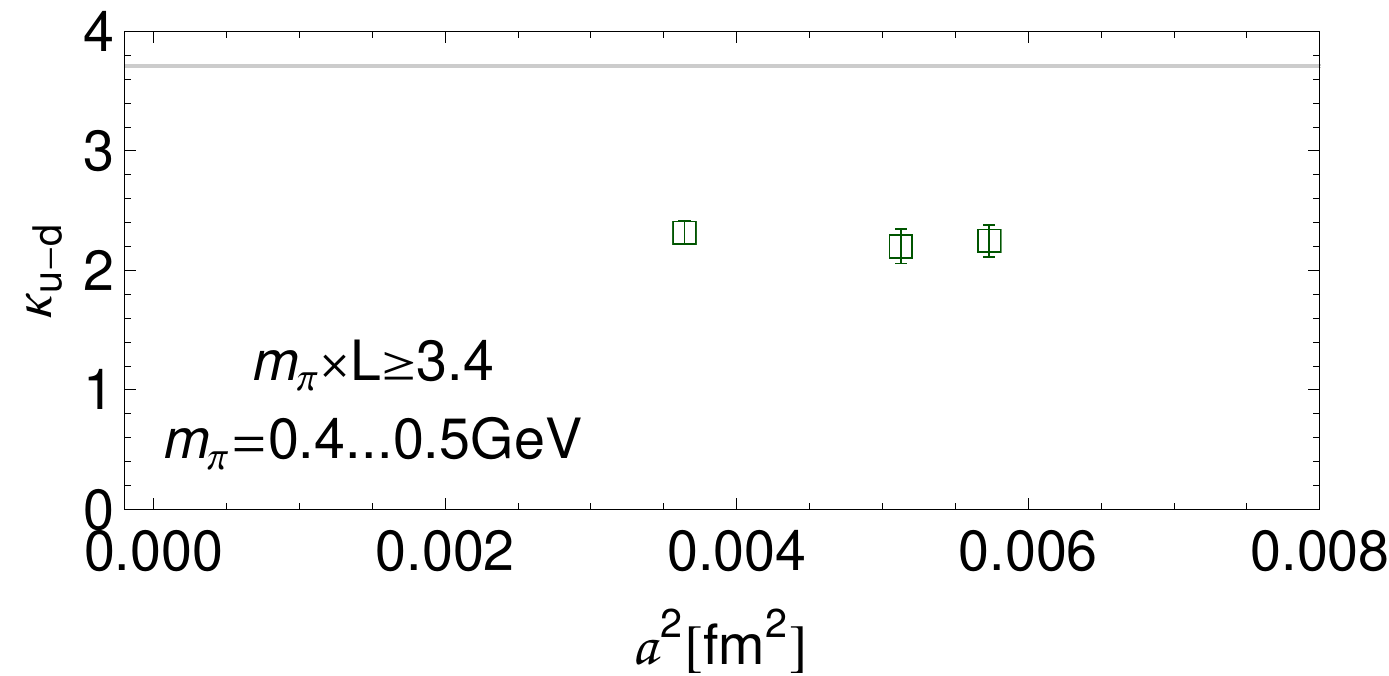}
     \end{minipage} 
         \hspace{0.2cm}
    \begin{minipage}{0.48\textwidth}
      \centering
          \includegraphics[angle=0,width=0.85\textwidth,clip=true,angle=0]{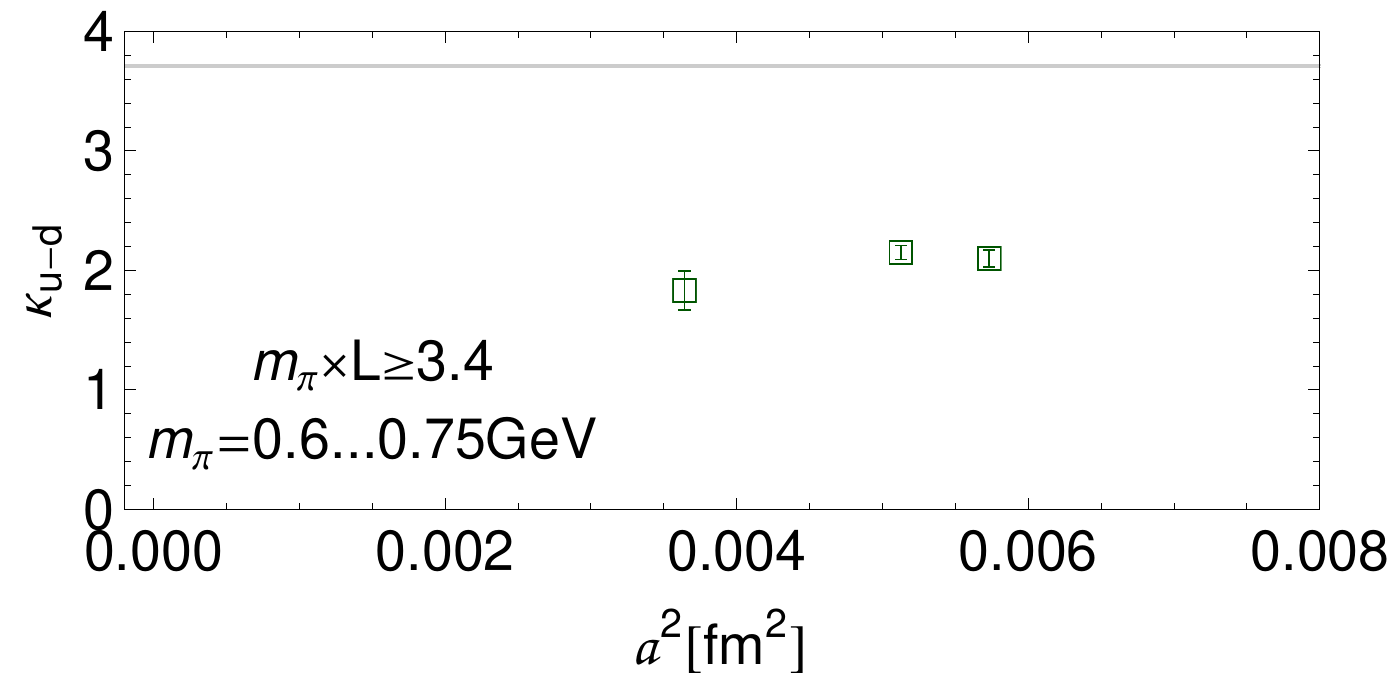}
     \end{minipage}
       \caption{Lattice spacing dependence of the isovector anomalous magnetic moment for fixed ranges of $m_\pi$, as obtained from fits to $F^{u-d}_2$ using Eq.~(\ref{eqF1poly}). The residual pion mass dependence in the given $m_\pi$-windows has been taken into account by restricted linear fits, and the data points have been shifted accordingly to the central $m_\pi$-values.
       The gray bands represent the value from experiment at the physical point.}
         \label{kappaVa}
 \end{figure}

We now turn to systematic uncertainties in the anomalous magnetic moment, 
following the same strategy as outlined above for the case of the isovector Dirac radius.
The $a^2$-dependence of $\kappa_{u-d}$ is shown in Fig.~\ref{kappaVa} for two ranges of $m_\pi$.
Within the uncertainties, the data points do not show any systematic trend as the lattice spacing decreases
and are fully compatible with constants in $a^2$.
As before, a linear extrapolation would not bring us any closer to the experimental value indicated by the thin gray band.

\begin{figure}[t]
    \begin{minipage}{0.48\textwidth}
        \centering
          \includegraphics[angle=0,width=0.85\textwidth,clip=true,angle=0]{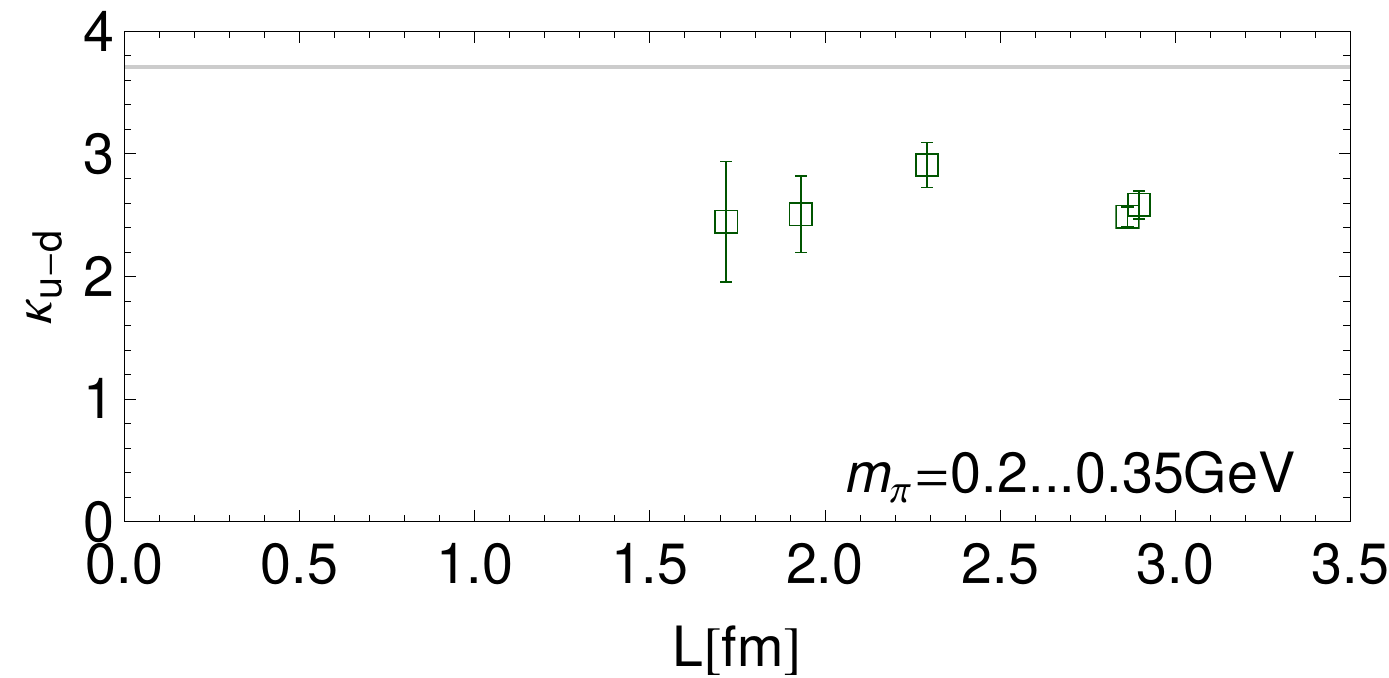}
     \end{minipage} 
         \hspace{0.cm}
    \begin{minipage}{0.48\textwidth}
      \centering
          \includegraphics[angle=0,width=0.85\textwidth,clip=true,angle=0]{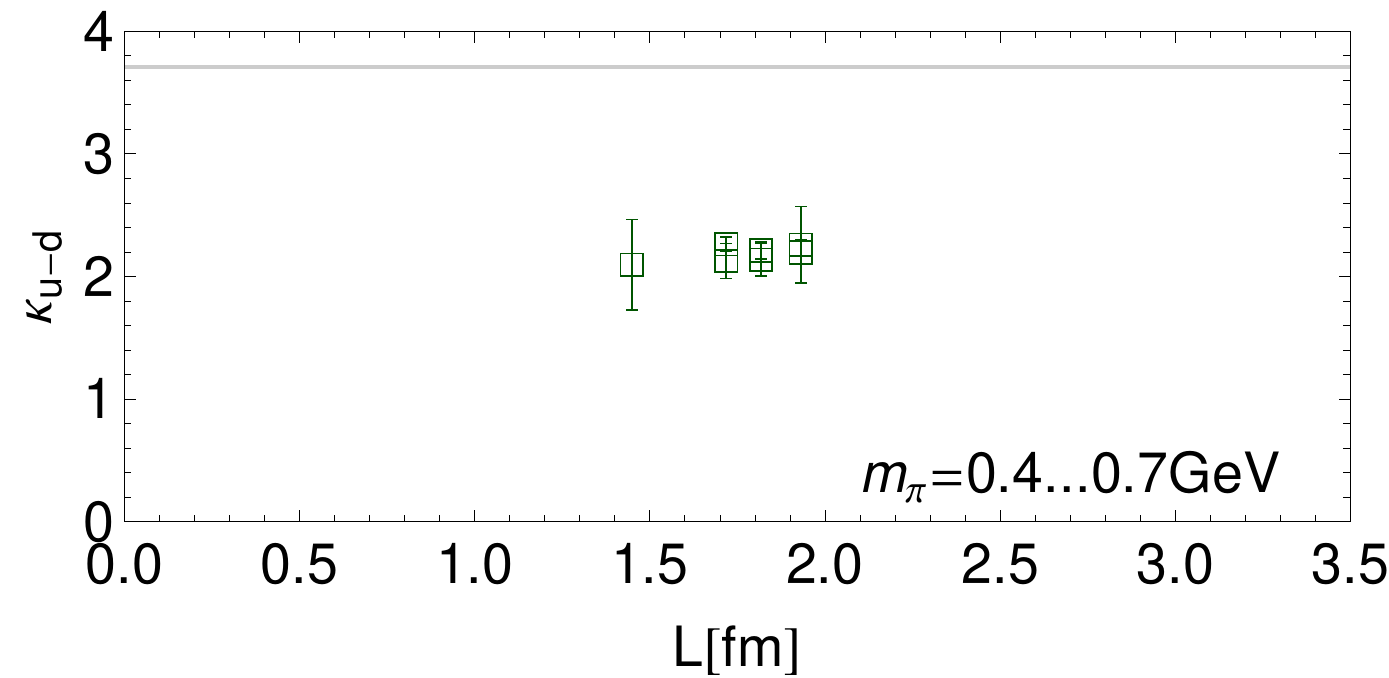}
     \end{minipage}
       \caption{Volume dependence of the isovector anomalous magnetic moment for fixed ranges of $m_\pi$, as obtained from fits to $F^{u-d}_2$ using Eq.~(\ref{eqF1poly}). The residual pion mass dependence in the given $m_\pi$-windows has been taken into account by restricted linear fits, and the data points have been shifted accordingly to the central $m_\pi$-values. The gray bands represent the value from experiment at the physical point.}
         \label{kappaVL}
 \end{figure}

Figure \ref{kappaVL} displays the dependence of $\kappa_{u-d}$ on the box length $L$ for selected ranges of $m_\pi$.
In contrast to $\langle r^2\rangle^{u-d}_1$ in Fig.~\ref{r1vsL}, the data points do not show any clear upward or downward 
trend as $L\rightarrow\infty$.
Since the uncertainties and fluctuations are somewhat larger, we will have to leave a more quantitative estimate 
of finite volume effects in $\kappa_{u-d}$ for future works.

\section{Summary and outlook}

On the basis of an extensive set of ensembles for two flavors of $\mathcal{O}(a)$-improved Wilson fermions and Wilson gluons,
we have computed and studied the Dirac, $F_1(Q^2)$, and Pauli, $F_2(Q^2)$, form factor of the nucleon.
Four different lattice spacings from $a=0.083\fm$ to $a=0.060\fm$,
spatial volumes of $V\sim(1.0,\ldots,3.0\fm)^3$, and a wide range of pion masses extending down to $m_\pi\sim180\MeV$, 
allowed us to investigate in some detail the continuum, infinite volume, and chiral limit.
As in previous studies, 
we do not yet see an overlap or agreement of the lattice data 
with results from experiment and phenomenology 
for the slopes of the isovector Dirac and Pauli form factors, 
nor for the normalization of the latter, i.e. the isovector anomalous magnetic moment $\kappa_{u-d}=F^{u-d}_2(Q^2=0)$,
even at the lowest accessible pion masses of $180\MeV< m_\pi< 300\MeV$.
Our results indicate that these discrepancies cannot be explained
by discretization or finite volume effects.
Contributions from excited states, another source of systematic uncertainties, 
were studied for a single ensemble by varying the sink time of the three point function,
also giving no hint for substantial effects larger than the statistical errors.
Concerning the pion mass dependence, our results for $F_1(Q^2)$ and $F_2(Q^2)$ at fixed values of $Q^2$
look mostly linear in $m_\pi$ or $m_\pi^2$ within the accessible ranges. Linear extrapolations to the
physical pion mass, however, would not lead to an agreement with experiment.

Our data for the ratio $(F^d_1/F^u_1)(Q^2)$, for pion masses below $\sim500\MeV$,
show a reasonable overlap with the phenomenological parametrization over the full range of $Q^2$ we could access.
It is interesting to observe that this ratio drops off by about $50\%$ reaching $Q^2\sim2\GeV^2$,
pointing towards a much narrower spatial distribution of up quarks in the proton than of down quarks.
Furthermore, we find that the Dirac radius of down quarks is systematically larger than for up quarks,
over the full range of available pion masses. 
Concerning the ratio of the Pauli to the Dirac form factor, $(F_2/(\kappa F_1))^{u,d}$, our results
are in general compatible with a rather flat $Q^2$-dependence as observed in experiment, 
although the statistics and the covered $Q^2$-ranges are at this point insufficient to permit a quantitative assessment.
With respect to results in the isosinglet channel or involving individual quark flavors, one has to keep in mind
that quark line disconnected contributions have been neglected.

For inter- and extrapolations in the momentum transfer, and in particular to extract the mean square radii and anomalous magnetic moments,
we have performed and studied different parametrizations of the $Q^2$-dependence of $F_1$ and $F_2$.
For the Dirac form factor, we find that a more flexible (polynomial) 2-parameter ansatz 
provides a numerically and physically much more convincing description compared to the commonly employed dipole fits.
This is borne out by a matching onto a basic vector meson exchange model:
In the case of the polynomial parametrization, we find that the extracted lowest vector meson masses
agree very well with the separately computed lattice vector meson ($\rho$) mass, over a very wide range of
pion masses from $1500\MeV$ down to $260\MeV$.
This indicates that the $Q^2$-dependence of the nucleon form factors on the lattice is 
to a significant extent governed by the exchange of the lowest vector meson resonances, that is the
$\rho$ in the isovector, and the $\omega$ in the isosinglet channel.

With respect to chiral extrapolations using chiral perturbation theory, we followed a somewhat different path than in the past.
Instead of attempting extrapolations of the lattice data down to the physical pion mass, we
investigated the applicability of the different ChPT-schemes by including the known
results from experiment and phenomenology at the physical point, and, only if necessary, 
lattice data for pion masses below $260\MeV$, in the chiral fits. 
Not precisely known low energy constants were varied over sufficiently wide ranges to assess the related uncertainties.
While our data points show for the first time the onset of a non-analytical chiral behavior at the lowest pion masses,
it still turns out to be difficult to achieve a consistent quantitative understanding of the $m_\pi$-dependences
using the different heavy baryon, small scale expansion (explicitly including
the $\Delta$ resonance), and covariant BChPT approaches.
This is in particular the case for the isovector Dirac radius, where traditional 
HBChPT 
predicts a too steep, and a covariant BChPT approach 
a too flat slope as the pion mass increases above $m_\pi^\phys$.
Also the extrapolations of the isovector anomalous magnetic moment and slope of the Pauli form factor are still challenging, 
as they clearly under- or overshoot the lattice data points for pion masses of $\gtrapprox300\MeV$.
In the chiral limit, we obtain a rather large $\kappa_{u-d}^0\sim4.8,\ldots,5.5$, in agreement with previous lattice studies.
For the anomalous magnetic moment in the isosinglet channel, we obtain a reasonable description of the lattice data up to
pion masses of $\sim500\MeV$, within rather broad ChPT extrapolation bands. 
In this case, we find a clearly negative value of $\kappa_{u+d}^0\sim-0.6,\ldots,-0.35$ at $m_\pi=0$.
Apart from $\kappa_{u+d}$, we did not attempt any chiral extrapolations in the isosinglet channel, since
the available 1-loop results from ChPT in this sector are clearly not applicable even at the physical pion mass.
In all considered cases, we find that the heavy-baryon limits of the covariant BChPT extrapolations appear to
break down at or even below $m_\pi^\phys$. 
This casts strong doubts on the applicability of the leading 1-loop heavy-baryon approaches in this region of the pion mass.

Recalling the importance of vector meson exchange contributions for the $Q^2$-dependence of the form factors that we observed before,
it could be interesting to include such contributions explicitly in the ChPT-description not only
of the $Q^2$-, but also the $m_\pi$-dependence of these observables.
Form factor calculations including explicit $\rho$, $\omega$ and $\phi$ resonances have been performed
in covariant BChPT in the so-called EOMS-scheme \cite{Fuchs:2003ir,Fuchs:2003sh,Schindler:2005ke}, leading to an
improved description of the $Q^2$-dependence of the nucleon form factors at the physical pion mass.
It will be interesting to study the applicability of such calculations at larger (lattice) pion masses,
and eventually to compare with the combined $m_\pi$- and $Q^2$-dependence of the lattice data.

Concerning future nucleon form factor studies on the lattice,
our current analysis underlines the importance to obtain results for pion masses 
below $200\MeV$ in sufficiently large volumes of $V\gtrapprox3.5\fm$, which represents a remarkable computational challenge.
Apart from being crucial for the chiral extrapolation and comparison with experiment, 
such calculations will be indispensable for a quantitative understanding of the volume dependence 
at our lowest pion mass of $\sim180\MeV$, where we begin to see finite size effects in, e.g., the data for the Dirac radius. 
Furthermore, our investigation and comparison of different ans\"atze for the $Q^2$-dependences
has shown that precise data points are required over a wide range of the momentum transfer in order
to limit additional parametrization uncertainties, in particular for the
extraction of the anomalous magnetic moments and the radii from the slopes at $Q^2=0$. 
In this respect, (partially) twisted boundary conditions for the quark fields in spatial directions
have already proven to be highly helpful to access very small non-zero values of the momentum transfer
in the case of the pion form factor, see, e.g., \cite{Boyle:2008yd,Frezzotti:2008dr,Nguyen:2011ek}.
First studies along these lines for the nucleon form factors are promising \cite{Gockeler:2008zz} 
and will be continued in the near future.
Regarding higher $Q^2>2\GeV^2$ (involving larger nucleon momenta), it will be important to carefully monitor 
fluctuations in the correlation functions and potential contaminations from excited state contributions.

\appendix

\section{Collection of numerical results}
\label{sec:tables}
Table \ref{tab:NumRes} shows our results for the mean square radii, anomalous magnetic moments, and vector meson masses $M_1^{u-d}$ and $M_1^{u+d}$, for all ensembles specified in Table \ref{tab:params}.
Definitions and details are given in sections \ref{sec:para} and \ref{sec:matching}.

\begin{table}[t]
\begin{tabular}{l|@{}*{8}{c}}
\hline
 \# & $\,\langle r^2\rangle^{u-d}_1[\fm^2]$ & $\langle r^2\rangle^{u+d}_1[\fm^2]$ & $\kappa_{u-d}$ & $\kappa_{u+d}$ & $\langle r^2\rangle^{u-d}_2[\fm^2]$ & 
 %$\langle r^2\rangle^{u+d}_2$ 
 $(\kappa\times\langle r^2\rangle_2)^{u+d}[\fm^2]$ & $M_1^{u-d}[\GeV]$ & $M_1^{u+d}[\GeV]$\\
 \hline  
 1 & \text{0.103(11)} & \text{0.125(8)} & \text{1.461(148)} & \text{-0.121(155)} & \text{0.145(41)} & \text{-0.147(111)} & \text{1.602(209)} &
   \text{1.629(133)} \\
 2 & \text{0.186(20)} & \text{0.207(15)} & \text{2.096(909)} & \text{. . .} & \text{0.320(273)} & \text{. . .} & \text{1.194(163)} &
   \text{1.307(102)} \\
 3 & \text{0.194(33)} & \text{0.236(29)} & \text{2.165(2.120)} & \text{. . .} & \text{. . .} & \text{. . .} & \text{1.135(190)} & \text{1.250(102)}
   \\
 4 & \text{0.100(11)} & \text{0.135(10)} & \text{1.507(273)} & \text{0.406(289)} & \text{0.179(79)} & \text{0.233(262)} & \text{1.542(236)} &
   \text{1.647(158)} \\
 5 & \text{0.127(14)} & \text{0.144(17)} & \text{1.801(1.562)} & \text{. . .} & \text{. . .} & \text{. . .} & \text{1.485(207)} & \text{0.962(119)}
   \\
 6 & \text{0.200(6)} & \text{0.267(5)} & \text{2.107(68)} & \text{-0.131(85)} & \text{0.343(23)} & \text{-0.195(90)} & \text{1.129(30)} &
   \text{1.131(50)} \\
 7 & \text{0.230(10)} & \text{0.322(9)} & \text{2.200(135)} & \text{-0.260(158)} & \text{0.367(45)} & \text{-0.242(179)} & \text{1.020(40)} &
   \text{0.986(65)} \\
 8 & \text{0.083(6)} & \text{0.111(5)} & \text{1.425(138)} & \text{0.164(119)} & \text{0.158(36)} & \text{0.040(67)} & \text{2.109(205)} &
   \text{1.885(155)} \\
 9 & \text{0.125(9)} & \text{0.141(9)} & \text{1.676(347)} & \text{-0.498(306)} & \text{0.215(93)} & \text{-0.401(290)} & \text{1.578(174)} &
   \text{1.428(141)} \\
 10 & \text{0.353(132)} & \text{0.516(173)} & \text{. . .} & \text{. . .} & \text{. . .} & \text{. . .} & \text{0.789(188)} & \text{0.659(191)} \\
 11 & \text{0.128(15)} & \text{0.170(16)} & \text{1.222(501)} & \text{. . .} & \text{. . .} & \text{. . .} & \text{1.692(193)} & \text{1.252(146)}
   \\
 12 & \text{0.160(5)} & \text{0.198(3)} & \text{1.924(49)} & \text{0.080(44)} & \text{0.279(17)} & \text{0.028(43)} & \text{1.325(48)} &
   \text{1.438(40)} \\
 13 & \text{. . .} & \text{. . .} & \text{7.885(3.210)} & \text{. . .} & \text{. . .} & \text{. . .} & \text{1.027(223)} & \text{1.017(165)} \\
 14 & \text{0.160(25)} & \text{0.223(15)} & \text{. . .} & \text{. . .} & \text{. . .} & \text{. . .} & \text{1.257(160)} & \text{1.009(37)} \\
 15 & \text{0.193(4)} & \text{0.259(3)} & \text{2.163(59)} & \text{-0.109(58)} & \text{0.374(21)} & \text{-0.104(63)} & \text{1.146(19)} &
   \text{1.128(27)} \\
 16 & \text{0.237(10)} & \text{0.326(11)} & \text{2.235(143)} & \text{-0.434(171)} & \text{0.381(50)} & \text{-0.457(209)} & \text{1.010(33)} &
   \text{0.972(66)} \\
 17 & \text{0.250(21)} & \text{0.445(19)} & \text{2.396(491)} & \text{. . .} & \text{0.482(148)} & \text{. . .} & \text{1.064(27)} &
   \text{1.008(55)} \\
 18 & \text{0.296(20)} & \text{0.417(14)} & \text{2.877(184)} & \text{0.005(273)} & \text{0.602(84)} & \text{. . .} & \text{0.863(84)} &
   \text{0.828(86)} \\
 19 & \text{0.319(15)} & \text{0.429(7)} & \text{2.466(80)} & \text{-0.400(91)} & \text{0.474(40)} & \text{-0.413(146)} & \text{0.817(30)} &
   \text{0.797(24)} \\
 20 & \text{0.330(60)} & \text{0.435(58)} & \text{3.475(2.301)} & \text{. . .} & \text{. . .} & \text{. . .} & \text{1.030(98)} & \text{0.870(136)}
   \\
 21 & \text{0.112(3)} & \text{0.138(2)} & \text{1.513(40)} & \text{0.039(38)} & \text{0.177(11)} & \text{-0.027(24)} & \text{1.660(99)} &
   \text{1.713(50)} \\
 22 & \text{0.146(6)} & \text{0.183(5)} & \text{1.792(74)} & \text{-0.074(90)} & \text{0.254(25)} & \text{-0.088(74)} & \text{1.337(58)} &
   \text{1.364(96)} \\
 23 & \text{0.169(11)} & \text{0.226(9)} & \text{1.789(164)} & \text{-0.182(200)} & \text{0.230(51)} & \text{-0.197(181)} & \text{1.353(137)} &
   \text{1.397(84)} \\
 24 & \text{0.221(14)} & \text{0.252(10)} & \text{2.031(369)} & \text{. . .} & \text{0.335(121)} & \text{. . .} & \text{1.156(124)} &
   \text{1.105(56)} \\
 25 & \text{0.198(25)} & \text{0.268(16)} & \text{2.098(1.508)} & \text{. . .} & \text{. . .} & \text{. . .} & \text{1.162(179)} & \text{1.009(12)}
   \\
 26 & \text{0.219(9)} & \text{0.314(7)} & \text{2.248(99)} & \text{-0.278(100)} & \text{0.399(38)} & \text{-0.240(125)} & \text{1.069(42)} &
   \text{0.977(37)} \\
 27 & \text{0.215(18)} & \text{0.389(24)} & \text{2.505(311)} & \text{. . .} & \text{0.428(98)} & \text{. . .} & \text{1.011(42)} & \text{0.982(48)}
   \\
 28 & \text{0.299(11)} & \text{0.438(8)} & \text{2.608(115)} & \text{-0.269(133)} & \text{0.548(60)} & \text{-0.146(234)} & \text{0.851(24)} &
   \text{0.766(14)}\\
   \hline
\end{tabular} 
\caption{Results for the mean square radii and anomalous magnetic moments, as well as the vector meson masses $M_1^{u-d}$ and $M_1^{u+d}$
obtained from a matching to the vector meson exchange ansatz, cf. section \ref{sec:matching}. Entries with very large uncertainties
have been replaced by ellipses.
The ensembles $1,\ldots,28$ are specified in Table \ref{tab:params}.
All results are based on the polynomial parametrizations of $F_1(Q^2)$ and $F_2(Q^2)$ discussed in section \ref{sec:para}.}
 \label{tab:NumRes}
\end{table}

\section{Chiral perturbation theory formulae}
\label{app}
Here we provide a collection of (parts of) SSE and covariant BChPT expressions for the mean square radii and the anomalous 
magnetic moments. For the details, we refer to Refs. \cite{Hemmert:2002uh,Gockeler:2003ay,Gail:2007} 
and the sections \ref{sec:chiral}, \ref{sec:kappachiral}
and \ref{sec:r2chiral} above.

\subsection*{Small scale expansion (SSE)}

\bea
 \Kappa_{u-d}(m_\pi)&=&
          - \frac{g_A^2\,m_\pi m_N}{4\pi F_\pi^2}
          +   \frac{2 c_A^2 \deltam m_N}{9\pi^2 F_\pi^2} 
          \Bigg\{\sqrt{1-\frac{m_\pi^2}{\deltam^2}}
          \ln \left(\frac{\deltam}{m_\pi}+\sqrt{\frac{\deltam^2}{m_{\pi}^2}-1}\right)
          \nonumber\\
          &&+ \ln\left(\frac{m_\pi}{2\deltam}\right) \Bigg\}
        +  \frac{4c_A c_V g_A m_N m_\pi^2}{9\pi^2 F_\pi^2} \ln\left(\frac{2\deltam}{\lambda} \right)
        +  \frac{4c_A c_V g_A m_N m_\pi^3}{27\pi F_\pi^2\deltam}
         \nonumber\\
         &&-   \frac{8 c_A c_V g_A \deltam^2 m_N}{27\pi^2 F_\pi^2}
               \Bigg\{\left(1-\frac{m_\pi^2}{\deltam^2}\right)^{3/2} 
               \ln \left(\frac{\deltam}{m_\pi}
               + \sqrt{\frac{\deltam^2}{m_{\pi}^2}-1}\right)
 %        \nonumber\\
         + \left(1-\frac{3m_\pi^2}{2\deltam^2}\right) 
          \ln\left(\frac{m_\pi}{2\deltam}\right) \Bigg\} \,.
\label{kappaSSE2}
\eea

\subsection*{BChPT}

In the following expressions, $m_N^0$ denotes the nucleon mass in the chiral limit, while $m_N$ represents the pion mass dependent nucleon mass, $m_N(m_\pi)$ \cite{Gail:2007}. 

\bea
  B_{c1} & = & -12d_6^r(\lambda\!=\!m_N^0)\,,  \label{eqr1BChPT2a0} \\
  (r^2_1)^{u-d,(3)} &=& -\frac{1}{16\pi^2 f_\pi^2 m_N^4}  
  \Bigg\{ 
  7g_A^2m_N^4 + 2(5g_A^2+1)m_N^4\ln\frac{m_\pi}{{m_N^0}} + m_N^4 \nonumber \\
  &&- 15g_A^2 m_\pi^2 m_N^2 + g_A^2m_\pi^2(15 m_\pi^2 -44 m_N^2)\ln\frac{m_\pi}{m_N} 
  \Bigg\} \nonumber \\
  &&+ \frac{g_A^2 m_\pi}{16\pi^2 f_\pi^2 m_N^4 \sqrt{4m_N^2-m_\pi^2}} \left(15 m_\pi^4 -74m_\pi^2m_N^2+70m_N^4\right)
  \arccos\left(\frac{m_\pi}{2m_N}\right)\,, 
  \label{eqr1BChPT2a}
\eea
\bea  
  (r^2_1)^{u-d,(4)} & = & -\frac{3c_6g_A^2m_\pi^2}
  {16\pi^2f_\pi^2 (m_N^0)^4\sqrt{4(m_N^0)^2-m_\pi^2}} \Bigg\{ m_\pi (m_\pi^2 -
  3(m_N^0)^2) \arccos\left(\frac{m_\pi}{2m^0_N}\right) \nonumber \\
  && +  \sqrt{4(m_N^0)^2-m_\pi^2}\left((m_N^0)^2  +
    ((m_N^0)^2-m_\pi^2)\ln\frac{m_\pi}{m^0_N}\right) \Bigg\}\,.
\label{eqr1BChPT2b}
\eea

\bea
  (\kappa_{u-d})^{(3)} & = &
   \frac{g_A^2m_\pi^2{m_N^0}}{8\pi^2f_\pi^2m_N^3}
   \bigg\{
    (3m_\pi^2 - 7m_N^2) \log\frac{m_\pi}{m_N}  - 3m_N^2 \bigg\} \nonumber\\
  &&- \frac{g_A^2 m_\pi {m_N^0}}{8\pi^2 f_\pi^2 m_N^3\sqrt{4m_N^2-m_\pi^2}} \left[
      3m_\pi^4-13m_N^2m_\pi^2+8m_N^4 \right] 
    \arccos\left (\frac{m_\pi}{2m_N} \right )\,,
    \label{kappaBChPT2} 
\eea
\bea  
  (\kappa_{u-d})^{(4)} & = &  -\frac{m_\pi^2}{32\pi^2f_\pi^2(m_N^0)^2}
     \bigg\{ 4g_A^2 (c_6+1)(m_N^0)^2-g_A^2(5c_6m_\pi^2+28(m_N^0)^2)\log\frac{m_\pi}{{m_N^0}} \nonumber\\
    && + 4(m_N^0)^2(2c_6g_A^2 + 7g_A^2 + c_6 -4c_4{m_N^0})\log\frac{m_\pi}{{m_N^0}}   \bigg\} \nonumber\\
  &&-\frac{g_A^2 c_6m_\pi^3}{32\pi^2f_\pi^2(m_N^0)^2\sqrt{4(m_N^0)^2-m_\pi^2}}(5m_\pi^2-16(m_N^0)^2)
      \arccos\left(\frac{m_\pi}{2{m_N^0}}\right)\,.
     \label{kappaBChPT3}
\eea

\bea
  (\kappa_{u+d})^{(3)} & = &  
  -\frac{9g_A^2 m_\pi^2 {m_N^0}}{8 \pi^2 f_\pi^2 m_N^3}
    \bigg\{
          m_N^2  + (m_N^2 - m_\pi^2) \ln \frac{m_\pi}{m_N} 
          +  \frac{m_\pi ( m_\pi^2 - 3 m_N^2)}{\sqrt{4m_N^2 - m_\pi^2}}
          \arccos\left ( \frac{m_\pi}{2m_N} \right ) 
          \bigg\}\,,
    \label{kappaBChPTupd2a}
\eea
\bea
  (\kappa_{u+d})^{(4)} & = &   \frac{3 g_A^2 m_\pi^2}{32\pi^2 f_\pi^2 (m_N^0)^2} 
  \bigg\{ 12 (m_N^0)^2 
       + \kappa_{u+d}^0\big[3m_\pi^2-4(m_N^0)^2\big] \ln\frac{m_\pi}{{m_N^0}}\nonumber\\
     &&- \kappa_{u+d}^0 \frac{m_\pi(3m_\pi^2 - 8 (m_N^0)^2)}{\sqrt{4(m_N^0)^2 -
    m_\pi^2} }\arccos\left( \frac{m_\pi}{2{m_N^0}} \right) \bigg\}\,.
     \label{kappaBChPTupd2b}
\eea

\bea
  (\kappa r_2^2)^{u-d,(3)} & = & \frac{g_A^2 {m_N^0}}{16\pi^2f_\pi^2m_N^5(m_\pi^2-4m_N^2)} \Bigg\{ -124m_N^6 +
  105m_\pi^2 m_N^4 -18m_\pi^4 m_N^2 \nonumber \\
  && +6(3m_\pi^6-22m_N^2 m_\pi^4 + 44m_N^4 m_\pi^2 -16m_N^6)\ln\frac{m_\pi}{m_N}\Bigg\}  \nonumber \\
  && + \frac{g_A^2 {m_N^0}}{8\pi^2f_\pi^2m_N^5
    m_\pi(4m_N^2 - m_\pi^2)^{3/2}} \Bigg\{ 9m_\pi^8-84m_N^2 m_\pi^6  \nonumber \\ 
    &&   +246 m_N^4 m_\pi^4 -216m_N^6m_\pi^2 + 16m_N^8\Bigg\} \arccos\left(\frac{m_\pi}{2m_N}\right)\,, 
      \label{kappaXr2BChPT2}
\eea
\bea
  (\kappa r_2^2)^{u-d,(4)} & = & -\frac{c_6 g_A^2m_\pi^3}{16\pi^2f_\pi^2(m_N^0)^4(4(m_N^0)^2-m_\pi^2)^{3/2}}\bigg(
    4m_\pi^4 -27 m_\pi^2(m_N^0)^2 +42(m_N^0)^4\bigg)\arccos\left(\frac{m_\pi}{2{m_N^0}}\right)  \nonumber \\
  && +  \frac{1}{16\pi^2f_\pi^2(m_N^0)^4(m_\pi^2-4(m_N^0)^2)} \Bigg\{ 16c_4(m_N^0)^7 + 52g_A^2(m_N^0)^6 \nonumber \\
  && - 4c_4 m_\pi^2 (m_N^0)^5 - 14c_6g_A^2 m_\pi^2 (m_N^0)^4 - 13g_A^2m_\pi^2(m_N^0)^4 \nonumber \\
  &&  +  8(3g_A^2-c_4{m_N^0})(m_\pi^2-4(m_N^0)^2)(m_N^0)^4\ln\frac{m_\pi}{{m_N^0}} +4c_6 g_A^2m_\pi^4(m_N^0)^2 \nonumber \\
  &&  - g_A^2(m_\pi^2-4(m_N^0)^2) (4c_6 m_\pi^4 - 3c_6m_\pi^2(m_N^0)^2+24(m_N^0)^4) \ln\frac{m_\pi}{{m_N^0}}\Bigg\}\,. 
  \label{kappaXr2BChPT3}
\eea

\begin{acknowledgments}
The numerical calculations have been performed
on the APEmille, apeNEXT systems and PAX cluster at NIC/DESY (Zeuthen),
the IBM BlueGene/L at EPCC (Edinburgh),
the IBM BlueGene/P at NIC/JSC (J\"ulich),
the QPACE systems of the SFB TR-55,
the SGI Altix and ICE systems at LRZ (Munich) and HLRN (Berlin/Hannover).
This work was supported in part
by the DFG (SFB TR-55) and
by the European Union (grants 238353, ITN STRONGnet and 227431, HadronPhysics2, and 256594).
SC acknowledges support from the Claussen-Simon-Foundation (Stifterband f\"ur die Deutsche Wissenschaft).
PH acknowledges support by the Heisenberg-programme of the DFG and would like to thank the 
DESY Theory Group for hospitality while this work was being completed.
WS wishes to thank Jiunn-Wei Chen at National Taiwan University and Hai-Yang Cheng and Hsiang-Nan Li 
at Academia Sinica for their hospitality and for valuable physics discussions and suggestions.
JZ is supported by the STFC grant ST/F009658/1.
\end{acknowledgments}

\bibliography{FFs2011}

\end{document}